\pdfoutput=1
\documentclass[oneside]{book}
\usepackage{mathptmx}
\usepackage{helvet}
\usepackage{courier}
\usepackage{type1cm}   
\usepackage{graphicx}
\usepackage{tikz}
\usepackage{pgfplots}
\usetikzlibrary{datavisualization} 
\usepackage{amssymb,amsmath}
\usepackage{array}
\usepackage[bottom]{footmisc}
\usepackage{enumitem}
\usepackage{esint}
\usepackage{rotating}

\newcommand{\rowcellC}[2][c]{\begin{tabular}[#1]{@{}c@{}}#2\end{tabular}}
\newcommand{\rowcellR}[2][c]{\begin{tabular}[#1]{@{}r@{}}#2\end{tabular}}
\newcolumntype{C}[1]{>{\centering\arraybackslash}m{#1}}
\newcolumntype{R}[1]{>{\raggedleft\arraybackslash}m{#1}}

\newcommand{\lan}{\left\langle}
\newcommand{\ran}{\right\rangle}
\newcommand{\bu}{\mathbf{u}}
\newcommand{\bx}{\mathbf{x}}
\newcommand{\wT}{\lan wT \ran}
\newcommand{\T}{\lan T \ran}
\newcommand{\J}{\lan J \ran}
\newcommand{\owT}{\overline{wT}}
\newcommand{\oT}{\overline{T}}
\newcommand{\oJ}{\overline{J}}
\newcommand{\dT}{\delta\hspace{-2pt}\lan T\ran}
\newcommand{\DT}{\delta \oT}
\newcommand{\Ly}{\mathcal L}
\newcommand{\cQ}{\mathcal Q}
\newcommand{\fT}{\mathcal F_T}
\newcommand{\fB}{\mathcal F_B}
\newcommand{\wt}{\widetilde}
\newcommand{\e}{\cdot10^}

\newtheorem{conj}{Conjecture}

\title{Internally heated convection and Rayleigh-B\'enard convection}
\author{David Goluskin}
\date{}

\begin{document}
\maketitle
\frontmatter


\newpage
\tableofcontents

\newpage
\thispagestyle{plain}
\addcontentsline{toc}{chapter}{Preface}
\section*{Preface}

The purpose of this SpringerBrief is to review heat transfer in layers of convective fluid. Six different configurations are considered---three that are versions of Rayleigh-B\'enard (RB) convection, which is driven by differential heating at the boundaries, and three that are driven by uniform internal heating. The essential features of all six models are derived mathematically. The experimental literature is reviewed in depth for the models of internally heated (IH) convection, which are much less studied than their RB counterparts. Experiments on RB convection are treated in less depth, as they have been thoroughly reviewed elsewhere.

Along with placing the various convective models within a conceptual framework that brings out their similarities, we give some minor results not published elsewhere. For instance, a few of the linear instability and energy stability thresholds given in tables \ref{tab: linear all} and \ref{tab: energy all} either have not been reported before or have been reported with less precision. One of the bounds proven in \S\ref{sec: bounds} is also new, as are the visualizations of simulations included as figure~\ref{fig: T examples}.

Chapter \ref{chap: intro} provides background and then defines the six configurations under study, the governing equations of our models, and their basic features. Chapter \ref{chap: stab} presents results that can be derived mathematically from the governing equations: linear and nonlinear stability thresholds of static states, along with proven bounds on mean temperatures and heat fluxes. For the IH cases only, chapter \ref{chap: exp} gives a quantitative survey of heat transport in both laboratory experiments and numerical simulations, followed by suggestions for future work.

The author is grateful to Charles R.\ Doering, Erwin P.\ van der Poel, Jared P.\ Whitehead, and Francis A.\ Kulacki for their many helpful comments on the manuscript. Thanks also to Martin W{\"o}rner for providing his original data and Francis A.\ Kulacki for providing not only his original data but also many hard-to-find references. More general thanks are due to Edward A.\ Spiegel, who taught the author much of what he knows about convection and many other topics.
\\
\\
\noindent David Goluskin\\
\noindent University of Michigan

\mainmatter

\chapter{A family of convective models}
\label{chap: intro}

This review is concerned entirely with convection---fluid motion driven by differential body forces. We focus on several simple configurations that lend themselves to theoretical and experimental study. Convection arises in many contexts and figures prominently in astrophysics, geophysics, and certain engineering applications. Astrophysical occurrences include stellar interiors \cite{Spiegel1971a, Kippenhahn1994, Featherstone2009, Gastine2014} and planetary atmospheres and interiors \cite{Heimpel2005, Kaspi2009, Jones2014, Soderlund2014}, while terrestrial occurrences include the Earth's outer core \cite{Fearn1981, Cardin1994, Aurnou2003, Calkins2012a}, mantle \cite{Schubert2001}, oceans \cite{Stern1975, Schott1999}, and atmosphere \cite{Emanuel1994}. Many of these systems have internal sources or sinks of buoyancy, including the Earth's mantle, the cores of large main-sequence stars, radiating atmospheres, and nearly any engineered system where chemical or nuclear reactions take place in a fluid environment. Among such engineering applications, particular attention has been paid to nuclear accident scenarios in which exothermic nuclear reactions drive convection in molten material \cite{Asfia1996, Nourgaliev1997, Grotzbach1999}.

We speak here in terms of thermal convection, where the body forces are gravitational and depend on the fluid's density, which, in turn, depends on its temperature. Other types of convection are not discussed but are often governed by similar dynamics. In compositional convection, for instance, chemical concentration takes the place of temperature. In electroconvection (e.g.\ \cite{Avsec1939, Storey2007}), electric charge takes the place of temperature, and electrical potential takes the place of gravitational potential.

Convection can be indefinitely sustained in each configuration studied here, and we focus on the time-averaged properties of sustained convection, especially heat transport. Transient phenomena are not addressed. The minimum requirement for ordinary fluid to convect is that warmer (less dense) fluid lie below cooler (more dense) fluid, and that this adverse temperature gradient be sufficiently destabilizing to overcome the viscous forces that damp fluid motion. For the convection to be indefinitely sustained against viscous dissipation, there is an additional requirement: an inexhaustible source of energy that drives the system away from equilibrium by endlessly adding heat somewhere other than the top of the domain or removing heat somewhere other than the bottom. This can be accomplished through the thermal boundary conditions, as when a pot of water is boiled on a stove, or it can be accomplished through internal heat sources or sinks, as when radioactive decay heats the Earth's mantle.

The present chapter lays out the basic features of six convective configurations---three that are driven solely by the boundary conditions, and three that are driven by internal heating. The configurations are defined in \S\ref{sec: configs}, and the Boussinesq equations that are used to model them are given in \S\ref{sec: eq}, followed in \S\ref{sec: nondim} by our chosen nondimensionalization. The basic commonalities and differences of the six configurations are then summarized: \S\ref{sec: static} discusses static states, \S\ref{sec: profiles} gives a qualitative look ahead to the experimental findings that are surveyed in chapter \ref{chap: exp}, and \S\ref{sec: relations} introduces the integral quantities and relations that govern heat transport.

\section{Six configurations}
\label{sec: configs}

The six configurations we study here share the same basic geometry: a horizontal fluid layer of height $d$. In its horizontal dimensions, the layer can be modeled as infinite, periodic, or bounded, though only the last case is realizable in physical experiments. In a theoretical investigation, we can allow the convection to have three-dimensional (3D) freedom, or we can make it two-dimensional (2D) by imposing uniformity in one of the horizontal dimensions.

The six configurations differ only in their top and bottom thermal boundary conditions and in the presence or absence of volumetric heating. Three of the cases are versions of Rayleigh-B\'enard (RB) convection, wherein the flow is driven solely by thermal boundary conditions that cause heat to enter across the bottom boundary and exit across the top one. The other three cases are instances of internally heated (IH) convection. Here, heat is added only by a constant, uniform source, and at least some of it exits across the top boundary. The only thermal boundary conditions we employ are fixed-temperature, fixed-flux, or perfectly insulating. Fixed temperatures model perfectly conductive boundaries, while fixed heat fluxes model boundaries that conduct heat poorly \cite{Sparrow1964}.

\setlength\tabcolsep{12pt}

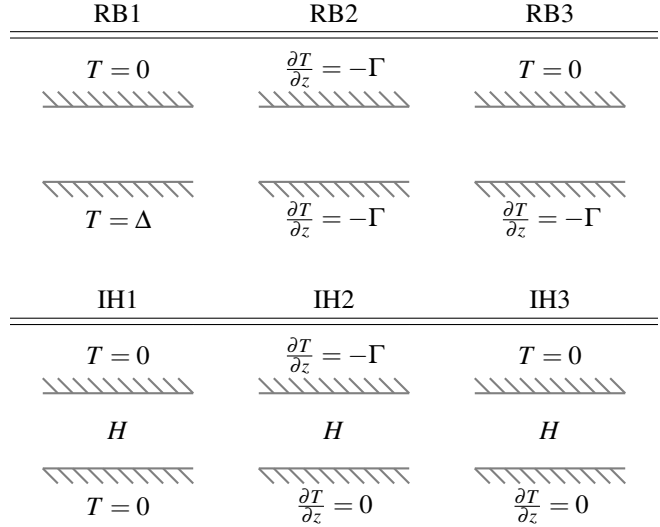
\begin{figure}
\begin{center}
\begin{tabular}{ccc}
RB1 & RB2 & RB3 \\ \hline \hline
\begin{tikzpicture}
\draw[white] (1,-1.4) -- (1,1.4);
\draw[gray,thick] (-1,.5) -- (1,.5);
\draw[gray,thick] (-1,-.5) -- (1,-.5);
\foreach \x in {-1,-.8,...,.8} \draw[gray,thick] (\x,-.5) -- (\x+.2,-.7);
\foreach \x in {-1,-.8,...,.8} \draw[gray,thick] (\x,.7) -- (\x+.2,.5);
\node at (0,1) {$T=0$};
\node at (0,-1) {$T=\Delta$};
\end{tikzpicture}
&
\begin{tikzpicture}
\draw[white] (1,-1.4) -- (1,1.4);
\draw[gray,thick] (-1,.5) -- (1,.5);
\draw[gray,thick] (-1,-.5) -- (1,-.5);
\foreach \x in {-1,-.8,...,.8} \draw[gray,thick] (\x,-.5) -- (\x+.2,-.7);
\foreach \x in {-1,-.8,...,.8} \draw[gray,thick] (\x,.7) -- (\x+.2,.5);
\node at (0,1) {$\frac{\partial T}{\partial z}=-\Gamma$};
\node at (0,-1) {$\frac{\partial T}{\partial z}=-\Gamma$};
\end{tikzpicture}
&
\begin{tikzpicture}
\draw[white] (1,-1.4) -- (1,1.4);
\draw[gray,thick] (-1,.5) -- (1,.5);
\draw[gray,thick] (-1,-.5) -- (1,-.5);
\foreach \x in {-1,-.8,...,.8} \draw[gray,thick] (\x,-.5) -- (\x+.2,-.7);
\foreach \x in {-1,-.8,...,.8} \draw[gray,thick] (\x,.7) -- (\x+.2,.5);
\node at (0,1) {$T=0$};
\node at (0,-1) {$\frac{\partial T}{\partial z}=-\Gamma$};
\end{tikzpicture}
\\[10pt]
IH1 & IH2 & IH3 \\ \hline \hline
\begin{tikzpicture}
\draw[white] (1,-1.4) -- (1,1.4);
\draw[gray,thick] (-1,.5) -- (1,.5);
\draw[gray,thick] (-1,-.5) -- (1,-.5);
\foreach \x in {-1,-.8,...,.8} \draw[gray,thick] (\x,-.5) -- (\x+.2,-.7);
\foreach \x in {-1,-.8,...,.8} \draw[gray,thick] (\x,.7) -- (\x+.2,.5);
\node at (0,1) {$T=0$};
\node at (0,-1) {$T=0$};
\node at (0,0) {$H$};
\end{tikzpicture}
&
\begin{tikzpicture}
\draw[white] (1,-1.4) -- (1,1.4);
\draw[gray,thick] (-1,.5) -- (1,.5);
\draw[gray,thick] (-1,-.5) -- (1,-.5);
\foreach \x in {-1,-.8,...,.8} \draw[gray,thick] (\x,-.5) -- (\x+.2,-.7);
\foreach \x in {-1,-.8,...,.8} \draw[gray,thick] (\x,.7) -- (\x+.2,.5);
\node at (0,1) {$\frac{\partial T}{\partial z}=-\Gamma$};
\node at (0,-1) {$\frac{\partial T}{\partial z}=0$};
\node at (0,0) {$H$};
\end{tikzpicture}
&
\begin{tikzpicture}
\draw[white] (1,-1.4) -- (1,1.4);
\draw[gray,thick] (-1,.5) -- (1,.5);
\draw[gray,thick] (-1,-.5) -- (1,-.5);
\foreach \x in {-1,-.8,...,.8} \draw[gray,thick] (\x,-.5) -- (\x+.2,-.7);
\foreach \x in {-1,-.8,...,.8} \draw[gray,thick] (\x,.7) -- (\x+.2,.5);
\node at (0,1) {$T=0$};
\node at (0,-1) {$\frac{\partial T}{\partial z}=0$};
\node at (0,0) {$H$};
\end{tikzpicture}
\end{tabular}
\end{center}
\caption{Schematics of the six configurations studied in the present work. They are distinguished by their thermal boundary conditions and by the presence or absence of a constant and uniform internal heat source ($H$). Gravity acts vertically downward. All quantities are dimensional; nondimensionalized versions of these schematics appear in tables \ref{tab: main 1} and \ref{tab: main 2}. In the IH2 configuration, $H$ and $\Gamma$ are related such that heat loss balances internal heat production (see text).}
\label{fig: configs}
\end{figure}

\setlength\tabcolsep{2pt}

The thermal boundary conditions of our three RB configurations are shown in the top row of figure \ref{fig: configs}. In the case we call RB1, which is the most-studied RB model, the boundary temperatures are fixed, with a temperature drop of $\Delta$ from the bottom boundary to the top one. (The temperature at the top boundary can be fixed at zero for convenience because, under the governing equations of our models, only temperature \emph{differences} affect the dynamics.) In RB2, the heat flux across each boundary is fixed by setting the temperature gradients there to the same value, $-\Gamma$. The RB3 case mixes the two previous cases, having a fixed-temperature condition on the top boundary and a fixed-flux condition on the bottom one. The configuration where these two boundary conditions are swapped need not be considered since it is related to RB3 by symmetry, at least with the governing equations of our models.

The thermal boundary conditions of our three IH configurations are shown in the bottom row of figure \ref{fig: configs}. Temperature is produced volumetrically at rate $H$ in all three cases, corresponding to heat production at rate $H/c_p$, where $c_p$ is the fluid's heat capacity. In IH1, the top and bottom temperatures are fixed at the same value, and the internally produced heat escapes across both boundaries. In IH2 and IH3, the bottom boundary is perfectly insulating, meaning the vertical temperature gradient must vanish there. The top boundary, across which all the internally produced heat escapes, has a fixed flux in IH2 and a fixed temperature in IH3. In IH2, the boundary flux must match the rate of internal heat production in order for convection to be statistically steady, so $\Gamma$ is determined by $H$, as described in \S\ref{sec: nondim}. Configurations where the boundary conditions in IH2 or IH3 are reversed, making the tops insulating, do not need to be considered; fluid in such configurations would remain static, as follows from the stability results of \S\ref{sec: energy stab}. Notice that the temperature is subject to Dirichlet conditions in RB1 and IH1, Neumann conditions in RB2 and IH2, and one condition of each type in RB3 and IH3.

For the velocity, we impose either no-slip or free-slip boundary conditions. No-slip conditions apply to most laboratory experiments and many engineering applications, while free-slip conditions are more appropriate in modeling certain astrophysical, geophysical, and plasma physical systems. Although we address both possibilities, some results are available only for no-slip boundaries. If side boundaries exist, we assume they are perfect thermal insulators. If side boundaries exist, we assume they are perfect thermal insulators.

The six models of figure \ref{fig: configs} can be extended in various ways. For instance, internal heating can be added to the RB configurations, creating hybrid models driven both by the boundary conditions and by internal heating. The thermal boundary conditions can also be made more complicated, perhaps to model thermal radiation or moderate conductivity. However, such models require at least one more control parameter than those of figure \ref{fig: configs}. When modeled by the Boussinesq equations, which are introduced in the next section, each of the six configurations we study is governed by \emph{only two} dimensionless control parameters, aside from any parameters used to describe the geometry. In a sense described at the end of \S\ref{sec: nondim}, they are the only models of convective layers for which this is true, hence they are a natural starting point.

In many sections of this SpringerBrief results are presented for all six configurations in figure \ref{fig: configs}, letting us highlight their similarities and differences. Section \ref{sec: bounds} and chapter \ref{chap: exp}, however, focus mainly on IH convection. This work is not meant to be a comprehensive review of RB convection, which has been reviewed several times in recent decades \cite{Siggia1994, Getling1998, Ahlers2009, Lohse2010a, Chilla2012}. Interest in RB convection goes well beyond heat transport, as the system has become a canonical model of nonlinear science, having provided early examples of instabilities, bifurcations, pattern formation, and chaos in spatially extended systems. IH convection, which has been the subject of numerous works but is still much less studied than its RB counterpart, was last reviewed in the 1980s \cite{Kulacki1985, Cheung1987}. Bringing this topic up to date requires not only that we review the more recent studies of IH convection, which are relatively few in number, but also that we reinterpret older studies in light of our contemporary understanding of RB convection.

Many of the laboratory experiments reviewed in chapter \ref{chap: exp} are captured well by one of our three IH models, but applications are rife with further complications, including compressibility, temperature-dependent material properties, complicated geometries, chemical and nuclear reactions, rotation, magnetism, and other thermal boundary conditions. Nonetheless, characterizing heat transport in the relatively simple models we consider is already a formidable challenge, and the task is far from complete.

\section{Boussinesq equations}
\label{sec: eq}

A mathematical model of thermal convection must include equations governing the velocity and temperature fields, along with a constitutive relation between temperature and density. The compressible Navier-Stokes equations \cite{Thompson1972} describe the velocity field accurately in a wide range of physical applications, but they are challenging equations to study analytically or integrate numerically, and they can be avoided when the density field does not deviate too strongly from hydrostatic equilibrium. Pressure-driven flows with weak density variations can simply be approximated as having constant densities, yielding the incompressible Navier-Stokes equations. In buoyancy-driven flows, however, density variations cannot be totally ignored because they create the buoyancy gradients needed to drive motion. The typical compromise is to employ the Boussinesq approximation, as we do here.

The Boussinesq approximation, which was first invoked by Oberbeck \cite{Oberbeck1879} and is also called the Oberbeck-Boussinesq approximation, involves two main assumptions. First, the fluid's density, $\rho$, is assumed to vary linearly with temperature, $T$, about some hydrostatic reference state denoted by $\rho_*$ and $T_*$. That is,
\begin{equation}
\rho(T) = \rho_*\big[1 - \alpha (T-T_*)\big], \label{eq: density}
\end{equation}
where $\alpha$ is the linear coefficient of thermal expansion. Second, the density variations are assumed to be sufficiently weak that they can be ignored everywhere except in the buoyancy force. The fluid is sometimes called incompressible since the velocity field is divergence-free, although compressibility does manifest in the buoyancy variations. Numerous justifications have been put forth for replacing the fully compressible Navier-Stokes equations with the simpler Boussinesq equations, typically invoking some combination of asymptotic expansions and \emph{ad hoc} assumptions. See Spiegel and Veronis \cite{Spiegel1962} for one physical justification and Rajagopal \emph{et al.}\ \cite{Rajagopal1996} for a discussion of various other justifications. The precise assumptions invoked vary, but in all versions there is a sense in which gradients of the fluid's properties should not be too steep. If the Boussinesq approximation is used in modeling a physical system, the assumptions under which the approximation holds should be checked if possible, either by physical measurement or by numerical simulation of compressible equations.

With constant gravitational acceleration $g$ acting in the $-\mathbf{\hat z}$ direction, applying the Boussinesq approximation to the compressible Navier-Stokes equations yields the Boussinesq equations \cite{Rayleigh1916, Chandrasekhar1981},
\begin{align}
\nabla \cdot \bu &=  0 \label{eq: inc dim}\\
\partial_t \bu + \bu \cdot \nabla \bu  &= 
	-\tfrac{1}{\rho_*}\nabla p + \nu \nabla^2 \bu + g \alpha T \mathbf{\hat z} \label{eq: u dim} \\
\partial_t T + \bu \cdot \nabla T& = \kappa \nabla^2 T + H, \label{eq: T dim}
\end{align}
where $\mathbf u=(u,v,w)$ is the fluid's velocity vector, $p$ its pressure, $\nu$ its kinematic viscosity, and $\kappa$ its thermal diffusivity. The temperature source term, $H$, is absent from RB convection but drives the convection in our IH models. The pressure term in (\ref{eq: u dim}) has absorbed hydrostatic terms of the buoyancy force coming from (\ref{eq: density}).

\section{Nondimensionalization}
\label{sec: nondim}

To nondimensionalize the Boussinesq equations, we scale distance by the layer height, $d$, time by the characteristic timescale of thermal diffusion, $d^2/\kappa$, and pressure by $\rho_*d^2/\kappa$. We scale temperature by a dimensional quantity, $\Delta$, that is defined differently in various configurations. In the RB1 case, $\Delta$ is the prescribed temperature difference between the boundaries. In the other cases,
\begin{equation}
\Delta := \begin{cases}
d\Gamma & \text{RB2, RB3, IH2} \\
\frac{d^2H}{\kappa} & \text{IH1, IH2, IH3.}
\end{cases}
\label{eq: Delta}
\end{equation}
Nondimensionalized by these $\Delta$, the temperature difference between the boundaries is unity in RB1; the fixed temperature fluxes are unity in RB2, RB3, and IH2; and the volumetric heating rate is unity in the IH cases. Both definitions in (\ref{eq: Delta}) apply to IH2 because we add the consistency condition $\Gamma=dH/\kappa$ in that case to ensure that heat production balances heat loss.

The Boussinesq equations (\ref{eq: inc dim})-(\ref{eq: T dim}) in dimensionless form are
\begin{align}
\nabla \cdot \mathbf u &= 0 \label{eq: inc} \\
\partial_t \mathbf u + \mathbf u \cdot \nabla \mathbf u  &= 
	-\nabla p + Pr \nabla^2 \mathbf u + Pr\,R\,T \mathbf{\hat z} \label{eq: u} \\
\partial_t T + \mathbf u \cdot \nabla T& = \nabla^2 T + Q, \label{eq: T}
\end{align}
where the symbols $\bu$, $T$, $p$, $\bx$, and $t$ henceforth represent dimensionless quantities. The vertical extent is $0\le z\le1$, and the temperature source term is
\begin{equation}
Q = \begin{cases}
0 & \text{RB} \\
1 & \text{IH.}
\end{cases}
\end{equation}
The dimensionless control parameters are the Rayleigh number, $R$, and Prandtl number, $Pr$, defined by
\begin{align}
R :=& \frac{g\alpha d^3\Delta}{\kappa\nu} \label{eq: R} \\
Pr :=& \frac{\nu}{\kappa}. \label{eq: sigma}
\end{align}
The definition of $R$ differs between cases when the definition (\ref{eq: Delta}) of $\Delta$ differs. 

The Rayleigh number may be thought of as the ratio of inertial forces to viscous forces. When it is large, the fluid is strongly driven by differential buoyancy forces. We regard $R$ as the primary control parameter since raising it typically makes the flow more complex. For $R$ to indeed be a control parameter, we needed to define the dimensional temperature scale, $\Delta$, in terms of quantities that are known \emph{a~priori}: the boundary conditions or heating rate. However, it is sometimes useful to define a different Rayleigh number using a temperature scale that is determined dynamically by the flow. This sort of Rayleigh number cannot serve as a control parameter but can be a useful diagnostic quantity. Thus, we will sometimes distinguish between the \emph{control} Rayleigh number, $R$, and \emph{diagnostic} Rayleigh numbers, $Ra$ or $\wt{Ra}$, defined in \S\ref{sec: diagnostic Rayleigh numbers}.

The Prandtl number is the rate at which the fluid diffuses momentum, relative to the rate at which it diffuses heat. Unlike the Rayleigh number, it is a material property of the fluid and does not depend on the geometry or boundary conditions. The Prandtl number is large in fluids that damp motion strongly and conduct heat poorly. In the Earth's mantle, for instance, $Pr$ is effectively infinite. The Prandtl number is small in fluids that damp motion weakly and conduct heat well, such as liquid metals and stellar plasmas. Air and water are intermediate examples, having Prandtl number close to 0.7 and 7, respectively, under atmospheric conditions.

In all six configurations of figure \ref{fig: configs}, the dynamics depend on only two control parameters, $Pr$ and $R$, aside from any parameters describing the geometry, such as aspect ratios. This is the minimum number of parameters we can hope for in the study of convection, except in those special cases where $Pr$ can be eliminated because it is effectively infinite. Additional parameters would be needed if the thermal boundary conditions were more complicated \cite{Sparrow1964, Hurle1967}, the internal heating law were more complicated \cite{Galdi1985, Straughan2012a}, or the boundary conditions and internal heating each introduced their own temperature scales \cite{Joseph1966a, Chapman1980a, Houseman1988, Ames1990, Sotin1999, Veltishchev2004, Kolmychkov2013}. We restrict ourselves to models that require only $Pr$ and $R$ since every additional parameter makes it much harder to understand parameter space. In fact, among the ways that uniform heating, fixed-temperature boundaries, and fixed-flux boundaries can be combined, our six configurations (and their symmetry-related siblings) seem to be the only ones governed by so few parameters.

\begin{table}[htbp]
\begin{center}
\begin{tabular}{| R{55pt} || C{91pt} | C{91pt} | C{91pt} | @{}m{0cm}@{}}
\hline
& RB1 & RB2 & RB3 \\ \hline \hline
Configuration
&
\begin{tikzpicture}
\draw[gray,thick] (-1,.5) -- (1,.5);
\draw[gray,thick] (-1,-.5) -- (1,-.5);
\foreach \x in {-1,-.8,...,.8} \draw[gray,thick] (\x,-.5) -- (\x+.2,-.7);
\foreach \x in {-1,-.8,...,.8} \draw[gray,thick] (\x,.7) -- (\x+.2,.5);
\node at (0,1) {$T=0$};
\node at (0,-1) {$T=1$};
\end{tikzpicture}
&
\begin{tikzpicture}
\draw[gray,thick] (-1,.5) -- (1,.5);
\draw[gray,thick] (-1,-.5) -- (1,-.5);
\foreach \x in {-1,-.8,...,.8} \draw[gray,thick] (\x,-.5) -- (\x+.2,-.7);
\foreach \x in {-1,-.8,...,.8} \draw[gray,thick] (\x,.7) -- (\x+.2,.5);
\node at (0,1) {$\frac{\partial T}{\partial z}=-1$};
\node at (0,-1) {$\frac{\partial T}{\partial z}=-1$};
\end{tikzpicture}
&
\begin{tikzpicture}
\draw[gray,thick] (-1,.5) -- (1,.5);
\draw[gray,thick] (-1,-.5) -- (1,-.5);
\foreach \x in {-1,-.8,...,.8} \draw[gray,thick] (\x,-.5) -- (\x+.2,-.7);
\foreach \x in {-1,-.8,...,.8} \draw[gray,thick] (\x,.7) -- (\x+.2,.5);
\node at (0,1) {$T=0$};
\node at (0,-1) {$\frac{\partial T}{\partial z}=-1$};
\end{tikzpicture}
& \\ [45pt]  \hline
\rowcellR{Static \\ temperature \\ profile}
&
\begin{tikzpicture}
\draw[white] (-1.2,1) -- (-1.2,-1.4);
\draw[white] (-1.2,.85) -- (1.2,.85);
\draw[gray,thick] (-1,.8) -- (1,.8);
\draw[gray,thick] (-1,-.8) -- (1,-.8);
\draw[black,dotted] (-.8,-.8) -- (-.8,.8);
\draw[black] (-.8,.8) -- (.8,-.8);
\draw[black,thick,-latex] (-.8,-.8) -- (-.8,-.3);
\draw[black,thick,-latex] (-.8,-.8) -- (-.3,-.8);
\node at (-.4,-.55) {$T$};
\node at (-1,-.4) {$z$};
\draw[decorate,decoration={brace,mirror,raise=.03in}] (-.8,-.8) -- (.8,-.8);
\node at (0,-1.1) {$\scriptstyle 1$};
\end{tikzpicture}
&
\begin{tikzpicture}
\draw[white] (-1.2,1) -- (-1.2,-1.4);
\draw[white] (-1.2,.85) -- (1.2,.85);
\draw[gray,thick] (-1,.8) -- (1,.8);
\draw[gray,thick] (-1,-.8) -- (1,-.8);
\draw[black] (-.8,.8) -- (.8,-.8);
\draw[decorate,decoration={brace,mirror,raise=.03in}] (-.8,-.8) -- (.8,-.8);
\node at (0,-1.1) {$\scriptstyle 1$};
\end{tikzpicture}
&
\begin{tikzpicture}
\draw[white] (-1.2,1) -- (-1.2,-1.4);
\draw[white] (-1.2,.85) -- (1.2,.85);
\draw[gray,thick] (-1,.8) -- (1,.8);
\draw[gray,thick] (-1,-.8) -- (1,-.8);
\draw[black,dotted] (-.8,-.8) -- (-.8,.8);
\draw[black] (-.8,.8) -- (.8,-.8);
\draw[decorate,decoration={brace,mirror,raise=.03in}] (-.8,-.8) -- (.8,-.8);
\node at (0,-1.1) {$\scriptstyle 1$};
\end{tikzpicture}
& \\ [55pt] \hline
\rowcellR{Turbulent \\ temperature \\ profile}
&
\begin{tikzpicture}
\draw[white] (-1.2,.85) -- (1.2,.85);
\draw[gray,thick] (-1,.8) -- (1,.8);
\draw[gray,thick] (-1,-.8) -- (1,-.8);
\draw[black,dotted] (-.8,-.8) -- (-.8,.8);
\draw[black, rounded corners=2pt] (-.8,.8) -- (0,.7) -- (0,-.7) -- (.8,-.8);
\end{tikzpicture}
&
\begin{tikzpicture}
\draw[white] (-1.2,.85) -- (1.2,.85);
\draw[gray,thick] (-1,.8) -- (1,.8);
\draw[gray,thick] (-1,-.8) -- (1,-.8);
\draw[black, rounded corners=2pt] (-.2,.8) -- (0,.6) -- (0,-.6) -- (.2,-.8);
\end{tikzpicture}
&
\begin{tikzpicture}
\draw[white] (-1.2,.85) -- (1.2,.85);
\draw[gray,thick] (-1,.8) -- (1,.8);
\draw[gray,thick] (-1,-.8) -- (1,-.8);
\draw[black,dotted] (-.8,-.8) -- (-.8,.8);
\draw[black, rounded corners=2pt] (-.8,.8) -- (-.6,.6) -- (-.6,-.6) -- (-.4,-.8);
\end{tikzpicture}
& \\ [55pt] \hline
Heat balance & 
\multicolumn{3}{c|}{$\tfrac{d}{dz}\oT\big|_{z=0}=\tfrac{d}{dz}\oT\big|_{z=1}$}
& \\ [16pt] \hline
$\oJ(z)$ &
$1+\wT$ &
\multicolumn{2}{c|}{1}
& \\ [16pt] \hline
$\langle J\rangle$ &
$1+\wT$ &
\multicolumn{2}{c|}{$1$}
& \\ [16pt] \hline
\rowcellR{Uniform\\$\wT$ bounds} & 
$0\le\wT<\infty$ & 
\multicolumn{2}{c|}{$0\le\wT<1$}
& \\ [16pt] \hline
\rowcellR{Uniform\\$\dT$ bounds} & 
$0<\dT<1$ & 
$-\tfrac{1}{\sqrt{3}}\le\dT\le\tfrac{1}{\sqrt{3}}$ &
$0<\dT\le\tfrac{1}{\sqrt{3}}$
& \\ [16pt] \hline
\rowcellR{Empirical\\relation} & 
$\dT\sim\tfrac{1}{2}$ &
\multicolumn{2}{c|}{$\dT\sim\tfrac{1}{2}\left(1-\wT\right)$}
& \\ [16pt] \hline
$N$ & 
$1+\wT$ &
\multicolumn{2}{c|}{$\dfrac{1}{1-\wT}$}
& \\ [16pt] \hline
\end{tabular}
\end{center}
\caption{Summary of the properties of RB convection discussed in the present chapter. All quantities are dimensionless, and the vertical extent is $0\le z\le1$. Notation is defined throughout the chapter. Briefly, $\overline *$ denotes an average over horizontal directions and time, $\lan *\ran$ denotes an average over volume and time, $\dT$ is the mean temperature of the fluid relative to that of the top boundary, and $J=-\partial_zT+wT$ is the sum of the conductive and convective vertical heat fluxes.}
\label{tab: main 1}
\end{table}


\begin{table}[htbp]
\begin{center}
\begin{tabular}{| R{55pt} || C{91pt} | C{91pt} | C{91pt} | @{}m{0cm}@{}}
\hline
& IH1 & IH2 & IH3 \\ \hline \hline
Configuration
&
\begin{tikzpicture}
\draw[white] (1,-1.15) -- (1,1.15);
\draw[gray,thick] (-1,.5) -- (1,.5);
\draw[gray,thick] (-1,-.5) -- (1,-.5);
\foreach \x in {-1,-.8,...,.8} \draw[gray,thick] (\x,-.5) -- (\x+.2,-.7);
\foreach \x in {-1,-.8,...,.8} \draw[gray,thick] (\x,.7) -- (\x+.2,.5);
\node at (0,1) {$T=0$};
\node at (0,-1) {$T=0$};
\node at (0,0) {$Q=1$};
\end{tikzpicture}
&
\begin{tikzpicture}
\draw[gray,thick] (-1,.5) -- (1,.5);
\draw[gray,thick] (-1,-.5) -- (1,-.5);
\foreach \x in {-1,-.8,...,.8} \draw[gray,thick] (\x,-.5) -- (\x+.2,-.7);
\foreach \x in {-1,-.8,...,.8} \draw[gray,thick] (\x,.7) -- (\x+.2,.5);
\node at (0,1) {$\frac{\partial T}{\partial z}=-1$};
\node at (0,-1) {$\frac{\partial T}{\partial z}=0$};
\node at (0,0) {$Q=1$};
\end{tikzpicture}
&
\begin{tikzpicture}
\draw[gray,thick] (-1,.5) -- (1,.5);
\draw[gray,thick] (-1,-.5) -- (1,-.5);
\foreach \x in {-1,-.8,...,.8} \draw[gray,thick] (\x,-.5) -- (\x+.2,-.7);
\foreach \x in {-1,-.8,...,.8} \draw[gray,thick] (\x,.7) -- (\x+.2,.5);
\node at (0,1) {$T=0$};
\node at (0,-1) {$\frac{\partial T}{\partial z}=0$};
\node at (0,0) {$Q=1$};
\end{tikzpicture}
& \\ [45pt]  \hline
\rowcellR{Static \\ temperature \\ profile}
&
\begin{tikzpicture}
\draw[white] (-1.2,1) -- (-1.2,-1.4);
\draw[white] (-1.2,.85) -- (1.2,.85);
\draw[gray,thick] (-1,.8) -- (1,.8);
\draw[gray,thick] (-1,-.8) -- (1,-.8);
\draw[black,dotted] (-.8,-.8) -- (-.8,.8);
\foreach \y in {-.8,-.79,...,.79}
   \draw[black] (-1/.4*\y*\y+.8,\y) -- (-1/.4*\y*\y-1/.4*.02*\y-1/.4*.0001+.8,\y+.01);
\draw[decorate,decoration={brace,mirror,raise=.03in}] (-.8,-.8) -- (.8,-.8);
\node at (0,-1.2) {$\tfrac{1}{8}$};
\draw[black,thick,-latex] (-.8,-.8) -- (-.8,-.3);
\draw[black,thick,-latex] (-.8,-.8) -- (-.3,-.8);
\node at (-.4,-.55) {$T$};
\node at (-1,-.4) {$z$};
\end{tikzpicture}
&
\begin{tikzpicture}
\draw[white] (-1.2,1) -- (-1.2,-1.4);
\draw[white] (-1.2,.85) -- (1.2,.85);
\draw[gray,thick] (-1,.8) -- (1,.8);
\draw[gray,thick] (-1,-.8) -- (1,-.8);
 \foreach \y in {-.8,-.79,...,.79}
    \draw[black] (-1/1.6*\y*\y-\y+.4,\y) -- (-1/1.6*\y*\y-1/1.6*.02*\y-1/1.6*.0001-\y-.01+.4,\y+.01);
\draw[decorate,decoration={brace,mirror,raise=.03in}] (-.8,-.8) -- (.8,-.8);
\node at (0,-1.2) {$\tfrac{1}{2}$};
\end{tikzpicture}
&
\begin{tikzpicture}
\draw[white] (-1.2,1) -- (-1.2,-1.4);
\draw[white] (-1.2,.85) -- (1.2,.85);
\draw[gray,thick] (-1,.8) -- (1,.8);
\draw[gray,thick] (-1,-.8) -- (1,-.8);
\draw[black,dotted] (-.8,-.8) -- (-.8,.8);
\foreach \y in {-.8,-.79,...,.79}
    \draw[black] (-1/1.6*\y*\y-\y+.4,\y) -- (-1/1.6*\y*\y-1/1.6*.02*\y-1/1.6*.0001-\y-.01+.4,\y+.01); 
\draw[decorate,decoration={brace,mirror,raise=.03in}] (-.8,-.8) -- (.8,-.8);
\node at (0,-1.2) {$\tfrac{1}{2}$};
\end{tikzpicture}
& \\ [55pt] \hline
\rowcellR{Turbulent \\ temperature \\ profile}
&
\begin{tikzpicture}
\draw[white] (-1.2,.85) -- (1.2,.85);
\draw[gray,thick] (-1,.8) -- (1,.8);
\draw[gray,thick] (-1,-.8) -- (1,-.8);
\draw[black,dotted] (-.8,-.8) -- (-.8,.8);
\draw[black, rounded corners=2pt] (-.8,.8) -- (-.4,.75) -- (-.4,-.65) -- (-.8,-.8);
\end{tikzpicture}
&
\begin{tikzpicture}
\draw[white] (-1.2,.85) -- (1.2,.85);
\draw[gray,thick] (-1,.8) -- (1,.8);
\draw[gray,thick] (-1,-.8) -- (1,-.8);
\draw[black, rounded corners=2pt] (-.2,.8) -- (.2,.6) -- (.2,-.8);
\end{tikzpicture}
&
\begin{tikzpicture}
\draw[white] (-1.2,.85) -- (1.2,.85);
\draw[gray,thick] (-1,.8) -- (1,.8);
\draw[gray,thick] (-1,-.8) -- (1,-.8);
\draw[black,dotted] (-.8,-.8) -- (-.8,.8);
\draw[black, rounded corners=2pt] (-.8,.8) -- (-.4,.6) -- (-.4,-.8);
\end{tikzpicture}
& \\ [55pt] \hline
Heat balance & 
$\tfrac{d}{dz}\oT\big|_{z=0}-\tfrac{d}{dz}\oT\big|_{z=1}=1$ &
\multicolumn{2}{c|}{$-\tfrac{d}{dz}\oT\big|_{z=1}=1$}
& \\ [16pt] \hline
$\oJ(z)$ &
{\small $\wT+\big(z-\tfrac{1}{2}\big)$} &
\multicolumn{2}{c|}{\small $z$}
& \\ [16pt] \hline
$\langle J\rangle$ &
$\wT$ &
\multicolumn{2}{c|}{$\tfrac{1}{2}$}
& \\ [16pt] \hline
\rowcellR{Uniform\\$\wT$ bounds} & 
$0\le\wT<\tfrac{1}{2}$ & 
$0\le\wT<\tfrac{1}{2}+\tfrac{1}{\sqrt 3}$ & 
$0\le\wT<\tfrac{1}{2}$
& \\ [16pt] \hline
\rowcellR{Uniform\\$\dT$ bounds} & 
$0<\dT\le\tfrac{1}{12}$ & 
\multicolumn{2}{c|}{$0<\dT\le\tfrac{1}{3}$}
& \\ [16pt] \hline
\rowcellR{Empirical\\relation} & 
$\dT\sim\oT_{max}$ &
\multicolumn{2}{c|}{$\dT\sim\tfrac{1}{2}-\wT$}
& \\ [16pt] \hline
$N$ & 
$\dfrac{1}{8\oT_{max}}$ &
\multicolumn{2}{c|}{$\dfrac{1}{1-2\wT}$}
& \\ [16pt] \hline
$\wt N$ & 
$\dfrac{1}{12\,\dT}$ &
\multicolumn{2}{c|}{$\dfrac{1}{3\,\dT}$}
& \\ [16pt] \hline
\end{tabular}
\end{center}
\caption{Summary of the properties of IH convection discussed in the present chapter. All quantities are dimensionless, and the vertical extent is $0\le z\le1$. Notation is defined throughout the chapter. Briefly, $\overline *$ denotes an average over horizontal directions and time, $\lan *\ran$ denotes an average over volume and time, $\dT$ is the mean temperature of the fluid relative to the top boundary, $J=-\partial_zT+wT$ is the sum of conductive and convective vertical heat fluxes, and $\oT_{max}$ is the maximum value of $\oT(z)$.}
\label{tab: main 2}
\end{table}

\section{Static states}
\label{sec: static}

Dimensionless schematics of the RB and IH configurations are shown in the first rows of tables \ref{tab: main 1} and \ref{tab: main 2}, respectively. The basic features of the six cases are summarized in the subsequent rows of both tables and are further laid out in the remainder of this chapter. The simplest solutions to the governing equations are static, with heat transported only by conduction. These are the unique asymptotic states when $R$ is sufficiently small (cf.\ \S\ref{sec: energy stab}), and they solve the Poisson or Laplace equation $\nabla^2T+Q=0$. Since we assume that side boundaries are nonexistent or perfectly insulating, the static temperature fields, $T_{st}$, vary only in $z$:
\begin{equation}
T_{st}(z) = \begin{cases}
1-z & \text{RB1, RB2, RB3} \\
\frac{1}{2}z(1-z) & \text{IH1} \\
\tfrac{1}{2}(1-z^2) & \text{IH2, IH3}.
\end{cases}
\label{eq: T_st}
\end{equation}
These purely conductive profiles are depicted in the second rows of tables \ref{tab: main 1} and \ref{tab: main 2}. They are parabolic with internal heating and linear without it. The static profiles in RB2 and IH2 are determined only up to additive constants, but these constants do not affect the dynamics, so we have fixed them for convenience.

In each configuration, our dimensional temperature scale, $\Delta$, is characteristic of the static state. This is why the dimensionless $T_{st}$ have no dependence on $R$. When we define \emph{diagnostic} Rayleigh number in \S\ref{sec: diagnostic Rayleigh numbers}, we will do so by replacing $\Delta$ with temperature scales characteristic of the convective states, rather than the static ones.

\section{Temperature fields in strong convection}
\label{sec: profiles}

\begin{figure}
\begin{center}
(a)\\ \includegraphics[width=4.55in]{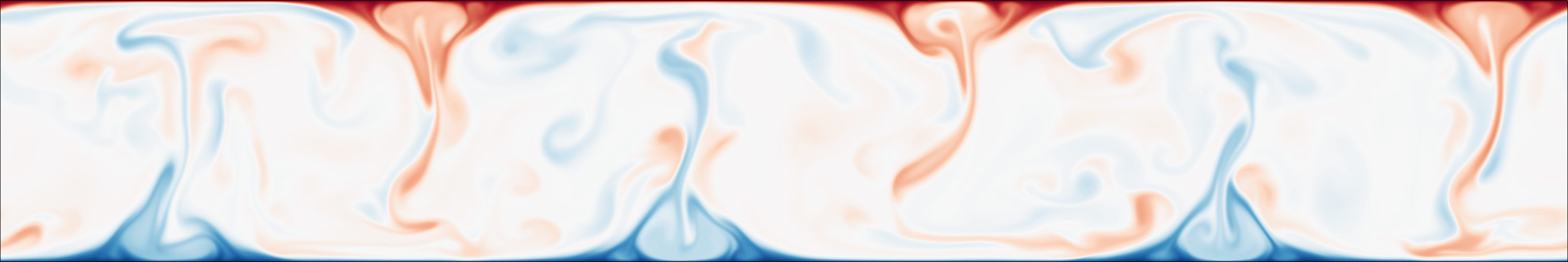} \\[10pt]
(b)\\ \includegraphics[width=4.55in]{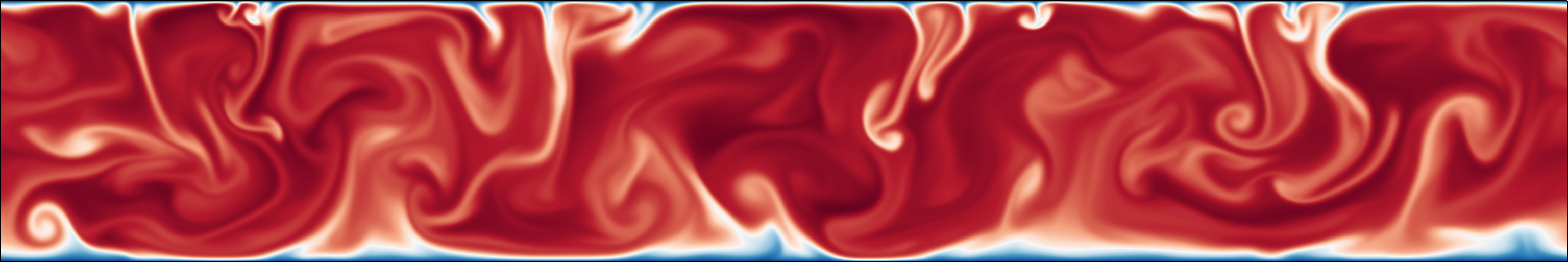} \\[10pt]
(c)\\ \includegraphics[width=4.55in]{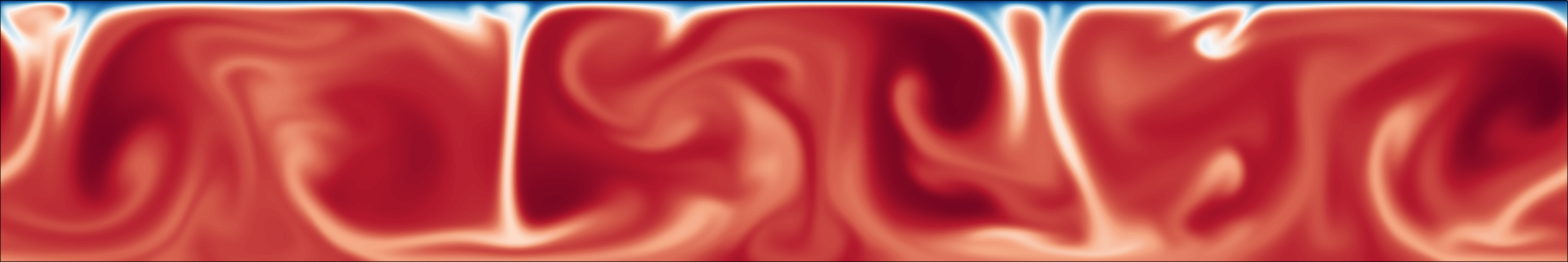}
\end{center}
\caption{Temperature fields from 2D simulations of (a) RB1, (b) IH1, and (c) IH3. Each simulation employed a horizontal period of 6, no-slip boundaries above and below, $Pr=1$, and $R/R_L=10^5$, where $R_L$ is the Rayleigh number at which the static becomes linearly unstable (cf.\ \S\ref{sec: linear stab}). The coolest fluid (blue) has a temperature of zero in each case, and the warmest fluid (red) has a temperature of 1, 0.017, and 0.044, respectively.}
\label{fig: T examples}
\end{figure}

Whereas the fluid remains static when $R$ is sufficiently small, it convects strongly when $R$ is large. Convection typically strengthens monotonically as $R$ is raised, though this is not universally true and can depend on how strength is quantified. (The non-monotonicity of convective transport in \cite{Goluskin2014} provides a counterexample.) Figure \ref{fig: T examples} shows instantaneous temperature fields from 2D simulations at large $R$. The RB1 field in figure \ref{fig: T examples}a is representative of all three RB cases: hot plumes rise from the bottom boundary, cold plumes descend from the top one, and both types of plumes contribute to upward heat transport. In the IH1 field of figure \ref{fig: T examples}b, cold plumes descend from the top, but the bottom boundary layer is cold and \emph{stably} stratified. This bottom layer emits no buoyant plumes, so any mixing between it and interior must be driven by shear, rather than buoyancy. The IH3 field in figure \ref{fig: T examples}c is representative of both IH2 and IH3: cold plumes descend from the top boundary, and there is no thermal boundary layer at the bottom.

The turbulent convection that occurs at large $R$ creates mean vertical temperature profiles very different from the static ones. Rough schematics of such profiles are shown in the third rows of tables \ref{tab: main 1} and \ref{tab: main 2}. Although these schematics include no secondary details, they illustrate the main differences between the various configurations. In each of the six cases, mixing by strong convection tends to flatten the temperature profile in the layer's interior. (We are not aware of a counterexample in 3D, though one exists in 2D under conditions that allow very strong winds to develop \cite{Goluskin2014, VanderPoel2014a}.) The roughly isothermal interiors are flanked by one or two thermal boundary layers, and these are what distinguish the various cases.

In all six of our models, temperature is unstably stratified at the top boundary. At the bottom boundary, the temperature is unstably stratified in the RB cases, stably stratified in IH1, and unstratified in IH2 and IH3. These facts are evident in the static temperature profiles (cf.\ tables \ref{tab: main 1} and \ref{tab: main 2}) and remain provably true of sustained convection, at least in a time-averaged sense. As convective heat transport rises, the mean temperature profiles undergo various changes. In RB1, where the temperature difference between the boundaries cannot change, the boundary layers steepen as the heat flux through the domain rises. In the other two RB cases, where the mean flux through the domain cannot change, the temperature difference between the boundaries decreases. In the IH cases, the temperature of the fluid, relative to that of the top boundary, drops as the convection strengthens. The produced heat leaves only across the top boundary in IH2 and IH3. It leaves across both boundaries in IH1, but the majority leaves across the top boundary, hence the top boundary layer is steeper than the bottom one.

Although the turbulent temperature profiles are fairly easy to understand qualitatively, it is very difficult in general to anticipate their quantitative features. This would be tantamount to accomplishing our main objective of characterizing the bulk heat transport.

\section{Mean heat fluxes and integral relations}
\label{sec: relations}

Heat in a convecting fluid is transported by two mechanisms simultaneously: conduction and convection. Conduction refers to the diffusion of heat down the temperature gradient, while convection refers to the advection of heat by fluid motion. The temperature equation (\ref{eq: T}) can be written in the standard form of a conservation law as $\partial_t T + \nabla\cdot\mathbf J=Q$, where $\mathbf J := \bu T - \nabla T$ is the total \emph{heat current} at a point. Evidently, the conductive current is $-\nabla T$ in our nondimensionalization, and the convective current is $\bu T$. The horizontal components of $\mathbf J$ vanish when averaged over the volume since our side boundaries are insulting or nonexistent. Here we focus on the heat current's vertical component, $J$, where
\begin{align}
J &:= J_{cond} + J_{conv} \\
&:= -\partial_zT + wT. \label{eq: J def}
\end{align}
Much of our effort is devoted to quantifying the relative contributions to vertical heat transport made by conduction and convection---that is, by $-\partial_zT$ and $wT$.

In our notation, an overbar, as in $\overline f$, denotes an average over horizontal surfaces and infinite time. Angular brackets, as in $\langle f\rangle$, denote an average over the entire volume and infinite time. When the dimensionless domain is bounded horizontally by $0\le x\le L_x$ and $0\le y\le L_y$,
\begin{align}
\overline f(z) &:= 
	\lim_{\tau\to\infty}\frac{1}{\tau}\frac{1}{L_xL_y}\int_0^\tau dt
	\int_0^{L_y}dy\int_0^{L_x}dx\,f(\bx,t), \label{eq: hor avg} \\
\lan f\ran &:= 
	\lim_{\tau\to\infty}\frac{1}{\tau}\frac{1}{L_xL_y}\int_0^\tau dt
	\int_0^1dz\int_0^{L_y}dy\int_0^{L_x}dx\,f(\bx,t). \label{eq: vol avg}
\end{align}
The above limits can be replaced with $\liminf$ or $\limsup$ to ensure their existence. For simplicity, we assume in our calculations that infinite-time averages commute with vertical averages and that horizontal averages commute with vertical derivatives, though these assumption can often be avoided. In the above notation, the mean heat flux across a horizontal surface is
\begin{equation}
\oJ(z) = -\oT '(z) + \owT(z), \label{eq: hor mean J}
\end{equation}
where the prime indicates differentiation in $z$. The mean vertical flux across the entire layer is
\begin{equation}
\langle J \rangle = \DT + \langle wT \rangle, \label{eq: mean J}
\end{equation}
where the mean conductive flux,
\begin{equation}
\DT:=\oT_B-\oT_T,
\label{eq: DT}
\end{equation}
is the difference between the mean bottom temperature, $\oT_B$, and mean top temperature, $\oT_T$. Expressions (\ref{eq: hor mean J}) and (\ref{eq: mean J}) are simply averages of the definition (\ref{eq: J def}) for $J$; configuration-specific constraints on $\oJ(z)$ and $\langle J \rangle$ are given in \S\ref{sec: additional constraints}.

\subsection{Heat balances}
\label{sec: heat balances}

Conservation of heat energy is expressed in the various cases by the heat balances
\begin{align}
\oT_T'&=\oT_B' && \text{RB1, RB2, RB3} \\
-\overline{T}_T'+\overline{T}_B'&=1 && \text{IH1} \label{eq: balance IH1} \\
-\oT_T'&=1 && \text{IH2, IH3.}
\end{align}
These balances, which are shown also in the fourth rows of tables \ref{tab: main 1} and \ref{tab: main 2}, are derived by averaging the temperature equation (\ref{eq: T}) over volume and time. Time derivatives vanish from such averages because the instantaneous volume averages of $\bu$ and $T$ are bounded uniformly in time (cf.\ \S\ref{sec: bounds}). In the RB cases, the balance reflects the fact that mean heat flux into the layer at the bottom ($-\oT_B'$) must equal the mean flux out of the layer at the top ($-\oT_T'$). In IH1, the rate of internal heat production (unity) is balanced by the combined outward fluxes of heat across the top boundary ($-\oT_T'$) and the bottom one ($\oT_B'$). In IH2 and IH3, where the bottom boundary is insulating, the internal production is balanced entirely by the outward flux across the top boundary ($-\oT_T'$).

\subsection{Constraints on net heat fluxes}
\label{sec: additional constraints}

Little can be said \emph{a~priori} about the variation with height of the mean heat flux components, $-\oT'(z)$ and $\owT(z)$, but we can derive constraints on their sum, $\oJ(z)$. Averaging the temperature equation (\ref{eq: T}) horizontally, vertically from 0 to $z$, and temporally gives
\begin{equation}
\oJ(z) := -\oT '(z) + \owT(z) = \begin{cases}
-\oT'_B & \text{RB1} \\
1 & \text{RB2, RB3} \\
-\oT'_B + z & \text{IH1} \\
z & \text{IH2, IH3.}
\end{cases} \label{eq: oJ constraint}
\end{equation}
The net vertical flux is the same at every height in the RB cases and increases linearly with height in the IH cases. In the four cases where a boundary flux is fixed, $\oJ(z)$ is known exactly. In RB1 and IH1, $\oJ(z)$ is determined only up to the mean heat flux at the bottom boundary, $-\oT'_B$. Below we give some alternate expressions for $\oJ(z)$ in these two cases, but they all involve quantities that, like $-\oT_B$, are not known \emph{a~priori}.

The volume-averaged heat flux, $\J$, is also constrained. In the four cases where $\oJ(z)$ is known exactly, $\J$ is found by vertically integrating (\ref{eq: oJ constraint}). In RB1 and IH1, the fixed-temperature boundary conditions ensure that the mean conductive fluxes are $\DT=1$ and $\DT=0$, respectively. The results are
\begin{equation}
\J := \DT + \wT = \begin{cases}
1+\wT & \text{RB1} \\
1 & \text{RB2, RB3} \\
\wT & \text{IH1} \\
\tfrac{1}{2} & \text{IH2, IH3.}
\end{cases} \label{eq: J constraint}
\end{equation}
In all six cases, we would like to know how the control parameters affect the convective flux, $\wT$. In RB1 and IH1, where $\DT$ is fixed, this is equivalent to knowing $\J$. In the other four cases, where $\J$ is fixed, it is equivalent to knowing $\DT$.

In the RB1 and IH1 cases, the expressions (\ref{eq: oJ constraint}) for $\oJ(z)$ can be rewritten by replacing $\oT'_B$ with other integral quantities. Relations between $\oT'_B$ and $\oT'_T$ are provided by the heat balances of \S\ref{sec: heat balances}, while relations between $\oT'_B$ and $\wT$ are found by equating the $\J$ expressions (\ref{eq: J constraint}) with the vertical integrals of the $\oJ(z)$ expressions (\ref{eq: oJ constraint}). The alternate expressions for $\oJ(z)$ found in this way are
\begin{equation}
\oJ(z) = \begin{cases}
\begin{array}{llll}
-\oT'_B 		&~=~ -\oT'_T 			&~=~ 1+\wT 							& \quad\text{RB1} \\[3pt]
-\oT'_B + z 	&~=~ -\oT'_T - (1-z) 	&~=~ \big(z-\tfrac{1}{2}\big) + \wT& \quad\text{IH1.}
\end{array}
\end{cases}
\label{eq: alternate oJ}
\end{equation}
The constraints on $\oJ(z)$ and $\J$, expressed in terms of volume integrals, are summarized in the fifth and six rows of tables \ref{tab: main 1} and \ref{tab: main 2}.

In the IH1 configuration, there is yet another useful way to interpret $\wT$: in terms of the \emph{fractions} of internally produced heat that flow outward across the top and bottom boundaries. Expressions for these fractions, which we call $\fT$ and $\fB$, follow from relations (\ref{eq: balance IH1}) and (\ref{eq: alternate oJ}),
\begin{align}
\fT &= -\oT'_T = \tfrac{1}{2}+\wT \label{eq: fT} \\
\fB &= \hspace{8pt}\oT'_B = \tfrac{1}{2}-\wT. \label{eq: fB}
\end{align}
The top and bottom fractions are both 1/2 in the static state, but convective transport breaks this symmetry.

\subsection{$\wT$ and $\dT$} 
\label{sec: wT and T}

Essential information about heat transport is captured by the volume integrals $\wT$ and $\dT$, where
\begin{equation}
\dT := \lan T-\oT_T\ran
\label{eq: dT}
\end{equation}
is the mean fluid temperature, relative to that of the top boundary. The above definition is needed only for RB2 and IH2, where the top temperature is not fixed. In the other configurations, $\dT\equiv\T$ since we have set $T_T\equiv0$. Neither $\wT$ nor $\dT$ is known \emph{a~priori} when the fluid is flowing. Instead, the quantities must be studied by physical and computational experiments, and they can sometimes be bounded analytically. All bounds and many experimental findings that we review in the following chapters can be stated in terms of $\wT$ or $\dT$. In the literature, however, results on $\wT$ are often stated differently but equivalently in terms of other quantities, including $\DT$, $\oT'_B$, $\oT_T'$, and the Nusselt number $N$ defined in \S\ref{sec: Nusselt numbers} below.

\subsubsection{\label{sec: param-indep} Uniform bounds}

The seventh and eighth rows of tables \ref{tab: main 1} and \ref{tab: main 2} give uniform bounds on $\wT$ and $\dT$---that is, bounds that are independent of $R$ and $Pr$. The bounds are derived in this chapter's appendix. The lower bounds on $\wT$ are tight in all six configurations. The upper bounds are thought to be tight among uniform bounds, except in the IH2 case, where we suspect a uniform upper bound of 1/2. In the IH cases, the upper bounds on $\dT$ are tight, and the lower bounds are thought to be. In the RB cases, on the other hand, it is not clear whether any of the bounds on $\dT$ are tight.

The mean convective flux, $\wT$, saturates its lower bound of zero in each configuration only in the static state. Physically, this is because $\wT$ is proportional to the work exerted by buoyancy, and when motion persists this work must be positive to balance viscous dissipation. Mathematically, the positivity of $\wT$ in sustained convection follows from relation (\ref{eq: u power int}) in the next chapter. The upper bounds on $\wT$ correspond to limits in which convective transport is infinitely stronger than conductive transport. In the RB1 case, where $\DT=1$, this limit is approached when $\wT$ grows without bound. In the four cases where the total heat flux, $\J$, is fixed, this limit is approached as $\wT\to\J$ and $\DT\to0$. The IH1 case is different in that $\DT=0$, so $\wT$ is solely responsible for the mean vertical flux. However, if one thinks of the outward transport of heat across both boundaries, rather than upward transport, then the upper limit $\wT\to1/2$ indeed means that convection fully takes over from conduction. The corresponding limits of the top and bottom flux fractions (\ref{eq: fT})-(\ref{eq: fB}) are $\fT\to1$ and $\fB\to0$.

The mean temperature relative to the that of the top boundary, $\dT$, is bounded above and below in all six cases, but the IH bounds differ in character from the RB1 bounds. The RB1 bounds given in table \ref{tab: main 1} ensure that the mean temperature of the fluid lies between those of the top and bottom boundaries. The same may be true of $\dT$ in RB2 and RB3, but the bounds derived in this chapter's appendix are too crude to show it. In RB convection, the mean fluid temperature is exactly halfway between the boundary temperatures in the static state. The same is often true when the fluid is flowing, at least with symmetric boundary conditions, but it seems no rigorous statements have been proven that reflect this observation. In the IH cases, $\dT$ saturates the upper bounds of table \ref{tab: main 2} only in the static states and is strictly smaller when the fluid is flowing. Its lower bound of zero, much like the uniform upper bounds on $\wT$, corresponds to convection being infinitely stronger than conduction. The $R$-dependent bounds of \S\ref{sec: bounds} show that $\dT$ could approach zero only as $R\to\infty$.

In IH convection, where $R$ is proportional to the dimensional heating rate, $H$, it might seem counterintuitive that raising $R$ tends to decrease $\dT$. However, the \emph{dimensional} mean temperature, $\dT\Delta$, indeed rises with $H$. The dimensionless quantity $\dT$ falls as convection strengthens because it has essentially been normalized by its static value.

\subsubsection{\label{sec: R-dep bounds} $R$-dependent bounds}

Since many of the uniform bounds on $\wT$ and $\dT$ given in tables \ref{tab: main 1} and \ref{tab: main 2} are tight, improving them requires finding bounds that depend explicitly on $R$ or $Pr$. Some $R$-dependent bounds have been proven for the configurations we are considering, as summarized in table \ref{tab: bounds}. Bounds that vary with $Pr$ have been proven recently for RB1 \cite{Choffrut2014} but not yet for other cases, although some bounds have been proven for the infinite-$Pr$ limit that are tighter than the corresponding uniform-in-$Pr$ results, as discussed in \S\ref{sec: bounds}.

\begin{table}
\begin{center}
\begin{tabular}{rcc}
& \rowcellC{$R$-dependent\\bound on $\wT$} &
\rowcellC{$R$-dependent\\bound on $\dT$} \\
\hline
RB1 & $\wT\le cR^{1/2}$\hspace{18pt} & none \\[4pt]
RB2, RB3 & $\wT\le 1-cR^{-1/3}$ & none \\[4pt]
IH1, IH2, IH3 & none & $\dT\ge cR^{-1/3}$
\end{tabular}
\end{center}
\caption{Proven bounds on $\wT$ and $\dT$ that hold at large $R$. The constant $c$ differs between cases. Details and references are given in \S\ref{sec: bounds}.}
\label{tab: bounds}
\end{table}

In RB convection, the $R$-dependent bounds that have been proven are all upper bounds on $\wT$. They approach the uniform upper bounds as $R\to\infty$ but are tighter at all finite $R$. The uniform lower bounds of $0\le\wT$ cannot be improved upon, if they are to hold for all solutions, since they is saturated by the static states. These states are unstable at large $R$, however, and $\wT$ typically grows with $R$ in experiments and simulations. There might exist better lower bounds that hold only for attracting states, rather than all solutions, but we lack the mathematical machinery to prove them.

In IH convection, the $R$-dependent bounds that have been proven are all lower bounds on $\dT$. They approach the uniform lower bounds of zero as $R\to\infty$ but are tighter at all finite $R$. No $R$-dependent upper bounds on $\wT$ have been proven in IH convection, but we argue in \S\ref{sec: conjectures} that such proofs should be possible. On the other hand, the uniform upper bounds on $\dT$ and lower bounds on $\wT$ are saturated by the static states, so any efforts to improve them run into the same obstacle as efforts to improve the lower bounds on $\wT$ in RB convection.

\subsubsection{\label{sec: expected relations} Empirical approximate relations}

In addition to the exact integral relations and bounds discussed above, experiments and simulations suggest some approximate relations for $\dT$ at large $R$. These relations are summarized in the ninth rows of tables \ref{tab: main 1} and \ref{tab: main 2}. Most can be expressed as relations between $\dT$ and $\wT$, at least approximately, but the underlying assumption in IH convection differs from that in RB convection.

In the RB cases, the underlying assumption is that the top and bottom boundary layers are roughly symmetric. This implies that the mean fluid temperature is about halfway between the top and bottom temperatures, as in the schematics of turbulent temperature profiles in table \ref{tab: main 1}. That is,
\begin{equation}
\dT \sim \tfrac{1}{2}\DT = \begin{cases}
\tfrac{1}{2} & \text{RB1} \\
\frac{1}{2}\left(1-\wT\right) & \text{RB2, RB3}
\end{cases}
\label{eq: RB assump}
\end{equation}
for large $R$. These relations are not expected to hold exactly when the top and bottom boundary conditions differ, but they could nonetheless be approached as $R\to\infty$. It might be possible to prove precise versions of the above statements, such as upper and lower bounds on $\dT$ that converge to $\DT$ as $R\to\infty$, but we are not aware of any such results.

In the IH cases, the underlying assumption is that mean temperature profile, $\oT(z)$, at large $R$ is roughly isothermal outside of thin thermal boundary layers, as in the schematics of turbulent temperature profiles in table \ref{tab: main 2}. Experimental support of this assumption is presented in chapter \ref{chap: exp}. Approximate isothermally in the IH cases implies that
\begin{equation}
\dT \sim \begin{cases}
\oT_{max} & \text{IH1} \\
\DT = \tfrac{1}{2}-\wT & \text{IH2, IH3}
\end{cases}
\label{eq: IH assump}
\end{equation}
for large $R$, where $\oT_{max}$ is the maximum mean temperature that $\oT(z)$ attains in the layer. In IH1, the assumption of an isothermal interior does not give a relation between $\dT$ and $\wT$, nor is any simple relation suggested by experiments.

\subsubsection{\label{sec: conjectures} Conjectured upper bounds on $\wT$ in IH convection}

In IH2 and IH3, the empirical observation that $\dT\sim\DT$ at large $R$ suggests that the two quantities might obey similar bounds. Since bounds of the form $\dT\ge cR^{-1/3}$ have been proven, it seems likely that bounds of the form $\DT\ge cR^{-1/3}$ could be proven also. The latter can be alternately stated as upper bounds on $\wT$:

\begin{conj}
In the IH2 and IH3 configurations, there exists a constant $c>0$ such that for all $Pr$ and sufficiently large $R$,
\[ \wT\le\tfrac{1}{2} - cR^{-1/3}. \]
\end{conj}

In IH1, experiments suggest that the growth of $\wT$ with $R$ is similarly bounded above (cf.\ chapter \ref{chap: exp}). However, since no empirical relation between $\dT$ and $\wT$ is apparent, the proven lower bound on $\dT$ does not suggest a form for an upper bound on $\wT$. We speculate that the upper bound should approach the uniform bound of $1/2$ algebraically as $R\to\infty$, but we cannot anticipate the algebraic power:

\begin{conj}
In the IH1 configuration, there exist constants $c>0$ and $\alpha>0$ such that for all $Pr$ and sufficiently large $R$,
\[ \wT\le\tfrac{1}{2} - cR^{-\alpha}. \]
\end{conj}

\subsection{\label{sec: Nusselt numbers} Nusselt numbers}

The relative strengths of convective and conductive heat transport are often quantified using dimensionless Nusselt numbers. In RB convection, the typically used definitions of Nusselt numbers can all be expressed in terms of $\wT$. In IH convection, dimensionless quantities resembling the RB Nusselt numbers can be defined in various ways by invoking $\wT$, $\dT$, or $\oT_{max}$. Here we consider two ways of defining Nusselt-number-like quantities, $N$ and $\wt N$. The quantity $N$ is determined by $\oT_{max}$ in IH1 and by $\wT$ in the other five cases, while the quantity $\wt N$ is determined in the IH cases by $\dT$.

\subsubsection{The Nusselt number $N$}
\label{sec: N def}

The definition of $N$ that we choose is one that has helped reveal parallels between various RB configurations when used in concert with the quantity $Ra$ defined in the next subsection \cite{Otero2002, Verzicco2008, Johnston2009, Wittenberg2010}. In every case other than IH1, our definition of $N$ can be expressed as the ratio of mean total heat flux to mean conductive heat flux, where both quantities are averages over volume and time in the developed flow,
\begin{equation}
N = \frac{\lan J\ran}{\lan J_{cond}\ran} = \frac{\DT+\wT}{\DT}
\qquad \text{RB1, RB2, RB3, IH2, IH3.} \label{eq: N def}
\end{equation}
The above definition would fail for IH1 because its denominator would be zero. Applying the various constraints on $\DT$ and $\wT$ (cf.\ tables \ref{tab: main 1} and \ref{tab: main 2}) to expression (\ref{eq: N def}) and adding an \emph{ad hoc} definition for the IH1 case, we obtain
\begin{equation}
N := \begin{cases}
1 + \wT & \text{RB1} \\
\dfrac{1}{\DT} = \dfrac{1}{1-\wT} & \text{RB2, RB3} \vspace{2pt} \\
\dfrac{1}{8\oT_{max}} & \text{IH1} \vspace{2pt} \\
\dfrac{1}{2\DT} = \dfrac{1}{1-2\wT} & \text{IH2, IH3.}
\end{cases} 
\label{eq: N def 2}
\end{equation}

The rationale for our definition of $N$ in the IH1 case, the only case where heat flows outward across both boundaries, is that we are considering outward heat fluxes instead of upward fluxes. The mean total outward flux is unity since it must balance heat production. To determine the mean outward conduction, we imagine dividing the layer at the height $z^*$ where the temperature profile $\oT(z)$ assumes its maximum value of $\oT_{max}$. The upward conduction above $z^*$ is proportional to $\oT_{max}$, as is the downward conduction below $z^*$. Thus, the ratio of total outward transport to conductive outward transport is inversely proportional to $\oT_{max}$. The $1/8$ factor makes $N$ unity in the static state. Although the analogy between $N$ in IH1 and in the other five cases is not perfect, the experiments reviewed in chapter \ref{chap: exp} reveal similarities in $N$ between all cases. In the IH3 case, the quantity we call $N$ has been considered under various names, perhaps first by Thirlby \cite{Thirlby1970}. In the IH1 case, our definition has apparently not been used, but many authors have considered $\oT_{max}$, as well as the so-called top and bottom Nusselt numbers discussed in \S\ref{sec: IH1 dT}.

In all cases except IH1, our definitions (\ref{eq: N def 2}) of $N$ can be expressed in terms of $\wT$ alone. One might wonder whether $N$ in the IH1 case would be better defined as inversely proportional to $1-2\wT$ instead of to $\oT_{max}$. This would be superficially identical to the IH2 and IH3 definitions of $N$, provided the latter are expressed in terms of $\wT$. However, the IH1 experiments discussed in \S\ref{sec: IH1} confirm that $\oT_{max}$ behaves very much like an inverse Nusselt number, while the quantity $1-2\wT$ does not. As described at the end of \S\ref{sec: additional constraints} above, $\wT$ in IH1 convection instead conveys the asymmetry between upward and downward heat fluxes.

\subsubsection{The Nusselt number $\wt N$}

Since $\dT$ is physically important in IH convection, it is natural to define a Nusselt-number-like quantity that is exactly related to $\dT$, rather than to $\wT$ or $\oT_{max}$. It works well to simply define $\wt N$ as inversely proportional to $\dT$,
\begin{equation}
\wt N: = \begin{cases}
\dfrac{1}{12\dT} & \text{IH1} \vspace{3pt} \\
\dfrac{1}{3\dT} & \text{IH2, IH3.}
\end{cases} 
\label{eq: N tilde def 2}
\end{equation}
In the IH2 and IH3 cases, this definition could be interpreted as
\begin{equation}
\wt N = \frac{\lan zJ\ran}{\lan zJ_{cond}\ran}
\qquad \text{IH2, IH3,} \label{eq: N tilde def}
\end{equation}
which is like the expression (\ref{eq: N def}) for $N$ with averages weighted proportionally to height. We do not define $\wt{N}$ for RB convection, although the mean temperature in those cases merits attention also, as discussed in \S\ref{sec: wT and T}.

\subsubsection{Basic properties}

In almost all cases it has been proven that $N\ge1$ and $\wt N\ge1$, with equality holding only in the static states. These facts follow from the uniform bounds on $\wT$ and $\dT$ discussed in \S\ref{sec: param-indep}. It remains to be proven that $N\ge1$ in the IH1 case, which would be true if $\oT_{max}$ could not exceed its static value of 1/8. In turbulent convection, it is typically expected that $N\to\infty$ and $\wt N\to\infty$ as $R\to\infty$. In the IH cases, this is tantamount to $\oT_{max}\to0$ or $\dT\to0$. Such limiting behavior has not been proven but is supported by the experimental results described in chapter \ref{chap: exp}.

\begin{table}
\begin{center}
\begin{tabular}{rcc}
& \rowcellC{$R$-dependent\\bound on $\wT$} &
\rowcellC{$R$-dependent\\bound on $\dT$} \\
\hline
RB1 & $N \le cR^{1/2}$ & none \\[4pt]
RB2, RB3 & $N\le cR^{1/3}$& none \\[4pt]
IH1, IH2, IH3 & none & $\wt N\le cR^{1/3}$ \\[4pt]
\end{tabular}
\end{center}
\caption{Proven $R$-dependent bounds on $N$ and $\wt N$ that hold at large $R$. The constants $c$ differ between cases. These are re-expressions of the bounds on $\wT$ and $\dT$ shown in table \ref{tab: bounds}.}
\label{tab: N bounds}
\end{table}

Table \ref{tab: N bounds} summarizes the $R$-dependent bounds that are known for $N$ and $\wt N$. These are simply restatements of the bounds on $\wT$ and $\dT$ given above in table \ref{tab: bounds}. In RB convection, the upper bounds on $N$ are equivalent to upper bounds on $\wT$. In IH convection, the upper bounds on $\wt N$ are equivalent to lower bounds on $\dT$. Upper bounds on $N$ have not yet been proven for IH convection. In IH2 and IH3, bounds of the form $N\le cR^{1/3}$ would follow from the upper bounds on $\wT$ that we have conjectured in \S\ref{sec: conjectures}. A bound of the same form for IH1 would require showing that $\oT_{max}$ decays no faster than $R^{-1/3}$. The quantity $\oT_{max}$ seems harder to access mathematically than volume averages like $\dT$ and $\wT$, which arise naturally in integral relations.

\subsection{\label{sec: diagnostic Rayleigh numbers} Diagnostic Rayleigh numbers}

The primary purpose of defining $N$ and $\wt N$ as we have is to identify similarities between the various configurations. The bounds in table \ref{tab: N bounds} suggest that we have almost succeeded, but the RB1 exponent is 1/2, while the others are 1/3. To bring the various cases completely into alignment, we must speak of the dependence of $N$ and $\wt N$ on \emph{diagnostic} Rayleigh numbers, $Ra$ and $\wt{Ra}$, instead of on the control Rayleigh number, $R$. These diagnostic parameters can be written simply as
\begin{align}
Ra: &= \begin{cases}
R & \text{RB1}\\
R/N & \text{RB2, RB3, IH1, IH2, IH3}
\end{cases}
\label{eq: Ra}
\\
\wt{Ra}: &= 
R/\wt N \hspace{18pt} \text{IH1, IH2, IH3.}
\label{eq: wt Ra}
\end{align}
In terms of these variables, the RB bounds in table \ref{tab: N bounds} all take the form $N\le cRa^{1/2}$, and the IH bounds take the form $\wt N\le c\wt{Ra}^{1/2}$. Moreover, the analogies brought out by considering $N$ and $Ra$ (or $\wt N$ and $\wt{Ra}$) are not limited to bounds; similarities emerge also in experimental data \cite{Verzicco2008, Johnston2009} and heuristic scaling arguments \cite{Goluskin2012}.

The different definitions of $R$, $Ra$, and $\wt{Ra}$ can be viewed as differences in the temperature scale used to define a Rayleigh number. The dimensional temperature scales $\Delta$ used to define $R$ in \S\ref{sec: nondim} are characteristic of the static states, whereas $Ra$ and $\wt{Ra}$ effectively replace $\Delta$ with temperature scales of the flowing fluid, $\Delta_{Ra}$ and $\Delta_{\wt{Ra}}$. In IH1, the temperature scale of $Ra$ is the maximum horizontally averaged temperature in the flowing fluid. In the other five cases it is the mean temperature difference between the boundaries in the flowing fluid. That is,
\begin{equation}
\Delta_{Ra} = \begin{cases}
\Delta & \text{RB1} \\
\DT \Delta & \text{RB2, RB3} \\
8\oT_{max}\Delta & \text{IH1} \\
2 \DT \Delta  & \text{IH2, IH3.}
\end{cases}
\end{equation}
Replacing $\Delta$ with $\Delta_{Ra}$ in the definition (\ref{eq: R}) of $R$ leads to the above definition (\ref{eq: Ra}) of $Ra$. In the IH cases, the temperature scale of $\wt{Ra}$ is the volume-averaged temperature of the flowing fluid,
\begin{equation}
\Delta_{\wt{Ra}} = \begin{cases}
12\, \dT \hspace{-1pt}\Delta  & \text{IH1} \\
3\, \dT \hspace{-1pt}\Delta & \text{IH2, IH3.}
\end{cases}
\end{equation}
Replacing $\Delta$ with $\Delta_{\wt{Ra}}$ in the definition (\ref{eq: R}) of $R$ leads to the above definition (\ref{eq: wt Ra}) of $\wt{Ra}$.

\section*{Appendix}
\addcontentsline{toc}{section}{Appendix}

In this appendix we prove various bounds on $\wT$ and $\dT$ that are uniform in the parameters, $R$ and $Pr$. Most of these results are standard, but it is difficult to trace their origins, and we do not try.

\subsection*{Extremum principles}
\addcontentsline{toc}{subsection}{Extremum principles}

In each configuration with $T=0$ on a boundary, there holds a minimum principle giving pointwise, instantaneous lower bounds on $T(\bx,t)$. For simplicity we assume that solutions to the Boussinesq equations exist and remain smooth. If $T(\bx,t)$ ever achieves a local minimum on the interior, then at that point $\bu\cdot\nabla T=0$ and $\nabla^2T\ge0$, and so $\partial_tT\ge0$. Thus, if the interior is initially warmer than the fixed boundary temperature of zero, it remains warmer for all time. Even if part of the interior is initially cooler than the boundary, it will be warmer than the boundary at large times. In the RB1 case, an analogous maximum principle holds relative to the warmer boundary, on which $T=1$. For all $\bx$ on the interior and sufficiently large $t$,
\begin{align}
T(\bx,t)>0 & \qquad \text{RB1, RB3, IH1, IH3} \\
T(\bx,t)<1 & \qquad \text{RB1.} 
\end{align}
In the RB2 and IH2 configurations, where fixed-flux conditions are imposed on both boundaries, neither maximum nor minimum principles hold pointwise.

\subsection*{Mean convective transport}
\addcontentsline{toc}{subsection}{Mean convective transport}

Uniform bounds on the mean convective flux, $\wT$, in our RB and IH configurations are summarized in tables \ref{tab: main 1} and \ref{tab: main 2}. Many of these bounds follow from the power integrals (\ref{eq: u power int})-(\ref{eq: T power int}), which are stated below for convenience.
\begin{align*}
\lan|\nabla\bu|^2\ran &= R\wT \\
\lan|\nabla T|^2\ran &= \begin{cases}
1+\wT & \text{RB1} \\
\DT=1-\wT & \text{RB2, RB3} \\
\dT & \text{IH1, IH2, IH3}.
\end{cases}
\end{align*}
In all six configurations, the $\bu$ power integral implies $\wT\ge0$. Since $\lan|\nabla\bu|^2\ran>0$ if convection persists, $\wT$ saturates its lower bound of zero if and only if the system approaches the static state as $t\to\infty$.

The uniform upper bounds on $\wT$ given in tables \ref{tab: main 1} and \ref{tab: main 2} follow in most cases from lower bounds on $\DT$. In RB1 there is no upper bound on $\wT$. We get $\DT>0$ from the $T$ power integral in RB2 and RB3 and from the minimum principle in IH3. This lower bound on $\DT$ gives $\wT<1$ in RB2 and RB3, where $\wT+\DT=1$, and it gives $\wT<1/2$ in IH3, where $\wT+\DT=1/2$. These upper bounds on $\wT$ are probably approached by certain solutions, including the turbulent attractors, as $R\to\infty$. If so, they are tight among uniform bounds. In IH1, the upper bound $\wT<1/2$ follows from the minimum principle since $1/2-\wT=\oT_B'>0$.

It is likely that $\DT>0$ in IH2 also, but we settle for the cruder estimate $\DT>-1/\sqrt 3$, derived as follows.
\begin{align*}
\left| \delta \oT \right| =& \left| \lan\partial_zT\ran \right| \\
\le& \lan\left|\partial_zT\right|\ran \\
\le& \lan|\partial_zT|^2\ran^{1/2} \\
\le& \lan|\nabla T|^2\ran^{1/2} \\
\le& \dT^{1/2} \\
\le& \tfrac{1}{\sqrt 3}.
\end{align*}
The third line above follows from the Cauchy-Schwarz inequality, the fifth line follows from the $T$ power integral, and the last line follows from the bound $\dT\le1/3$ that is proven below. The inequality is in fact strict since $\dT<1/3$, except in the static state. Since $\wT+\DT=1/2$ in IH2, the lower bound on $\DT$ gives $\wT<1/2+1/\sqrt 3$.

In the IH1 configuration, we have also discussed $\oT_{max}$, the maximum value that $\oT(z)$ attains. For this quantity, the lower bound $\oT_{max}>0$ follows from the minimum principle, and the upper bound $\oT_{max}<1/\sqrt 3$ follows from a calculation very similar to the one in the previous paragraph. This upper bound is likely not tight; it might well be that $\oT_{max}$ never exceeds its static value of $1/8$.

\subsection*{Mean temperature}
\addcontentsline{toc}{subsection}{Mean temperature}

Uniform bounds on the mean fluid temperature relative to that of the top boundary, $\dT$, are summarized in tables \ref{tab: main 1} and \ref{tab: main 2}. In the IH cases, the lower bounds \mbox{$\dT>0$} follow from the $T$ power integral. The upper bounds are proven by integrating $z^2$ against the $T$ equation (\ref{eq: T}), and using \eqref{eq: fT} in the IH1 case, to find
\begin{equation}
\dT=\begin{cases}
\frac{1}{12} - \lan \left(z-\tfrac{1}{2}\right)wT\ran & \text{IH1} \\
\frac{1}{3} - \lan zwT\ran & \text{IH2, IH3}.
\end{cases}
\end{equation}
Incompressibility gives $\overline w=0$ and thus $\lan wT\ran=\lan w\theta\ran$ and $\lan zwT\ran=\lan zw\theta\ran$, where $\theta$ is the deviation of $T$ from its static profile. Integrating the temperature fluctuation equation (\ref{eq: theta}) against $\theta$ gives $\lan \left(z-\tfrac{1}{2}\right)w\theta\ran = \lan|\nabla\theta|^2\ran\ge0$ in IH1 and $\lan zw\theta\ran = \lan|\nabla\theta|^2\ran\ge0$ in IH2 and IH3. Therefore, $\dT\le1/12$ in IH1, and $\dT\le1/3$ in IH2 and IH3.

In the RB cases, none of the uniform bounds on $\dT$ are likely to be tight. The extremum principles give $0<\dT<1$ in RB1 and $0<\dT$ in RB3. The upper bound for RB3 and the upper and lower bounds for RB2 follow from the inequality $\left|\dT\right|\le1/\sqrt 3$ that is derived below.
\begin{align*}
|\dT| =& \left|\lan z\partial_zT\ran\right| \\
\le& \lan|z\partial_zT|\ran \\
\le& \lan z^2\ran^{1/2}\lan\partial_zT^2\ran^{1/2} \\
\le& \tfrac{1}{\sqrt{3}}\lan|\nabla T|^2\ran^{1/2} \\
\le& \tfrac{1}{\sqrt{3}}\DT^{1/2} \\
\le& \tfrac{1}{\sqrt{3}}.
\end{align*}
The first line of the derivation follows from integration by parts, the third line follows from the Cauchy-Schwarz inequality, the fifth line follows from the $T$ power integral, and the last line follows from the bound $\DT=1-\wT\le1$ that is proven above.

\chapter{Stabilities and bounds}
\label{chap: stab}

The preceding chapter has defined six convective configurations and described, for each case, how various integral quantities characterize bulk heat transport. The quantities of central importance include the mean vertical transport by convection, $\wT$, in RB and IH convection and the mean temperature of the fluid relative to that of the top boundary, $\dT$, in IH convection. We would like to predict the values that these quantities assume for various Rayleigh numbers, Prandtl numbers, and confining geometries. This is equivalent to predicting the parameter-dependence of the Nusselt numbers we have defined; $N$ is determined by $\oT_{max}$ in IH1 and by $\wT$ in the other five cases, and $\wt N$ is determined by $\dT$ in the three IH cases. The task before us is very difficult in general and would require a greatly improved understanding of fluid turbulence, so we are limited to partial results. The present chapter presents facts about heat transport that can be determined purely by mathematical analysis of the Boussinesq equations, while the next chapter addresses simulations and laboratory experiments.

There are two main ways of studying heat transport analytically: examining simple particular solutions that can be written down exactly or asymptotically, and deriving bounds on integral quantities that apply to all solutions. The first method yields much stronger results but is useful only at small Rayleigh numbers, where the system either remains static or assumes a simple flow. Here we consider only the static states. Heat transport in static states is purely conductive and easy to understand, so the main task is to determine the parameters at which such states are stable. To this end, we can find a Rayleigh number, $R_L$, above which a static state is linearly unstable, and a Rayleigh number, $R_E$, below which we can prove that it is the unique globally stable state. These results work together with the second method of analysis---bounding $N$ or $\wt N$ above by functions $N_b(R)$ or $\wt N_b(R)$---to constrain the dependence of Nusselt numbers on $R$. Sections \ref{sec: linear stab}-\ref{sec: bounds} outline the calculations and values of $R_L$, $R_E$, and $\wt N_b(R)$, respectively, for the various configurations.

\begin{figure}
\begin{center}
\includegraphics[width=220pt]{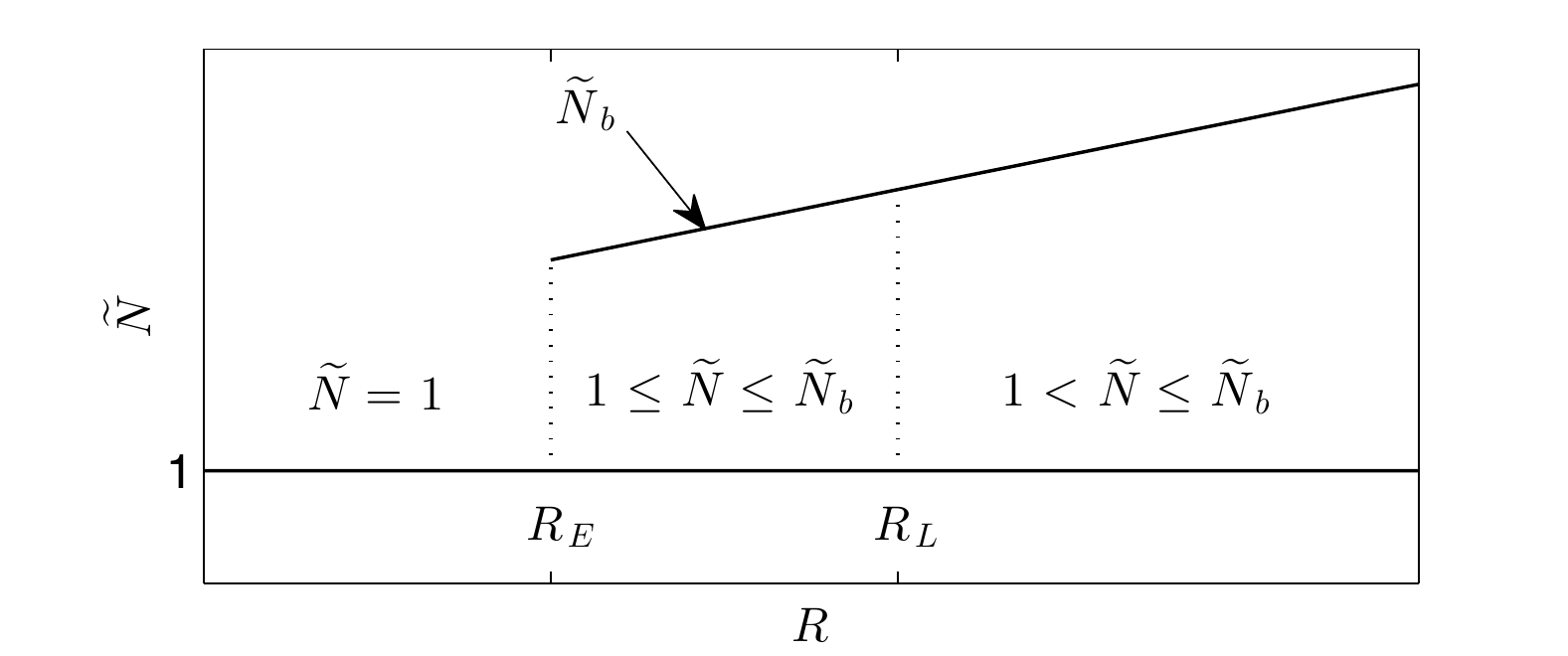}
\end{center}
\caption{Schematic of what this chapter's analytical results say about the dependence of $\wt N$ on $R$ in IH convection. The numerical values of $R_E$, $R_L$, and $\wt N_b(R)$ vary between configurations. In RB convection, the analogous diagram for $N$ lacks a middle region since $R_E=R_L$.}
\label{fig: stability schematic}
\end{figure}

The schematic of figure \ref{fig: stability schematic} shows how $R_L$, $R_E$, and $\wt N_b(R)$ combine to give some knowledge of $\wt N$ in IH convection. In the lowest-$R$ regime, where $R<R_E$, we know that $\wt N=1$. This is because the system asymptotically approaches the static state, so its Nusselt number, which we have defined as an infinite-time limit, must be that of the static state. In the subcritical regime, where $R_E<R<R_L$, the static state is linearly stable, but sustained flow might also be possible, so all we can say is that $1\le \wt N\le \wt N_b(R)$ in this regime. In the larger-$R$ regime where $R_L<R$, the static state is linearly unstable, so any physically realizable state must have sustained flow and thus a Nusselt number strictly greater than unity. That is, $1<\wt N\le \wt N_b(R)$ for attracting states, although $\wt N=1$ remains possible if unstable states are included. A schematic like figure \ref{fig: stability schematic} for the other Nusselt number, $N$, in IH convection would lack the upper bound since $R$-dependent bounds on $\wT$ have not yet been proven (cf.\ \S\ref{sec: conjectures}).

Figure \ref{fig: stability schematic} represents a scenario where $R_E$ is strictly smaller than $R_L$. In RB convection there is no subcritical regime since $R_E=R_L$. This allows for asymptotic solutions in the weakly supercritical regime, giving more precise expressions for Nusselt numbers there. In IH convection, on the other hand, subcritical convection is not ruled out because $R_E<R_L$. Stronger methods of analysis may be able to prove stability thresholds larger than $R_E$, but not necessarily as large as $R_L$. Subcritical convection is indeed possible in IH1 \cite{Tveitereid1977} and IH3 \cite{Schwiderski1972, Tveitereid1976, Busse2014}. In IH2, the possibility of subcritical convection remains open.

For no-slip conditions on the velocity, figure \ref{fig: stability schematic} is made concrete in each configuration by table \ref{tab: no-slip summary}, which gives values for $R_L$, $R_E$, $N_b(R)$, and $\wt N_b(R)$. The bounds have been simplified by assuming that $R$ is asymptotically large, and they are stated in terms of the diagnostic Rayleigh numbers, $Ra$ and $\wt Ra$, that we have defined in \S{sec: diagnostic Rayleigh numbers}. (Recall that $Ra$ equals $R$ in RB1 but equals $R/N$ in the other five cases, and that $\wt Ra=R/\wt N$ in IH convection.) The similarities between the various bounds are evident. The present chapter explains how the values in table \ref{tab: no-slip summary} are calculated and gives values for some other boundary conditions on the velocity.

\begin{table}
\begin{center}
\begin{tabular}{R{15pt} | C{50pt} C{50pt} C{45pt} C{45pt} C{45pt} C{45pt}}
		& RB1		& RB2	& RB3		& IH1		& IH2		& IH3 \\
\hline
$R_E$	& 1707.76	& 720	& 1295.78	& 26\,926.6	& 1429.86	& 2737.16 \\ 
$R_L$	& 1707.76	& 720	& 1295.78	& 37\,325.2	& 1440~~~~~~& 2772.27 \\
$N_b$	& $0.027\;Ra^{1/2}$	& $0.28\;Ra^{1/2}$	& $0.28\;Ra^{1/2}$ &
none & none & none \\
$\wt N_b$	&&&&
$0.025\;\wt{Ra}^{1/2}$ & $0.13\;\wt{Ra}^{1/2}$ & $0.094\;\wt{Ra}^{1/2}$
\end{tabular}
\end{center}
\caption{For no-slip boundary conditions: a Rayleigh number above which the static state is linearly unstable ($R_L$), a Rayleigh number below which the static state is globally stable ($R_E$), and upper bounds ($N_b$ and $\wt N_b$) on the Nusselt numbers $N$ and $\wt N$ that are valid for asymptotically large $R$. References for these values are given throughout the chapter. In the IH cases, $R$-dependent bounds on $N$ have not been proven. In the RB cases, $\wt N$ is not defined.}
\label{tab: no-slip summary}
\end{table}

The linear and nonlinear stability analyses that we apply to the static state can be applied to other particular solutions as well. Such analyses must be carried out asymptotically or numerically, however, since none of the finite-amplitude particular solutions can be expressed in closed form. The weakly nonlinear regime of IH convection has been theoretically examined in a few studies \cite{Roberts1967, Thirlby1970, Schwiderski1972, Tveitereid1976}. Such analyses of particular solutions reveal much about bifurcations and pattern formation, but they do not yield robust information about heat transport. This is because the results often depend strongly on geometry, and also because each particular solution typically is stable over only a narrow range of parameters.

Sections \ref{sec: linear stab} and \ref{sec: energy stab} address each static state's linear and nonlinear stability, respectively. Section \ref{sec: bounds} outlines a proof of $R$-dependent lower bounds on the mean temperature for all three IH configurations. These bounds, which amount to upper bounds on $\wt N$, are then compared with upper bounds on $N$ in RB convection.

The results laid out in this chapter constitute most of what can be deduced mathematically about $N$ and $\wt N$. These results are rather meager in that they tell us neither the actual values assumed in the ranges $1\le N\le N_b(R)$ and $1\le\wt N\le\wt N_b(R)$, nor how the Prandtl number and geometry affect these values. Such questions must wait until the next chapter because substantial answers, so far, come only from simulations and laboratory experiments.

\section{Linear instability of static states}
\label{sec: linear stab}

In each RB and IH configuration, we can find a Rayleigh number, $R_L$, above which the static state is linearly unstable. The Prandtl number of the fluid does not affect this threshold. In most cases it has been proven that the linear instability is stationary, meaning that the non-static states that bifurcate at the point of instability are steady, rather than time-dependent. The method of calculating $R_L$ is similar in every case and is well known from the study of the canonical RB1 system. We outline this methods here and give references for further details.

\subsection{Linear stability eigenproblem}

We want to study the stability of the static state, wherein $\bu=\mathbf 0$ and $T=T_{st}(z)$ for the various $T_{st}(z)$ profiles given in expression (\ref{eq: T_st}). It is convenient to decompose the temperature field into its static part and a fluctuation, $\theta$,
\[T(\bx,t)=T_{st}(z)+\theta(\bx,t). \]
Since $\bu$ and $T$ evolve according to the Boussinesq equations (\ref{eq: inc})-(\ref{eq: T}), fluctuations evolve according to
\begin{align}
\nabla \cdot \bu &= 0 \label{eq: inc pert} \\
\partial_t \bu + \bu \cdot \nabla \bu  &= 
	-\nabla p + Pr \nabla^2 \bu + Pr\,R\,\theta \mathbf{\hat z} \label{eq: u pert}  \\
\partial_t\theta + \bu\cdot\nabla\theta &= \nabla^2\theta - T_{st}'w, \label{eq: theta} 
\end{align}
where the prime denotes $\tfrac{d}{dz}$. The static state enters the fluctuation equations only through its gradient, $T_{st}'$, reflecting the fact that Boussinesq dynamics are affected only by relative temperature differences, not absolute temperatures. This gradient is constant in RB convection but varies linearly in IH convection when the heating is uniform:
\begin{equation}
T'_{st}(z) = \begin{cases}
-1 & \text{RB1, RB2, RB3} \\
-z + \tfrac{1}{2} & \text{IH1} \\
-z & \text{IH2, IH3},
\end{cases}
\label{eq: T'_st}
\end{equation}
where we recall that $0\le z\le1$. The boundary conditions on $\theta$ are the homogenous analogs of the conditions on $T$:
\begin{align}
\text{RB1, IH1:}\hspace{23.5pt} \theta|_{z=0},\hspace{10.5pt}\theta|_{z=1}&=0 \label{eq: theta BC1} \\
\text{RB2, IH2:}\hspace{15pt} \partial_z\theta|_{z=0},~\partial_z\theta|_{z=1}&=0 \label{eq: theta BC2} \\
\text{RB3, IH3:}\hspace{15pt} \partial_z\theta|_{z=0},\hspace{10.5pt}\theta|_{z=1}&=0. \label{eq: theta BC3}
\end{align}
The fluctuation dynamics of RB1 and IH1 are distinguished only by differing $\oT_{st}'(z)$. The same is true of RB2 and IH2, and of RB3 and IH3.

We study the stability of the zero solution of the fluctuation equations (\ref{eq: inc pert})-(\ref{eq: theta}), which is equivalent to the stability of the static state. The nonlinear terms in the fluctuation equations can be neglected when finding linear stability thresholds. As is standard \cite{Chandrasekhar1981}, we find a closed pair of equations governing the linear evolution of $w$ and $\theta$ by taking $\mathbf{\hat z}\cdot\nabla\times\nabla\times(\ref{eq: u pert})$. Omitting time derivatives gives equations for the marginally stable states that are stationary, meaning they do not vary in time, are governed by
\begin{align}
\nabla^4 w &= -R \nabla_H^2 \theta \label{eq: w marginal}  \\
\nabla^2\theta &= T_{st}'w, \label{eq: theta marginal} 
\end{align}
where $\nabla_H^2:=\partial_x^2+\partial_y^2$ is the horizontal Laplacian operator. The validity of considering only stationary instabilities is discussed at the end of this subsection.

The Rayleigh number, $R_L$, at which the static state becomes linearly unstable is the smallest $R$ for which there is a marginally stable state---that is, the smallest $R$ for which equations (\ref{eq: w marginal})-(\ref{eq: theta marginal}) have a nonzero solution. This is a (generalized) eigenproblem whose spectrum of eigenvalues is continuous and bounded below. Assuming there are no horizontal boundaries, we can Fourier transform the eigenproblem in $x$ and $y$, decomposing it into an independent eigenproblem for each horizontal wavevector $(k_x,k_y)$, where $k_x$ and $k_y$ are real. If the horizontal periods of a mode are $L_x$ and $L_y$, then $k_x=2\pi/L_x$ and $k_y=2\pi/L_y$. The resulting decomposed eigenproblems take the form \cite{Rayleigh1916, Chandrasekhar1981}
\begin{align}
\hat w^{(4)} -2k^2\hat w''+k^4\hat w &= R k^2 \hat\theta \label{eq: w marginal k}  \\
\hat\theta''-k^2\hat\theta &= T_{st}'\hat w, \label{eq: theta marginal k}
\end{align}
where $\hat w(z)$ and $\hat\theta(z)$ are complex in general, and $k^2:=k_x^2+k_y^2$. We call $k$ the horizontal wavenumber.

The sixth-order linear system (\ref{eq: w marginal k})-(\ref{eq: theta marginal k}) requires six boundary conditions. The conditions (\ref{eq: theta BC1})-(\ref{eq: theta BC3}) on $\theta$ apply also to $\hat\theta$. The first two $\hat w$ conditions are that $\hat w=0$ at both boundaries, and the other two depend on whether each boundary is no-slip or free-slip:
\begin{align}
\text{no-slip:}~~& \hat w'(0),~\;\hat w'(1)\;=0 \label{eq: no-slip} \\
\text{free-slip top:}~~& \hat w'(0),~\;\hat w''(1)=0 \label{eq: free-slip top}  \\
\text{free-slip bottom:}~~& \hat w''(0),~\hat w'(1)\;=0 \label{eq: free-slip bottom} \\
\text{free-slip:}~~& \hat w''(0),~\hat w''(1)=0. \label{eq: free-slip}
\end{align}
Throughout this work, we give results for all four pairs of velocity conditions when possible. In some cases, analytical bounds and experimental results are available only for no-slip boundaries, which are the most natural in the laboratory. Condition (\ref{eq: free-slip top}) is experimentally realizable in a container with an open top. Condition (\ref{eq: free-slip bottom}) might appear unrealizable since it describes a container with an open bottom, but it also describes dynamically equivalent systems with an open top. For instance, IH convection with an open bottom, when viewed upside down, has the same dynamics as internally \emph{cooled} convection with an open top.

For each $k^2$, equations (\ref{eq: w marginal k})-(\ref{eq: theta marginal k}) and their boundary conditions form a linear eigenproblem in $R$ with a \emph{discrete} spectrum that is easier to compute than the continuous spectrum of (\ref{eq: w marginal})-(\ref{eq: theta marginal}). The $R_L$ at which the static state loses stability is the smallest eigenvalue of (\ref{eq: w marginal k})-(\ref{eq: theta marginal k}), minimized over all admissible $k^2$. If all horizontal wavenumbers are possible,
\begin{equation}
R_L = \inf_{k^2>0}R^{(0)}(k), \label{eq: R_L def}
\end{equation}
where
\begin{equation}
R^{(0)}(k):=\min\left\{ R ~\big|~ \text{(\ref{eq: w marginal k})-(\ref{eq: theta marginal k}) has a nonzero solution} \right\}. 
\end{equation}
The definition of $R_L$ requires an infimum rather than a minimum because the infimum sometime occurs in the limit $k^2\to0$, in which case no minimum is achieved. Perturbations with $k=0$ are not admissible since a horizontally uniform $\hat w$ would violate incompressibility. The value (or limit) of $k$ at which $R_L$ occurs is called the critical wavenumber of linear instability, $k_L$.

Because the eigenproblem (\ref{eq: w marginal})-(\ref{eq: theta marginal}) is derived assuming a stationary instability, the resulting $R_L$ is the value at which a steady state bifurcates from the static one. In cases where it is proven that all marginally stable states are indeed stationary, $R>R_L$ is not only sufficient but necessary for instability of the static state. Stationarity has been proven for all RB cases \cite{Pellew1940}. In the IH3 case it follows for free-slip boundaries from an argument of Spiegel (see footnote 4 of \cite{Veronis1962}) and for no-slip boundaries from a theorem of Herron \cite{Herron2003}. The latter method of proof may suffice to show stationarity in the remaining IH configurations. Until that is done, we can say in those cases only that $R>R_L$ is sufficient for instability.

\subsection{Solutions of the linear stability eigenproblem}
\label{sec: lin stab solutions}

In RB convection, where $T_{st}'=-1$, the eigenfunctions solving (\ref{eq: w marginal k})-(\ref{eq: theta marginal k}) are combinations of trigonometric and hyperbolic functions. The minimum eigenvalue at a given wavenumber, $R^{(0)}(k)$, must satisfy an expression involving the eigenfunctions. This expression can be solved for $R^{(0)}(k)$, analytically in a few cases and numerically in the others \cite{Rayleigh1916, Chandrasekhar1981}. In IH convection, where $T_{st}'$ varies linearly with $z$, the analogous approach would involve hypergeometric functions, so it is simpler to solve the eigenproblem (\ref{eq: w marginal k})-(\ref{eq: theta marginal k}) numerically. We have done this for all six configurations by the general method described in \cite{Trefethen2000}: discretizing the differential operators using a spectral collocation method and computing the spectra of the resulting matrices. Our computed values of $R_L$ agree with or add precision to the values in the literature. As explained shortly, the exact values of $R_L$ in the RB2 and IH2 cases can be calculated also by asymptotic expansion

\begin{figure}
\begin{center}
(a)
\includegraphics[width=153pt]{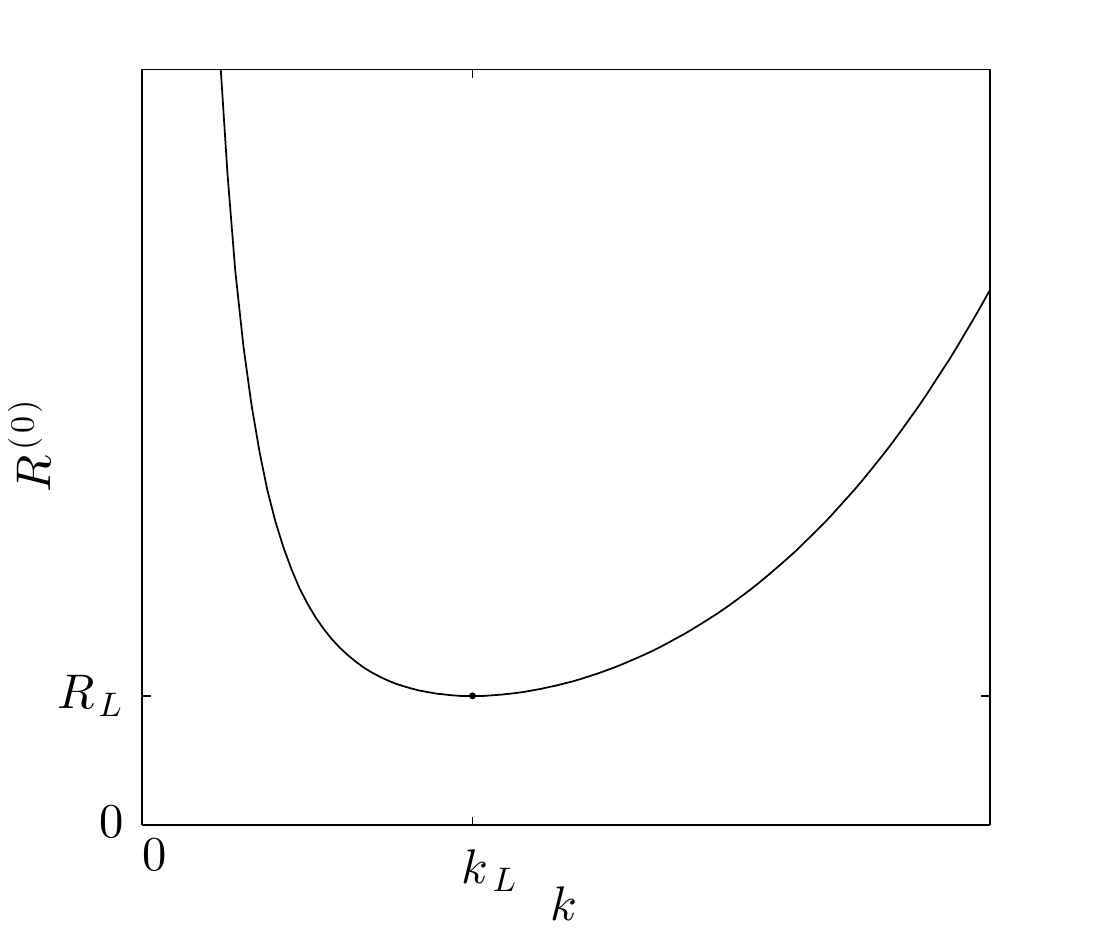}
(b)
\includegraphics[width=153pt]{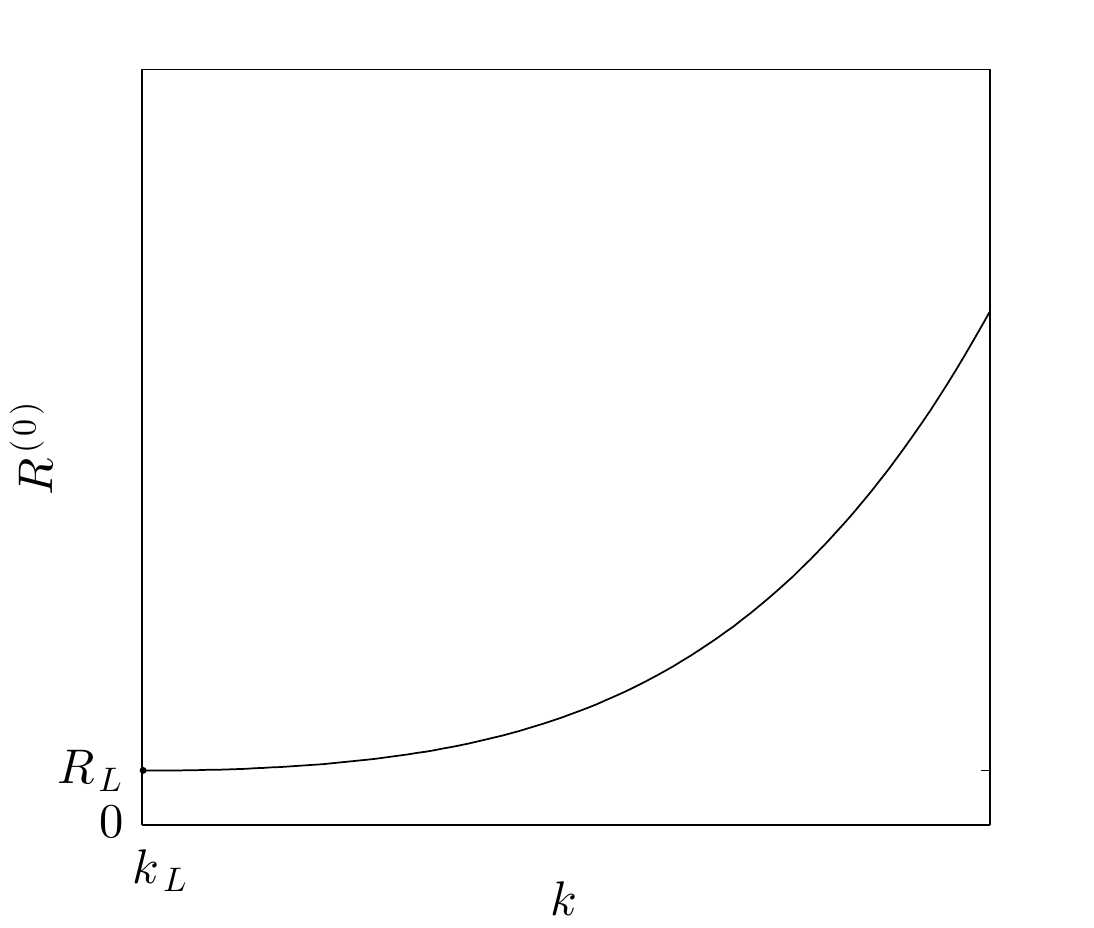}
\end{center}
\caption{Schematic diagrams of how the first marginally stable eigenvalue, $R^{(0)}$, depends on the horizontal wavenumber, $k$, in (a) the RB1, RB3, IH1, and IH3 cases, and in (b) the RB2 and IH2 cases, where heat fluxes are fixed at both boundaries.}
\label{fig: R0 vs k}
\end{figure}

The $R^{(0)}(k)$ curve can assume one of two qualitative shapes in our models, depending on the thermal boundary conditions. Figure \ref{fig: R0 vs k}a shows what the $R^{(0)}(k)$ curves look like in the four cases where the temperature is fixed at one or both boundaries: $R^{(0)}$ is minimized at a finite $k$ and grows unboundedly both as $k\to0$ and as $k\to\infty$. Figure \ref{fig: R0 vs k}b shows what the $R^{(0)}(k)$ curves look like when the temperature flux is fixed at both boundaries: $R^{(0)}$ approaches its infimum as $k\to0$ and grows unboundedly as $k\to\infty$.

For all of our RB and IH configurations and all four pairs of velocity conditions (\ref{eq: no-slip})-(\ref{eq: free-slip}), table \ref{tab: linear all} gives the smallest Rayleigh number, $R_L$, at which the static state undergoes a stationary, horizontally periodic instability, along with the instability's horizontal wavenumber, $k_L$. Most of these values have been known for a long time. The RB1 case was analyzed first, with free-slip boundaries in Rayleigh's seminal analysis of 1916 \cite{Rayleigh1916} and then with other velocity conditions \cite{Jeffreys1928, Low1929, Pellew1940}. Later, values of $R_L$ for some of our velocity conditions were reported for RB2 and RB3 \cite{Sparrow1964}, IH1 \cite{Debler1959, Veronis1962, Sparrow1964, Watson1968, Kulacki1975a}, IH3 \cite{Debler1959, Roberts1967, Mckenzie1974, Kulacki1975a}, and then IH2 \cite{Ishiwatari1994, Goluskin2015}.

\begin{table}
\begin{center}
\begin{tabular}{R{10pt} R{65pt} C{35pt} C{41pt}}
&					& $R_L$ 	& $k_L$ \\
\hline
RB1 \\
& no-slip			& 1707.76 	& 3.1163 \\
& free-slip top	& 1100.65	& 2.6823 \\
& free-slip bottom	& 1100.65	& 2.6823 \\
& free-slip		& 657.511	& 2.2214 \\
RB2 \\
& no-slip			& 720		& 0 \\
& free-slip top	& 320		& 0 \\
& free-slip bottom	& 320		& 0 \\
& free-slip		& 120		& 0 \\
RB3 \\
& no-slip			& 1295.78	& 2.5519 \\
& free-slip top	& 816.744	& 2.2147 \\
& free-slip bottom	& 668.998	& 2.0856 \\
& free-slip		& 384.693	& 1.7576 \\
IH1 \\
& no-slip			& 37\,325.2	& 3.9989 \\
& free-slip top	& 16\,669.8	& 3.0131 \\
& free-slip bottom	& 37\,949.4	& 4.0867 \\
& free-slip		& 16\,992.2	& 3.0277 \\
IH2 \\
& no-slip			& 1440		& 0 \\
& free-slip top	& 576		& 0 \\
& free-slip bottom	& 720		& 0 \\
& free-slip		& 240		& 0 \\
IH3 \\
& no-slip			& 2772.27	& 2.6293 \\
& free-slip top	& 1612.62	& 2.2611 \\
& free-slip bottom	& 1650.55	& 2.1429 \\
& free-slip		& 867.766	& 1.7897
\end{tabular}
\end{center}
\caption{Rayleigh number, $R_L$, above which each configuration's static state is linearly unstable, and the horizontal wavenumber, $k_L$, of the linear perturbation that is marginally stable at $R_L$. The RB2 and IH2 values are exact (cf.\ \S\ref{sec: long-wave}), while the other values are numerical approximations that are accurate to the precision shown.}
\label{tab: linear all}
\end{table}

In most configurations, $R_L$ is smallest when both boundaries are free-slip, largest when both boundaries are no-slip, and somewhere in between when one boundary is free-slip and the other is no-slip. The IH1 configuration provides a surprising exception: $R_L$ is smallest when only the top is free-slip and largest when only the bottom is free-slip. 

In principle, knowing $R_L$ is relevant to heat transport because $R>R_L$ suggests that sustained convection will occur. In confined geometries, however, not all $k$ are admitted, and the flow can be only approximately periodic in the horizontal directions. These effects raise $R_L$ by an amount particular to the confining geometry, and only when $R$ exceeds this larger value is convection guaranteed.

The RB2 and IH2 configurations are special in that analytical expressions for $R_L$ can be found for any velocity boundary conditions. This is because the critical wavenumber is zero (cf.\ table \ref{tab: linear all}), so $R_L$ can be calculated by long-wavelength asymptotics.

\subsection{Long-wavelength asymptotics for RB2 and IH2}
\label{sec: long-wave}

In both RB2 and IH2, the infimum in the definition (\ref{eq: R_L def}) of $R_L$ occurs when $k\to0$, so
\begin{equation}
R_L = \lim_{k\to0}R^{(0)}(k).
\end{equation}
It has apparently not been proven analytically that fixed-flux boundary conditions imply $k_L=0$, so the finding must be verified on a case-by-case basis by numerically computing $R^{(0)}(k)$. This has been done for RB2 and IH2, so $R_L$ can be found exactly by expanding the linear stability eigenproblem (\ref{eq: w marginal k})-(\ref{eq: theta marginal k}) in small $k^2$. (We cannot simply set $k=0$ because the limit is singular.) The eigenmode with eigenvalue $R^{(0)}(k)$ has scaling $O(\hat\theta)=k^2O(\hat w)$ when $k\ll1$ \cite{Chapman1980}. We thus let $\hat w=k^2\hat W$ and seek solutions where $\hat W$ and $\hat\theta$ are both $O(1)$. The rescaled eigenproblem is
\begin{align}
\hat W^{(4)} & = R \hat\theta + 2k^2\hat W'' - k^4 \hat W \label{eq: W scaled}  \\
\hat\theta'' &= k^2\left(\hat\theta + T_{st}'\hat W\right). \label{eq: theta scaled} 
\end{align}
The no-flux conditions on $\hat\theta$ require the vertical integral of $\hat\theta''$ to vanish, and this furnishes a consistency condition on the righthand side of (\ref{eq: theta scaled}):
\begin{equation}
\int_0^1 \left(\hat\theta + T_{st}'\hat W\right) dz = 0. \label{eq: solvability}
\end{equation}

Equations (\ref{eq: W scaled})-(\ref{eq: solvability}) suffice to determine $R_L$, but it is possible to also retain the nonlinear terms in a long-wavelength expansion of the fluctuation equations. This has been carried out for RB2 \cite{Chapman1980}, IH2 \cite{Ishiwatari1994}, and some other convective models with fixed boundary fluxes \cite{Chapman1980a, Childress2004}. The simpler linear calculation we describe here is contained in these nonlinear analyses.

Equations (\ref{eq: W scaled})-(\ref{eq: theta scaled}) are solved asymptotically by expanding in $k^2$:
\begin{align}
\hat W(z) &= W_0(z) + k^2W_2(z) + \cdots \\
\hat \theta(z) &= \theta_0(z) + k^2\theta_2(z) + \cdots \\
R &= R_L + k^2R_2 + \cdots.
\end{align}
The $R$ expansion anticipates that $R_0=R_L$---in other words, that $k_L=0$. We and others have confirmed this for RB2 and IH2 by computing the $R^{(0)}(k)$ curves numerically.

The procedure for asymptotically solving equations (\ref{eq: W scaled})-(\ref{eq: theta scaled}) to any order $k^{2n}$ is entirely systematic. Assuming all lower-order terms are known, the polynomial $\theta_{2n}(z)$ is found by integrating equation (\ref{eq: theta scaled}) at $O(k^{2n})$, then the polynomial $W_{2n}(z)$ is found by integrating equation (\ref{eq: W scaled}) at $O(k^{2n})$, and finally $R_{2n}$ is found from the consistency condition (\ref{eq: solvability}) at $O(k^{2n})$. Since $R_L=R_0$ here, we only need to carry out these three steps at leading order to find $R_L$. This has been done for RB2 in \cite{Chapman1980} and for IH2 in \cite{Ishiwatari1994, Goluskin2015}. The results of the three steps are that
\begin{equation}
R_L = \frac{-1}{\int_0^1T_{st}'P(z)dz} = \begin{cases}
\dfrac{1}{\int_0^1P(z)dz} & \text{RB2} \vspace{5pt} \\
\dfrac{1}{\int_0^1zP(z)dz} & \text{IH2},
\end{cases}
\end{equation}
where $P(z)$ is the unique fourth-order polynomial that has a leading coefficient of 1/24 and satisfies the $\hat w$ boundary conditions. For our domain of $0\le z\le1$, these $P(z)$ are given in \cite{Goluskin2015}, for instance. The values of $R_L$ for various velocity conditions appear in table \ref{tab: linear all} above.

\section{Energy stability of static states}
\label{sec: energy stab}

Knowing the Rayleigh number, $R_L$, above which a static state is linearly unstable would be well complemented by knowing the critical Rayleigh number, $R_c$, below which it is the globally attracting state of the fully nonlinear dynamics. This is difficult in general, so we settle for finding a so-called \emph{energy} Rayleigh number, $R_E$, that is a lower bound on $R_c$. Since linear instability implies nonlinear instability, we can anticipate that $R_E\le R_c\le R_L$. RB convection is special in that $R_E=R_c=R_L$ \cite{Joseph1965}. IH convection is more complicated in that $R_E<R_L$ for the largest known $R_E$, as depicted in figure \ref{fig: stability schematic}. The values of $R_E$ and $R_L$ serve as upper and lower bounds on $R_c$, respectively, that hold uniformly for all $Pr$. The exact values of $R_c$ can depend on $Pr$ and are not yet known. The upper bounds on $R_c$ can be tightened by finding particular subcritical solutions, as has been done for IH1 \cite{Tveitereid1977} and IH3 \cite{Schwiderski1972, Tveitereid1976}, since any $R$ at which subcritical convection persists must be larger than $R_c$. Proving a lower bound tighter than $R_E$ is a more daunting challenge.

\subsection{Lyapunov stability and the energy method}

The global stability of the static state is equivalent to the global stability of the zero solution to the fluctuation equations (\ref{eq: inc pert})-(\ref{eq: theta}). The nonlinear terms in those equations that could be ignored in the linear stability analysis must now be included. The typical method of proving global stability, due to Lyapunov, requires finding a functional of the state variables that is positive definite and whose evolution is negative definite. That is, we must find a functional $\mathcal L[\bu,\theta]$ such that
\begin{align}
\Ly[\bu,\theta] &\ge 0 \label{eq: Lyap 1} \\
\tfrac{d}{dt}\Ly[\bu,\theta] &\le0. \label{eq: Lyap 2}
\end{align}
To show also that the static state attracts all initial conditions, it suffices for the above inequalities to be strict whenever $\bu$ or $\theta$ is nonzero. In the convective systems we are studying, the best we can hope for is to find an $\Ly$ where (\ref{eq: Lyap 1})-(\ref{eq: Lyap 2}) hold for $R$ below some finite value, $R_\Ly$. That is,
\begin{equation}
R_\Ly := \sup\left\{ R ~|~ \Ly \text{ satisfies (\ref{eq: Lyap 1})-(\ref{eq: Lyap 2})}\right\}.
\end{equation}
The critical Rayleigh number, $R_c$, is the largest $R$ at which \emph{any} Lyapunov functional exists,
\begin{equation}
R_c := \sup_\Ly \;R_\Ly. \label{eq: R_c}
\end{equation}

There is no universally successful method for constructing Lyapunov functionals, let alone the optimal $\Ly$ that is valid for $R$ up to $R_c$. It is even difficult to confirm that an optimal $\Ly$ is indeed optimal, except when $R_c=R_L$, as in RB convection. All we can do in general is make educated guesses for $\Ly$, determine the corresponding values of $R_\Ly$, and declare the largest $R_\Ly$ we can find to be a lower bound on $R_c$. In fluid dynamical systems like ours, even this guess-and-check procedure cannot be carried out for general $\Ly$ because it is too difficult to determine whether the second Lyapunov condition (\ref{eq: Lyap 2}) holds. In most nonlinear analyses of fluid stability, this trouble is avoided by considering only a particular subset of possible Lyapunov functionals for which it is tractable to check the second Lyapunov condition. This approach is called the \emph{energy method}.

The energy method in fluid mechanics \cite{Serrin1959, Joseph1976, Straughan2004} is a special case of Lyapunov's method. In one definition of the energy method that is neither the narrowest nor the broadest definition possible, the Lyapunov functional, which is called the energy, has two special features:
\begin{enumerate}[leftmargin=.35in]
\item The energy is quadratic in the state variables.
\item The energy is conserved by the nonlinear terms of the fluctuation equations (\ref{eq: u pert})-(\ref{eq: theta}), meaning that these terms do not contribute to the expression for the time-evolution of the energy.
\end{enumerate}
The energy method is so named because quadratic quantities are often proportional to physical energies. Here we follow Joseph \cite{Joseph1965} in considering energies of the form
\begin{equation}
E_\gamma[\bu,\theta](t) : = \tfrac{1}{2}\fint\left( \tfrac{1}{Pr\,R}|\bu|^2
	+ \gamma\theta^2\right)d\bx, \label{eq: E_gamma}
\end{equation}
where $\fint$ denotes an instantaneous volume average. The constant $\gamma>0$ is called a \emph{coupling parameter}, and each positive value defines an energy that is a valid Lyapunov functional for $R$ up to some $R_{E_\gamma}$. This value of $R_{E_\gamma}$ is maximized by some optimal choice of $\gamma$, where it achieves the critical Rayleigh number of energy stability, $R_E$:
\begin{equation}
R_E := \max_{\gamma>0}\sup\left\{ R ~|~ E_\gamma
	\text{ satisfies (\ref{eq: Lyap 1})-(\ref{eq: Lyap 2})} \right\}. \label{eq: R_E}
\end{equation}
The value of $R_E$ is the best lower bound on $R_c$ that we find by the energy method, though it is likely still smaller than $R_c$. Deriving a better lower bound on $R_c$ would require going beyond the energy method to search over a larger class of Lyapunov functionals, and this presents technical challenges. Progress beyond the energy method has been made for a few shear flow models \cite{Kaiser2005, Chernyshenko2014} but not yet for a convective system.

\subsection{Energy stability eigenproblem}

The functional $E_\gamma$ suffices to show that the static state is globally stable whenever it satisfies conditions (\ref{eq: Lyap 1})-(\ref{eq: Lyap 2}). The first condition holds whenever all the parameters are positive, so it remains only to determine the parameters for which $\frac{d}{dt}E_\gamma\le0$. Adding the volume averages of $\frac{1}{Pr\,R}\,\bu\cdot(\ref{eq: u pert})$ and $\gamma\,\theta\times(\ref{eq: theta})$ and then integrating by parts gives
\begin{equation}
\tfrac{d}{dt}E_\gamma = -\fint\left[
	\tfrac{1}{R}|\nabla\bu|^2+\gamma|\nabla\theta|^2-\left(1-\gamma T'_{st}\right)w\theta\right]d\bx. \label{eq: dE/dt}
\end{equation}
The static state is globally attracting if the righthand side of (\ref{eq: dE/dt}) is negative definite. Like the linear stability threshold, the satisfaction of this condition depends on $R$ but not on $Pr$. Only static states have this feature; other solutions and their stabilities depend also on $Pr$.

The calculus of variations yields a necessary and sufficient condition for the righthand side of (\ref{eq: dE/dt}) to be negative definite. In particular, $E_\gamma$ is a Lyapunov functional when $R$ is smaller than all eigenvalues, $R$, of the (generalized) eigenproblem \cite{Straughan1990, Ames1990, Straughan2004}
\begin{align}
\hat w^{(4)}-2k^2\hat w''+k^4\hat w &= \label{eq: w energy}
	\tfrac{1}{2}Rk^2\left(1-\gamma T'_{st}\right)\hat\theta \\
\gamma\left(\hat\theta''-k^2\hat\theta\right) &= -\tfrac{1}{2}\left(1-\gamma T'_{st}\right)\hat w.
	\label{eq: theta energy}
\end{align}
The boundary conditions are the same as in the linear stability eigenproblem of \S\ref{sec: lin stab solutions}, and again $\hat w(z)$ and $\hat\theta(z)$ can be complex, and $k$ is the horizontal wavenumber. Expression (\ref{eq: R_E}) for $R_E$ can thus be restated as
\begin{equation}
R_E = \max_{\gamma>0}\inf_{k^2>0}\min\left\{ R ~\big|~ \text{(\ref{eq: w energy})-(\ref{eq: theta energy}) has a nonzero solution} \right\}. \label{eq: R_E 2}
\end{equation}
It is a special feature of the energy method, and not of Lyapunov's method in general, that the nonlinear stability analysis can be reduced to the solution of a linear eigenproblem, much like the linear stability analysis.

\subsection{Solutions of the energy stability eigenproblem}

In RB convection, where $T'_{st}=-1$, there is no need to solve the energy stability eigenproblem (\ref{eq: w energy})-(\ref{eq: theta energy}) because it is identical to the linear stability eigenproblem (\ref{eq: w marginal k})-(\ref{eq: theta marginal k}), so long as the energy is defined with $\gamma=1$. This energy is thus a valid Lyapunov functional for all $R$ up to $R_L$. (The agreement of the two eigenproblems reflects the symmetry of the linear stability operator; see \cite{Galdi1985, Straughan2004}.) We expect $\gamma=1$ to be the optimal coupling parameter since $R_E$ should not exceed $R_L$, and indeed this can be shown directly \cite{Joseph1965}. These observations justify our earlier assertion that $R_E=R_c=R_L$ in RB convection, making subcritical instability impossible.

\begin{table}
\begin{center}
\begin{tabular}{R{10pt} R{65pt} C{40pt} C{50pt} C{35pt} C{35pt}}
&				& $R_E$ 	& \% below $R_L$	& $\gamma^*$	& $k_E$ \\
\hline
IH1 \\
& no-slip			& 26\,926.6	& 27.9\,\%			& 8.8831		& 3.6174 \\
& free-slip top	& 12\,620.2	& 24.3\,\%			& 7.9626		& 2.9014 \\
& free-slip bottom	& 24\,722.8	& 34.9\,\%			& 9.1975 	& 3.3664 \\
& free-slip		& 10\,618.1	& 37.5\,\%			& 8.8516		& 2.5498 \\
IH2 \\
& no-slip			& 1429.86	& 0.704\,\%			& 1.9720		& 0 \\
& free-slip top	& 573.391	& 0.453\,\%			& 1.7838		& 0 \\
& free-slip bottom	& 714.929	& 0.704\,\%			& 2.2185		& 0 \\
& free-slip		& 239.055	& 0.394\,\%			& 1.9843		& 0 \\
IH3 \\
& no-slip			& 2737.16	& 1.27\,\%			& 2.0678		& 2.6355 \\
& free-slip top	& 1594.42	& 1.13\,\%			& 1.9185		& 2.2661 \\
& free-slip bottom	& 1624.26	& 1.59\,\%			& 2.3702		& 2.1512 \\
& free-slip		& 855.674	& 1.39\,\%			& 2.1821		& 1.7958
\end{tabular}
\end{center}
\caption{Rayleigh number ($R_E$) below which the energy method proves that each IH configuration's static state is globally attracting, the percentage of $R_L$ by which $R_E$ falls short of $R_L$, the optimal coupling parameter ($\gamma^*$) used to define the energy that is a valid Lyapunov functional for all $R<R_E$, and the horizontal wavenumber ($k_E$) at which the infimum in (\ref{eq: R_E 2}) occurs for the optimal energy. The IH2 values are numerical approximations to the exact analytical expressions \eqref{eq: RE IH2}. The IH1 and IH3 values are computed numerically and are accurate to the precision shown.}
\label{tab: energy all}
\end{table}

In IH convection, $R_E$ must be calculated by performing the double optimization of expression (\ref{eq: R_E 2}), which requires solving the eigenproblem (\ref{eq: w energy})-(\ref{eq: theta energy}). In all IH cases the strict inequality $R_E<R_L$ holds. Table \ref{tab: energy all} gives values of $R_E$ for the various IH configurations, along with the percent differences between $R_L$ and $R_E$, and the arguments, $\gamma^*$ and $k_E$, that yield the maxima and infima in expression (\ref{eq: R_E 2}).

The relative magnitudes of the gaps between $R_E$ and $R_L$ are on the order of 1\% in IH2 and IH3, where the bottom is insulating, but are much larger in IH1, where heat escapes across both boundaries. We cannot say whether the larger gaps in IH1 are necessitated by subcritical solutions or are only mathematical artifacts of the optimal energies being poor approximations of the truly optimal Lyapunov functionals. The most energy-unstable wavenumber, $k_E$, is fairly close to $k_L$ in IH1 and IH3, and $k_E=k_L=0$ in IH2. The optimal coupling parameters, $\gamma^*$, are all significantly larger than unity, which is their optimal value in the RB cases.

For the IH1 and IH3 cases, we numerically computed the values of $R_E$, $\gamma^*$, and $k_E$ that appear in table \ref{tab: energy all} using the same spectral collocation method that we used to compute $R_L$. A similar energy stability analysis was carried out by Kulacki \& Goldstein \cite{Kulacki1975}. The values of $R_E$ that they reported are smaller than our own, perhaps because their coupling parameters were not quite optimal.\footnote{What we call the IH1 case is designated in \cite{Kulacki1975} by the parameter $Bi_0=\infty$, and what we call the IH3 case is designated by the parameters $Bi_0=0$ and $Bi_1=\infty$. Their Rayleigh numbers are converted to our scaling upon multiplication by 64.}
An energy stability analysis was carried out more recently for the IH3 configuration with no-slip boundaries \cite{Straughan1990}, and those findings agree exactly with our own.

In the IH2 case, the energy stability eigenproblem can be solved exactly using long-wavelength  asymptotics, much like the linear stability eigenproblem (cf.\ \S\ref{sec: long-wave}). This is possible because the infimum of expression (\ref{eq: R_E 2}) is reached as $k^2\to0$, an observation that has not been proven but has been confirmed numerically \cite{Goluskin2015}. The asymptotic calculations, which are detailed in \cite{Goluskin2015}, give the exact expressions
\begin{equation}
R_E=\begin{cases}
2880\big(6\sqrt{35}-35\big) & \text{no-slip} \\[2pt]
360\big(9\sqrt{385}-175\big) & \text{free-slip top} \\[2pt]
1440\big(6\sqrt{35}-35\big) & \text{free-slip bottom} \\[2pt]
1440\big(8\sqrt{7}-21\big) & \text{free-slip}
\end{cases}
\label{eq: RE IH2}
\end{equation}
Numerical approximations of the above values appear in table \ref{tab: energy all} above.

From the standpoint of scientific and engineering applications, the value of knowing $R_E$ in IH convection is that we know convection cannot be sustained when $R<R_E$. When $R$ lies between $R_E$ and $R_L$, little is known about when convection can occur, apart from some instances of subcritical convection that have been computed in IH1 \cite{Tveitereid1977} and IH3 \cite{Schwiderski1972, Tveitereid1976}. This ambiguous regime between $R_E$ and $R_L$ is small in IH2 and IH3, and thus of not much practical importance, but it is much larger in IH1. In any event, we cannot claim to fully understand the static states until we know when subcritical convection is possible---that is, until we know the true value of $R_c$ for every $Pr$. Lower bounds on $R_c$ could be improved by looking beyond the energy method to find better Lyapunov functionals, and upper bounds could be improved by numerically computing steady states that exist in the subcritical regimes.

\section{Bounds depending on the Rayleigh number}
\label{sec: bounds}

Our main goal is to predict the parameter-dependence of integral quantities like $\wT$, $\dT$, and $\oT_{max}$. Much of this effort is equivalent to seeking the functions $N(R,Pr)$ and $\wt N(R,Pr)$, where these Nusselt numbers are defined as in tables \ref{tab: main 1} and \ref{tab: main 2}. (Such functions are multivalued in general since multiple locally attracting solutions can coexist at a given set of parameters.) The stability analyses of \S\S\ref{sec: linear stab}-\ref{sec: energy stab} are useful because they give necessary and sufficient conditions for $N$ and $\wt N$ to equal unity. In particular, $R<R_E$ guarantees that both quantities equal unity, and $R>R_L$ guarantees that both are greater than unity. At large $R$, where convection is strong and complicated, exact expressions for $N(R,Pr)$ and $\wt N(R,Pr)$ are not available. Instead, we seek to bound these quantities analytically.

The only parameter-dependent bounds that have been proven for RB or IH configurations can be stated as upper bounds on how quickly $N$ or $\wt N$, respectively, can grow as $R$ is raised. Upper bounds on $N$ have not been proven in IH convection but seem likely to hold (cf.\ \S\ref{sec: Nusselt numbers}). We cannot improve the lower bounds of unity since known techniques cannot distinguish realizable solutions from the unstable static states.

In this section we outline a proof of lower bounds on the mean temperature, $\dT:=\lan T-\oT_T\ran$, in IH convection. (Recall that $\dT\equiv\lan T\ran$ in IH1 and IH3 but not in IH2, and that lower bounds on $\dT$ are equivalent to upper bounds on $\wt N$.) Our exposition combines existing results for IH1 \cite{Lu2004} and IH2 \cite{Goluskin2015} and a new result for IH3. The proof employs the background method \cite{Doering1992, Constantin1996}, which requires no assumptions beyond the governing equations. Like similar variational methods \cite{Howard1963, Busse1969, Howard1972}, the background method makes progress by relaxing the constraints on $\bu$ and $T$. Instead of enforcing the full Boussinesq equations, we enforce only incompressibility, the boundary conditions, and a few integral relations that follow from the governing equations. This yields bounds that hold for an enlarged class of $\bu$ and $T$ that includes solutions of the Boussinesq equations.

Two main integral relations are typically enforced when the background method is applied to convection. They are called the power integrals and are derived by taking $\lan\bu\cdot(\ref{eq: u})\ran$ and $\lan T\times(\ref{eq: T})\ran$ and integrate by parts to find
\begin{align}
\lan|\nabla\bu|^2\ran &= R\wT \label{eq: u power int} \\
\lan|\nabla T|^2\ran &= \begin{cases}
1+\wT & \text{RB1} \\
\DT=1-\wT & \text{RB2, RB3} \\
\dT & \text{IH1, IH2, IH3}. \label{eq: T power int}
\end{cases}
\end{align}
Time derivatives have vanished from the above relations in the infinite-time limit since the volume integrals of $|\bu|$ and $|T|$ are bounded uniformly in time. This boundedness is proven as a byproduct of the background-method analysis itself \cite{Doering1995, Constantin1996}. The absence of $Pr$ from the relaxed constraints on $\bu$ and $T$ precludes our analysis from producing bounds that depend on $Pr$.

In all six RB and IH configurations, the bounds that have been proven by the background method amount to bounds on the thermal dissipation, $\lan|\nabla T|^2\ran$, though they are often stated in terms of quantities like $\lan wT\ran$ or $\dT$ that are related to $\lan|\nabla T|^2\ran$ by (\ref{eq: T power int}). The thermal dissipation is bounded above in RB1 and below in the other five cases. These results constitute upper bounds on $N$ in RB convection and upper bounds on $\wt N$ in IH convection.

\subsection{Proof by the background method}

We now prove for all three IH configurations that the dimensionless mean temperature, $\dT$, decays no faster than $R^{-1/3}$. This is equivalent to the \emph{dimensional} mean temperature, $\dT\hspace{-1pt}\Delta$, growing with the rate of volumetric heating, $H$, no slower than $H^{2/3}$. We assume a no-slip top in the IH2 case but need not do so in the IH1 or IH3 cases.

To apply the background method, we decompose the temperature into a so-called background profile, $\tau(z)$, and the remaining part, $\Theta(\bx,t)$:
\begin{equation}
T(\bx,t) = \tau(z) + \Theta(\bx,t). \label{eq: background decomp}
\end{equation}
The bound we obtain depends on the $\tau(z)$ we choose. The background profile does not generally solve the governing equations, in which case $\Theta$ does not evolve according to the fluctuation equations (\ref{eq: inc pert})-(\ref{eq: theta}).

The $\tau(z)$ we choose must satisfy several conditions. First, it must be continuous. Second, it must satisfy the same boundary conditions as $T$ since this lets $\Theta$ satisfy the corresponding homogenous conditions. In practice, however, only the fixed-temperature conditions on $\tau(z)$ need to be enforced. This is because fixed-flux conditions on $\tau(z)$ can be met by boundary layers whose influence vanishes as we send their thicknesses to zero \cite{Goluskin2015}. These limiting bounds are the same as those reached by simply ignoring the fixed-flux conditions on $\tau(z)$, so we do the latter in our calculations. Finally, $\tau(z)$ must be chosen to make a particular quantity nonnegative, as explained below. We will see that for all admissible $\tau(z)$,
\begin{equation}
\dT \ge 2\lan\tau-\tau_T\ran - \lan\tau'^2\ran. \label{eq: T bound}
\end{equation}
We choose simple $\tau(z)$ that make our calculations analytically tractable, thereby yielding analytical bounds that are valid at all $R$. Optimizing $\tau(z)$ numerically at a given $R$ would give a tighter bound (as in \cite{Plasting2003}), but the bound would apply only at that value of~$R$.

To see where the inequality (\ref{eq: T bound}) comes from, and when it holds, we expand the power integral (\ref{eq: T power int}) for the IH cases using the decomposition (\ref{eq: background decomp}) to find
\begin{equation}
\dT = \lan|\nabla T|^2\ran = \lan\tau'^2\ran + 2\lan\tau'\Theta'\ran
	+ \lan|\nabla\Theta|^2\ran, \label{eq: gradT expansion}
\end{equation}
where primes denote $z$-derivatives. Our goal is to bound $\dT$ below. (We could equally well speak of bounding $\lan|\nabla T|^2\ran$ below or bounding $\wt N$ above.) The $\lan\tau'\Theta'\ran$ term in the above expression is difficult to bound, so we eliminate it using a third and final integral relation. Integrating $\tau(z)$ against the temperature equation (\ref{eq: T}) gives the needed relation \cite{Lu2004},
\begin{equation}
\lan\tau'\Theta'\ran = \lan\tau-\tau_T\ran - \lan\tau'^2\ran + \lan\tau'w\Theta\ran,
\end{equation}
where the top temperature $\tau_T$ may be nonzero only in the IH2 case. Eliminating $\lan\tau'\Theta'\ran$ from expression (\ref{eq: gradT expansion}) gives
\begin{equation}
\dT = 2\lan\tau-\tau_T\ran - \lan\tau'^2\ran + 
	\lan|\nabla\Theta|^2\ran + 2\lan\tau'w\Theta\ran. \label{eq: T equality}
\end{equation}
From the above equality it follows that the lower bound (\ref{eq: T bound}) would hold if we could show $\lan|\nabla\Theta|^2\ran + 2\lan\tau'w\Theta\ran\ge0$. This is an impossible task for arbitrary $w$, however, since the velocity enters only in the sign-indefinite term. Apparently, the temperature power integral (\ref{eq: T power int}) alone is not sufficiently constraining. We need the additional constraint of the velocity power integral (\ref{eq: u power int}), which tells us that $a\left( \frac{1}{R}\lan|\nabla\bu|^2\ran -\wT \right)=0$ for any $a$. Adding this relation to (\ref{eq: T equality}) shows that the lower bound (\ref{eq: T bound}) on $\dT$ would follow from the nonnegativity of the quadratic functional
\begin{equation}
\cQ[\bu,\Theta;\tau(z,R)] := \tfrac{a}{R}\lan|\nabla\bu|^2\ran + \lan|\nabla\Theta|^2\ran + \lan(2\tau'-a)w\Theta\ran. \label{eq: Q}
\end{equation}
We must choose a $\tau(z)$ for which we can verify that $\cQ\ge0$ for all admissible $\bu$ and~$\Theta$.

More generally, the background method is carried out by finding an expression, equal to the quantity to be bounded, that takes the form $\mathcal B+\cQ$, where $\mathcal B$ is a functional of the the background field alone, while $\cQ$ depends also on the other fields. The key idea is that $\mathcal B$ will be a lower bound when we can show that $\cQ$ is nonnegative (or an upper bound when we can show that $\cQ$ is nonpositive). In the present analysis, $\mathcal B:=2\lan\tau-\tau_T\ran - \lan\tau'^2\ran$, and $\cQ$ is as defined in (\ref{eq: Q}). 

\begin{figure}
\begin{center}
\includegraphics[width=150pt]{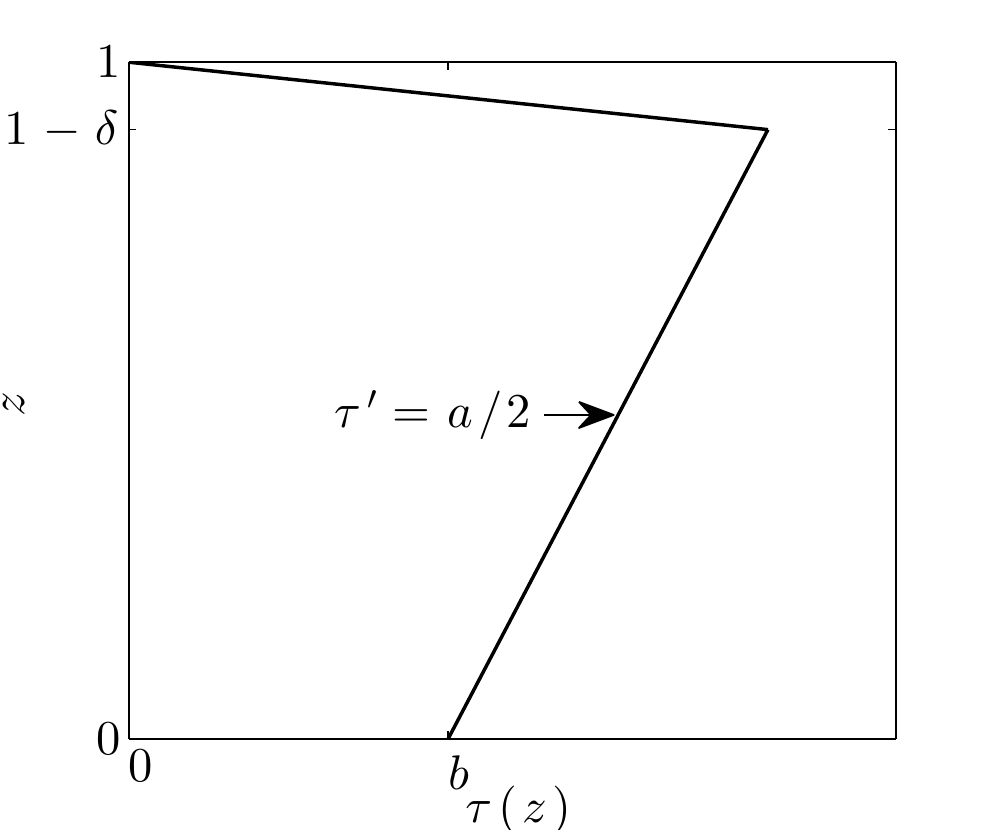}
\end{center}
\caption{Schematic of the class of background profiles, $\tau(z)$, that we consider. The parameters $a$, $b$, and $\delta$ are optimized, within some constraints, to maximize the lower bounds on the mean temperature.}
\label{fig: tau schematic}
\end{figure}

Two objectives compete in the choice of $\tau(z)$: making the lower bound (\ref{eq: T bound}) as large as possible, and maintaining the nonnegativity of $\cQ$ that is needed for that bound to be valid. Here, we optimize $\tau(z)$ only among profiles consisting of two linear pieces. Such profiles can all be written in the ansatz
\begin{equation}
\tau(z) = \begin{cases}
\left[\tfrac{b}{\delta}+\tfrac{a}{2}\left(\tfrac{1}{\delta}-1\right)\right](1-z) & 
	1-\delta\le z\le1 \\
b+\tfrac{a}{2}z & 0\le z\le1-\delta, \label{eq: linear tau}
\end{cases}
\end{equation}
where the geometric meanings of parameters $a$, $b$, and $\delta$ are shown in figure \ref{fig: tau schematic}. We will see that the top piece of $\tau(z)$ is a boundary layer whose thickness, $\delta$, goes to zero as $R\to\infty$. The bottom piece of $\tau(z)$ has a slope that is half the value of the yet-unspecified constant $a$, a known trick \cite{Constantin1996, Lu2004} for making the sign-indefinite term of $\cQ$ vanish outside the boundary layer.

For our three-parameter family of background profiles (\ref{eq: linear tau}), the lower bound (\ref{eq: T bound}) becomes
\begin{equation}
\dT \ge 
	b(2-\delta) + \tfrac{a}{2}(1-\delta) - 
	\left( \tfrac{a^2}{4}+ab \right)\left(\tfrac{1}{\delta}-1\right) - \tfrac{b^2}{\delta}.
	\label{eq: T bound 2}
\end{equation}
With a no-slip top in IH2 and any velocity conditions in IH1 or IH3, it can be shown that $\cQ\ge0$ is satisfied when $\delta$ is no larger than \cite{Lu2004, Goluskin2015}
\begin{equation}
\delta^4 = \begin{cases}
64\,a\,R^{-1} & \text{IH1, IH3} \\[2pt]
32\,a\,R^{-1} & \text{IH2}.
\end{cases} \label{eq: delta condition}
\end{equation}
We choose this $\delta$ because the tightest bounds result from choosing $\delta$ as large as possible. 

In the IH2 and IH3 cases, we are free to choose the parameters $\delta$, $a$, and $b$ to maximize the lower bound (\ref{eq: T bound 2}) subject to (\ref{eq: delta condition}). In IH1, the lower boundary condition requires that $b=0$, so we are free to choose only $\delta$ and $a$. This maximization is carried out for IH1 and IH2 in \cite{Lu2004} and \cite{Goluskin2015}, respectively, and the procedure for IH3 is analogous. The resulting optimal parameters are
\begin{align}
\delta^* &=
\begin{cases}
\hspace{22pt}4R^{-1/3} \\
\hspace{5.5pt}12^{1/3}R^{-1/3} \\
2\cdot3^{1/3}R^{-1/3}
\end{cases} 
&
a^* &=
\begin{cases}
\hspace{20pt}4R^{-1/3} \\
\tfrac{3\cdot12^{1/3}}{8}R^{-1/3} \\
\hspace{10pt}\tfrac{3^{4/3}}{4}R^{-1/3}
\end{cases}
&
b^* &=
\begin{cases}
0 & \text{IH1} \\
\tfrac{5\cdot12^{1/3}}{16}R^{-1/3} & \text{IH2} \\
\hspace{4pt}\tfrac{5\cdot3^{1/3}}{8}R^{-1/3} & \text{IH3},
\end{cases}
\label{eq: optimal parameters}
\end{align}
for which the lower bound (\ref{eq: T bound 2}) becomes
\begin{equation}
\dT \ge \begin{cases}
\hspace{35pt}R^{-1/3} - \hspace{34pt}4R^{-2/3} & \text{IH1} \\
\tfrac{9}{8}\left(\tfrac{3}{2}\right)^{1/3}R^{-1/3} - \tfrac{89}{64}\left(\tfrac{3}{2}\right)^{2/3}R^{-2/3} & \text{IH2} \\
\hspace{20pt}\tfrac{3^{7/3}}{8}R^{-1/3} - \hspace{13.5pt} \tfrac{89\cdot3^{2/3}}{64}R^{-2/3} & \text{IH3}.
\end{cases}
\end{equation}

At large $R$, the leading-order terms of the bounds dominate:
\begin{equation}
\dT \gtrsim \begin{cases}
~~~~~~~~R^{-1/3} & \text{IH1} \\
1.28\;R^{-1/3} & \text{IH2} \\
1.62\;R^{-1/3} & \text{IH3}.
\end{cases}
\label{eq: asymptotic T bounds}
\end{equation}
When Lu \emph{et al.}\ \cite{Lu2004} proved the above bound for IH1, they also raised the prefactor from 1 to 1.09 by generalizing the ansatz of $\tau(z)$ to include a bottom boundary layer, although this required solving an algebraic equation numerically. Their proof carries through for the IH3 case also, so they in fact proved the asymptotic lower bound of $1.09\;R^{-1/3}$ for both IH1 and IH3. By dropping the condition $\tau(0)=0$ in IH3, where it is not needed, we have raised the prefactor to 1.62. Optimizing $\tau(z)$ beyond our limited ansatz would lower the prefactors of the bounds, but results of numerically optimizing $\tau(z)$ in the RB1 case suggest that the scaling of the bounds would not change \cite{Plasting2003}.

\subsection{Similarities between RB and IH bounds}

A main virtue of the way we have defined the Nusselt numbers $N$ and $\wt N$ and the diagnostic Rayleigh number $Ra$ and $\wt{Ra}$ is that bounds for the various configurations all have the same scaling when expressed using these quantities. Recalling that the definitions (\ref{eq: N tilde def 2}) of $\wt N$ are inversely proportional to $\dT$, and that $\wt{Ra}=R/\wt N$ in IH convection, we see that the asymptotic bounds (\ref{eq: asymptotic T bounds}) on $\dT$ become
\begin{equation}
\wt N \lesssim \begin{cases}
0.025\;\wt{Ra}^{1/2} & \text{IH1} \\
0.132\;\wt{Ra}^{1/2} & \text{IH2} \\
0.094\;\wt{Ra}^{1/2} & \text{IH3}.
\end{cases}
\label{eq: asymptotic N bounds}
\end{equation}
These upper bounds on $\wt N$ appear in table \ref{tab: no-slip summary} at the start of this chapter, along with the best known bounds on $N$ in the RB configurations. The RB bounds have different prefactors but the same exponent, scaling proportionally to $Ra^{1/2}$. The RB1 prefactor in table \ref{tab: no-slip summary} comes from the improvement on \cite{Constantin1996} by Plasting \& Kerswell \cite{Plasting2003}, who also showed that their bound could not be improved without additional constraints. The prefactor in the other two RB cases comes from \cite{Otero2002}, where the analysis was aimed at RB2 but carries through for RB3 also. 

Upper bounds with an exponent of $1/2$ are the best available for three-dimensional convection in general, but bounds with smaller exponents have been proven in special cases. Here too, analogies hold between various configurations if results are stated in terms of our diagnostic parameters. When the boundaries are free-slip, and either $Pr=\infty$ or the flow is two-dimensional, upper bounds with exponents of $5/12$ have been proven for RB1 \cite{Whitehead2011a, Whitehead2012} and IH1 \cite{Whitehead2011, Whitehead2012}. When $Pr=\infty$ with no-slip boundaries, the best known bounds on $N$ scale like $Ra^{1/3}(\log\log Ra)^{1/3}$ in RB1 \cite{Otto2011} and like $Ra^{1/3}(\log Ra)^{1/2}$ in RB2 and RB3 \cite{Whitehead2014}, and the best known bound on $\wt N$ scale like $\wt{Ra}^{1/3}(\log \wt{Ra})^{1/3}$ in IH1 \cite{Whitehead2011}. Bounds with exponents smaller than $1/2$ are yet to be reported for the other RB or IH configurations.

Now that we have seen how the background method works, we can understand why it is challenging in the IH cases to prove upper bounds on $\wT$. This quantity is related to $\lan|\nabla\bu|^2\ran$ by the velocity power integral (\ref{eq: u power int}) but is not is not related \emph{a priori} to $\lan|\nabla T|^2\ran$ in IH convection. However, the parameter-dependent bounds that have been proven for convective models all amount to bounds on $\lan|\nabla T|^2\ran$ and rely on a background decomposition of the temperature field. Bounding $\lan|\nabla\bu|^2\ran$ instead suggests a background decomposition of the velocity field, which has been carried out for shear flows (e.g.\ in \cite{Constantin1994}) but not for convection.

\chapter{Internally heated convection experiments and simulations}
\chaptermark{Internally heated experiments and simulations}
\label{chap: exp}

The first two chapters have summarized features of heat transport in IH convection and RB convection that can be ascertained analytically from the Boussinesq equations. In this final chapter we summarize findings on IH convection from physical and computational experiments. Analogous results for RB convection are described only minimally, as the experimental literature on RB convection is vast and has been reviewed elsewhere (e.g.\ \cite{Siggia1994, Getling1998, Ahlers2009, Lohse2010a, Chilla2012}).

Precise laboratory experiments on IH convections are inherently more difficult to carry out than similar experiments on RB convection. Both require maintaining the chosen thermal boundary conditions, but IH experiments also require producing heat internally in a controlled way. If our simple models are to apply, the heat production should be constant and uniform. In most experiments, the internal heating has been achieved by Joule heating, where the working fluid is an electrolytic solution that is heated by passing current through it. Two sets of experiments \cite{Ralph1977, Lee2007} used a different method, wherein heating elements were distributed throughout the domain. Although neither method heats uniformly, it is possible that rapid mixing by strong convection limits the influence of non-uniformity. This is supported by the fairly good agreement between non-uniformly heated experiments and uniformly heated simulations.

Numerical simulations of IH convection avoid unknown variations in heating rate or material properties. However, most numerical studies were carried out several decades ago and were limited to 2D and fairly small $R$. The larger values of $R$ accessible with modern computers have been simulated only a few times, and much of the parameter space that could now be reached has yet to be explored.

Experimental findings before 1985 are collected in the review of Kulacki \& Richards \cite{Kulacki1985}, who discuss findings on IH1, IH3, and some similar configurations. The slightly later review of Cheung \& Chawla \cite{Cheung1987} adds various scaling arguments for heat transport. Nourgaliev \emph{et al.}\ \cite{Nourgaliev1997} summarize heat fluxes in these same early experiments, as well as in experiments with curved geometries and cooled side walls.

A number of experiments have examined heat transport quantitatively, and a number of others have focused on qualitative pattern formation near the onset of convection. Here we cite studies of both types but focus on quantitative findings, and we restrict ourselves to experiments that closely resemble one of the IH configurations defined in figure \ref{fig: configs}. Convection with internal heating has been studied also with various complications that we do not confront, such as cooled side walls \cite{Steinberner1978, Bergholz1980, Asfia1996, Nourgaliev1997, DiPiazza2000, Arcidiacono2001, Arcidiacono2001b, Horvat2001, Shi2003, Liu2006, Chen2009, Filippov2011}, non-uniform heating \cite{Riahi1984, Straughan1990, Tasaka2005, Kondratenko2008}, self-gravitating spheres \cite{Roberts1965, Roberts1968, Busse1975, Busse1975a, Ingersoll1978, Riahi1988}, and hybrid configurations driven both internally and by the boundary conditions \cite{Sparrow1964, Joseph1966a, Joseph1968a, Mckenzie1974, Straus1976, Clever1977, Chapman1980a, Houseman1988, Ames1990, Sotin1999, Veltishchev2004, Hartlep2006, Berlengiero2012, Kolmychkov2013}.

Section \ref{sec: IH3} addresses IH3, the last of the three IH configurations defined in figure \ref{fig: configs}. Section \ref{sec: IH1} addresses IH1, which is in some ways more complicated than IH3. We are not aware of any heat transport findings on the IH2 configuration, though 2D simulations have been carried out to study scale selection \cite{Hewitt1980, Ishiwatari1994}. Section \ref{sec: future} suggests directions for future work.

\section{The IH3 configuration}
\label{sec: IH3}

The internally heated configuration we call IH3, which is bounded above by a perfect conductor and below by a perfect insulator, has been the subject of numerous laboratory experiments \cite{Tritton1967, Schwiderski1971, DelaCruzReyna1970, Fiedler1970, Ralph1974, Kulacki1975, Kulacki1977, Ralph1977, Tasaka2005a, Lee2007, Takahashi2010}, as well as computational studies both in 2D \cite{Thirlby1970, Mckenzie1974, Emara1980, Ishiwatari1994} and in 3D \cite{Thirlby1970, Tveitereid1976, Houseman1988, Schubert1993, Ichikawa2006, CartlandGlover2009, CartlandGlover2013}. Many of these investigations have focused on pattern formation and scale selection, which we do not discuss here. Our interest is in quantities relevant to heat transport, including the mean vertical temperature profile, $\oT(z)$, and the mean temperature difference between the boundaries, $\DT$. No data are available on the mean fluid temperature, $\dT$.

Computational studies of IH3 convection that report $\oT(z)$ or $\DT$ are all several decades old. Most are limited to the steady states that are stable at modest $R$ \cite{Mckenzie1974, Thirlby1970, Tveitereid1976, Schubert1993}. At larger $R$, unsteady 2D simulations have been carried out using a turbulence closure model \cite{Farouk1988} and by direct numerical simulation (DNS) \cite{Emara1980}, although some runs in the latter study appear under-resolved. As far as we know, unsteady IH3 convection has not been simulated in 3D. The largest $R$ that has been reached in 2D DNS of IH3 \cite{Emara1980} could be greatly exceeded in 3D DNS on modern parallel computers.

Laboratory experiments, most of which were carried out in the 1970s, furnish nearly everything we know about $\oT(z)$ and $\DT$ in IH3 convection at large $R$ \cite{Fiedler1970, Ralph1974, Kulacki1975, Kulacki1977, Ralph1977, Lee2007}. These findings are subject to the uncertainties inherent to IH experiments, so there is cause to repeat them numerically. The largest $R$ reached in past laboratory experiments of IH3 could now be approached by 3D DNS, albeit in a smaller spatial domain.

\subsection{Temperature profiles}
\label{sec: IH3 profiles}

Several authors have reported mean vertical temperature profiles. It is simple to obtain $\oT(z)$ in numerical studies by averaging steady flows horizontally \cite{Thirlby1970, Mckenzie1974, Tveitereid1976} or averaging unsteady flows both horizontally and temporally \cite{Emara1980, Farouk1988}. In laboratory experiments, vertical profiles have been obtained by measuring temperatures at fixed points and averaging only over time \cite{Ralph1974}. If the flow is horizontally isotropic in a statistical sense, and time averages are sufficiently long, then the same mean profile would be obtained whether or not horizontal averages are also taken. This is generally expected to be true at large $R$ when side walls are absent or negligible. Some transient profiles have been reported also \cite{Kulacki1975, Kulacki1977}, but these do not bear directly on the infinite-time averages we seek.

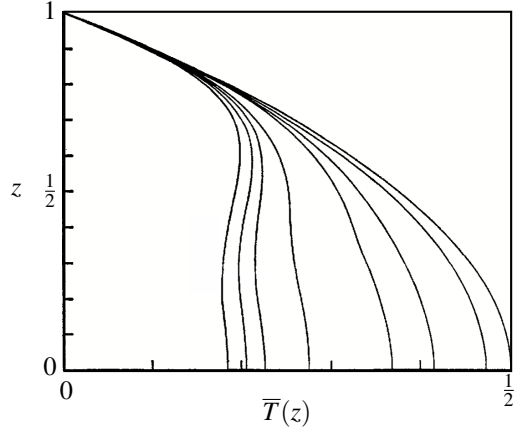
\begin{figure}
\begin{center}
\begin{tikzpicture}
\node[anchor=south,inner sep=0] (0,0) {\includegraphics[width=170pt]{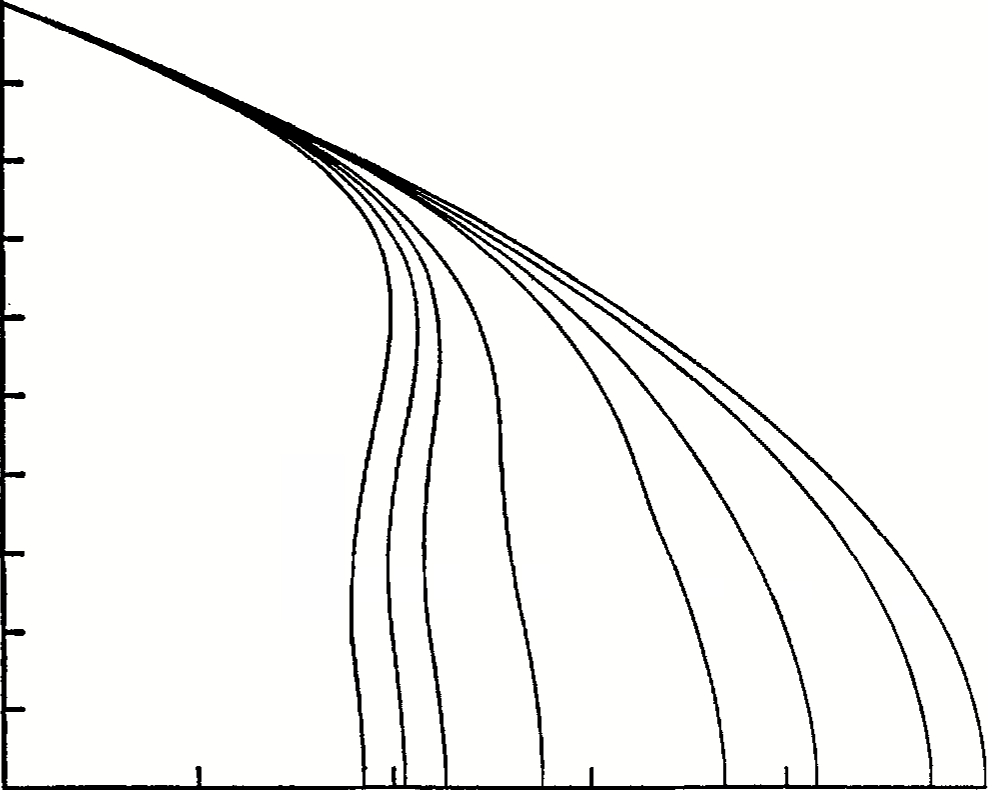}};
\draw[thick] (-2.96,0) rectangle (2.98,4.78);
\draw (-2.93,-.25) node {0};
\draw (2.95,-.25) node {$\tfrac{1}{2}$};
\draw (0,-.55) node {$\oT(z)$};
\draw (-3.15,.07) node {0};
\draw (-3.15,2.4) node {$\tfrac{1}{2}$};
\draw (-3.15,4.8) node {1};
\draw (-3.6,2.4) node {$z$};
\end{tikzpicture}
\end{center}
\caption{Numerically computed mean temperature profiles, $\oT(z)$, for steady 2D convection between no-slip boundaries. The Prandtl number is 6.8. The rightmost profile is that of the static state. The others, from right to left, are for $R=10^3\times (3, 4, 5, 10, 20, 30, 52)$. The figure is adapted from figure 2 of Thirlby \cite{Thirlby1970}}
\label{fig: IH3 profiles}
\end{figure}

Figure \ref{fig: IH3 profiles} shows $\oT(z)$ profiles for 2D steady states computed by Thirlby \cite{Thirlby1970} for $Pr=6.8$ and relatively small $R$. As $R$ is raised and convection strengthens, the dimensionless temperature decreases, and the interior becomes closer to isothermal. When convection is sufficiently strong, the maximum value of $\oT(z)$ occurs inside the layer, rather than at the bottom boundary. At still larger $R$, where convection is stronger and unsteady, the experimentally measured profiles of Ralph \& Roberts \cite{Ralph1974} follow similar trends but are closer to isothermal in the interior, lacking the pronounced temperature inversion found in the steady states of figure \ref{fig: IH3 profiles}. The unsteady motions responsible for homogenizing temperature outside the thermal boundary layer are evident in the IH3 temperature field of figure \ref{fig: T examples}c: plumes emerge from the unstably stratified upper boundary layer and strongly mix fluid in the rest of the domain.

\subsection{Mean temperature differences}
\label{sec: IH3 DT}

The difference by which the dimensionless temperature at any point exceeds its average value at the top boundary, $T-\oT_T$, tends to decrease as IH convection strengthens. This fact underlies the two related but distinct measures of convective strength discussed in \S\ref{sec: relations}: $\dT$, which is the average of $T-\oT_T$ over time and the entire volume, and $\DT$, which is its average over time and the bottom boundary. (Recall that $\DT$ is also the mean vertical conduction, and in IH3 it is tied to the mean vertical convection, $\wT$, by the relation $\DT+\wT=1/2$.) Both $\DT$ and $\dT$ are maximal in the static state and most likely approach zero in convective flows as $R\to\infty$.

A primary question for experimentalists is \emph{how quickly} $\DT$ and $\dT$ fall as $R$ is raised, along with how this answer is affected by the velocity boundary conditions, Prandtl number, and geometry. If $\DT$ and $\dT$ vary approximately as powers of $R$ when other parameters are held constant, data can be captured by fits of the form
\begin{align}
\DT&\sim a\,R^{-\alpha} & \dT&\sim b\,R^{-\beta}.
\label{eq: IH3 fits}
\end{align}
In the IH3 case, the Nusselt numbers and diagnostic Rayleigh numbers defined in \S\S\ref{sec: Nusselt numbers}-\ref{sec: diagnostic Rayleigh numbers} are
\begin{align}
N &= 1/2\DT & Ra &= R/N \\
\wt N &= 1/3\dT & \wt{Ra} &= R/\wt N.
\label{eq: IH3 N and Ra}
\end{align}
Various authors have considered quantities like $N$ in past studies of IH3, and Fiedler \& Wille \cite{Fiedler1970} considered both $N$ and $Ra$ together. Restated in terms of the diagnostic variables, the fits of expression (\ref{eq: IH3 fits}) become
\begin{align}
N&\sim c\,Ra^\gamma & \wt N&\sim d\,\wt{Ra}^\delta,
\label{eq: N fits}
\end{align}
were $\gamma=\alpha/(1-\alpha)$, $\delta=\beta/(1-\beta)$, $c=(2a)^{-1/(1-\alpha)}$, and $d=(3b)^{-1/(1-\beta)}$.

\def\arraystretch{1.3}
\begin{table}[t]
\begin{center}
\begin{tabular}{rC{45pt}C{90pt}R{50pt}R{55pt}}
& $Pr$ & $R$ & $\DT$ fit & $N$ fit \\
\hline
\multicolumn{1}{l}{Laboratory experiments}\\
\hline
Fiedler \& Wille \cite{Fiedler1970} &
	$6-7$ & $10^4-10^7$ & $1.90\,R^{-0.228}$ &
	$0.177\,Ra^{0.295}$ \\
Ralph \& Roberts \cite{Ralph1974} &
	$6-7$ & $2.3\e5-6.0\e9$ & $2.62\,R^{-0.25}\hspace{3pt}$ &
	$0.110\,Ra^{0.33}\hspace{3pt}$ \\
Kulacki \& Nagle \cite{Kulacki1975} &
	$6.2-6.6$ & $1.5\e5-2.5\e9$ & $3.28\,R^{-0.239}$ &
	$0.0845\,Ra^{0.314}$ \\
Kulacki \& Emara \cite{Kulacki1977} &
	$2.7-6.9$ & $\hspace{3pt}1.89\e3-2.17\e{12}$ & $2.53\,R^{-0.227}$ &
	$0.123\,Ra^{0.294}$ \\
Ralph \emph{et al.} \cite{Ralph1977} &
	$6-7$ & $\hspace{10pt}10^9-7\e9$ & $a\,R^{-0.24}\hspace{3pt}$ & $c\,Ra^{0.32}\hspace{3pt}$ \\
Lee \emph{et al.} \cite{Lee2007} &
	$0.71-0.74$ & $\hspace{3pt}9.9\e9-3.3\e{11}$ & $2.84\,R^{-0.247}$ &
	$0.0996\,Ra^{0.328}$ \\
\hline
Simulations (2D DNS)\\
\hline
\rowcellR{Mckenzie \emph{et al.} \cite{Mckenzie1974}\\(free-slip, steady)} &
	$\infty$ & $1.2\e4-7.0\e5$ & $a\,R^{-0.26}\hspace{3pt}$ &
	$c\,Ra^{0.35}\hspace{3pt}$ \\
\rowcellR{Emara \& Kulacki \cite{Emara1980}\\(free-slip top)} &
	6.5 & $5\e4-5\e8$ & $1.07\,R^{-0.182}$ & $0.397\,Ra^{0.222}$\\
\rowcellR{Emara \& Kulacki \cite{Emara1980}\\(no-slip)} &
	6.5 & $5\e3-5\e8$ & $2.38\,R^{-0.223}$ & $0.134\,Ra^{0.287}$\\
\rowcellR{Olwi \cite{Olwi1995}\\(no-slip, steady)} &
	6.5 & $10^4-10^8$ & $3.07\,R^{-0.255}$ & $0.0876\,Ra^{0.342}$
\end{tabular}
\end{center}
\caption{Summary of IH3 experiments and simulations reporting approximate power-law dependence of $\DT$ on $R$. Internal heating was achieved by electric current in the first four experiments and by heating elements in the last two. The Prandtl number range $6-7$ is an estimate for experiments that used aqueous solutions but did not report $Pr$ measurements \cite{Fiedler1970, Ralph1974, Ralph1977}.}
\label{tab: IH3 fits}
\end{table}
\def\arraystretch{1}

The mean temperature difference $\DT$ has been measured in a number of experiments. Table \ref{tab: IH3 fits} summarizes past fits of the form $\DT\sim a\,R^{-\alpha}$, along with their corresponding re-expressions as fits of the form $N\sim c\,Ra^\gamma$. Ranges of $Pr$ and $R$ are also given. The stated ranges of $R$ are those over which the data have been fit. The Prandtl number would ideally be held constant as $R$ is changed, but slight variations are unavoidable in the laboratory. Numerical studies do not suffer from this uncertainty, but the simulation results in table \ref{tab: IH3 fits} nonetheless must be regarded with care since they all are 2D and seem to be somewhat under-resolved at larger $R$.

The decay rates of $\DT$ reported for the six laboratory experiments in table \ref{tab: IH3 fits} fall between $\alpha=0.227$ and $\alpha=0.25$. This means that the dimensional temperature difference between the boundaries, $\DT\Delta$, \emph{grows} with the volumetric heating at rates between $H^{0.75}$ and $H^{0.773}$.

When the $\DT$ fits in table \ref{tab: IH3 fits} are restated in the form $N\sim c\,Ra^\gamma$, the exponents fall between $\gamma=0.294$ and $\gamma=0.33$. This range agrees very well with the analogous range of $\gamma$ measured in RB1 experiments, where fits still take the form $N\sim c\,Ra^\gamma$, but with $N:=1+\wT$ and $Ra:=R$ (cf.\ \S\ref{sec: Nusselt numbers}). The RB1 exponents summarized in table 1 of \cite{Grossmann2000} lie between 0.25 and 0.33, excluding the very small $Pr$ values for which corresponding IH3 data are unavailable. Exponents larger than 0.33 have sometimes been measured in RB1 experiments at very large $R$ \cite{Chavanne1997, He2012}, but no IH3 experiments have reached such $R$ values. The similarity between measured values of $\gamma$ in RB1 and IH3 is one of the analogies brought out by our chosen definitions of $N$ and $Ra$.

No data have been reported on the volume-averaged quantity $\dT$, so we cannot say exactly what exponents would emerge from fits of the form $\dT\sim bR^{-\beta}$ or $\wt N\sim d\,\wt{Ra}^\delta$. We can reasonably estimate the exponents, however, since the temperature profiles that have been reported are close to isothermal outside their boundary layers. This suggests that the values of $\dT$ and $\DT$ become ever closer as $R\to\infty$, in which case $\alpha\approx\beta$ and $\gamma\approx\delta$ for sufficiently large $R$. This speculation remains to be tested since volume averages like $\dT$ are difficult to measure in the laboratory. They are easy to extract from simulations, however, and we hope that future numerical studies will report $\dT$. 

Whereas we have data on $\DT$ but not on $\dT$---or, equivalently, on $N$ but not on $\wt N$---the state of affairs for analytical bounds is just the opposite. We have conjectured in chapter \ref{chap: intro}, but have not proven, than $N$ obeys an upper bound of the form $c\,Ra^{1/2}$. The experimental exponents $\beta$ in table \ref{tab: IH3 fits} are all smaller than 1/2 and thus consistent with this conjecture. On the other hand, we \emph{have} proven in chapter {chap: stab} that $\wt N$ can grow no faster than $0.093\,\wt{Ra}^{1/2}$, but no data on $\dT$ have been reported for the IH3 configuration.

\section{The IH1 configuration}
\label{sec: IH1}

The internally heated configuration we call IH1, which is bounded above and below by perfect conductors of equal temperature, has been studied in the laboratory \cite{Kulacki1972, Jahn1974, Mayinger1975, Ralph1977, Jaupart1984, Jaupart1986, Lee2007}, as well as numerically both in 2D \cite{Jahn1974, Mayinger1975, Peckover1974, Straus1976, Tveitereid1977, Emara1980, Goluskin2012} and in 3D \cite{Grotzbach1988, Worner1997, Goluskin2015b}. Almost all of these studies have reported quantitatively on heat transport in some way.

Numerical computations of IH1 include both steady states \cite{Peckover1974, Straus1976, Tveitereid1977} and DNS. Whereas DNS of the IH3 configuration has been limited to a single 2D study, DNS of the IH1 configuration has been carried out up to fairly large $R$ in both 2D \cite{Jahn1974, Emara1980, Goluskin2012} and 3D \cite{Grotzbach1988, Worner1997, Goluskin2015b}.

\subsection{Temperature profiles}
\label{sec: IH1 profiles}

Mean vertical temperature profiles have been reported in a number of studies. Numerical studies provide profiles, $\oT(z)$, that are averaged horizontally and, if the simulations are unsteady, over time as well \cite{Peckover1974, Mayinger1975, Straus1976, Emara1980, Grotzbach1988, Worner1997, Goluskin2012, Goluskin2015b}. In the laboratory, profiles measured pointwise by temperature probes are averaged only over time \cite{Ralph1977, Lee2007}, while profiles gleaned from interferograms are instantaneous but effectively averaged over a horizontal direction \cite{Kulacki1972, Mayinger1975}.

\begin{figure}
\begin{center}
(a)
\begin{tikzpicture}
\node[anchor=south,inner sep=0] (0,0) {\includegraphics[width=218pt]{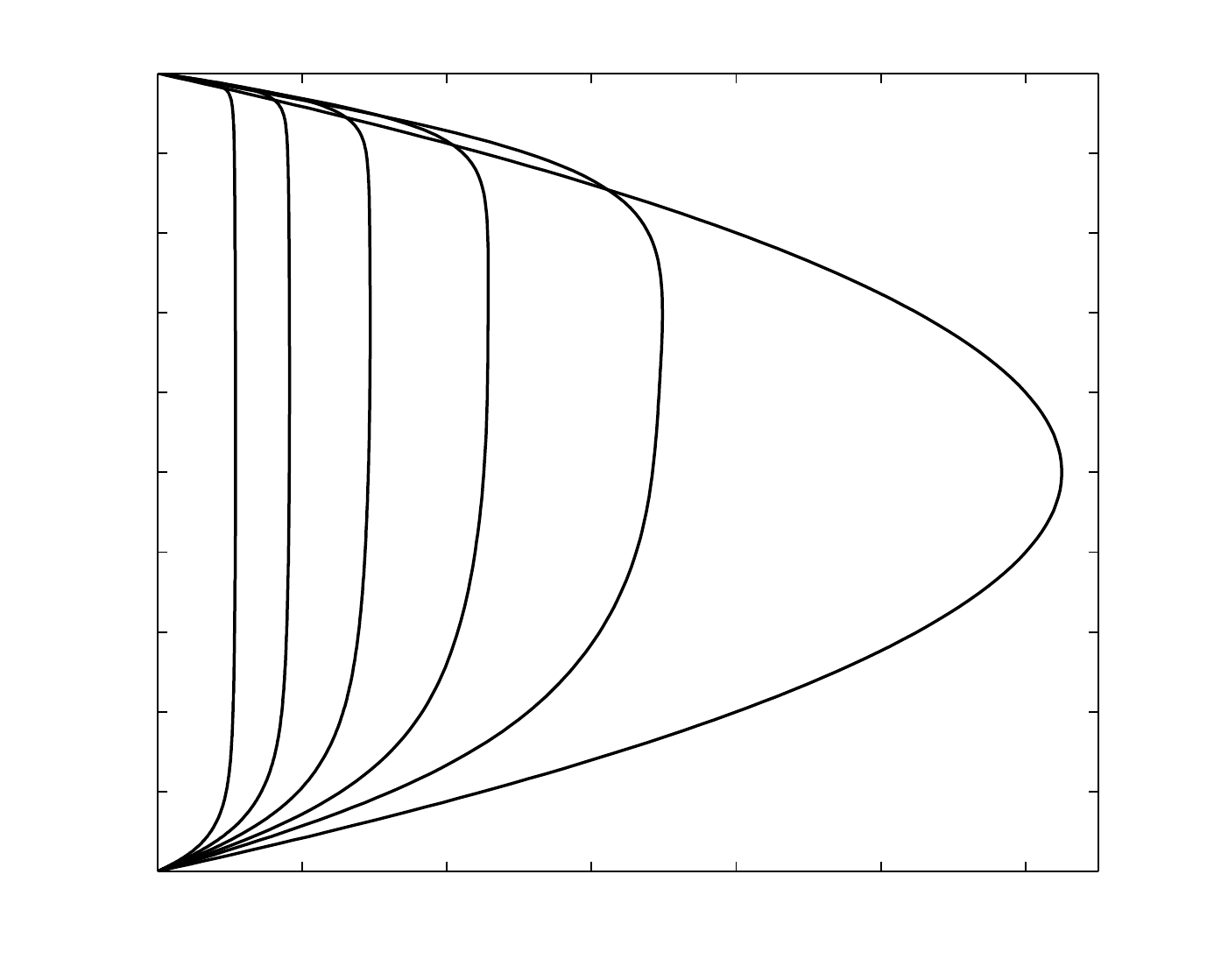}};
\draw (-2.8,.5) node {0};
\draw (1.75,.5) node {0.1};
\draw (.15,0.15) node {$\oT(z)$};
\draw (-3.1,.7) node {0};
\draw (-3.1,3.21) node {$\tfrac{1}{2}$};
\draw (-3.1,5.7) node {1};
\draw (-3.6,3.21) node {$z$};
\end{tikzpicture}
\\
(b)
\quad \includegraphics[width=168pt]{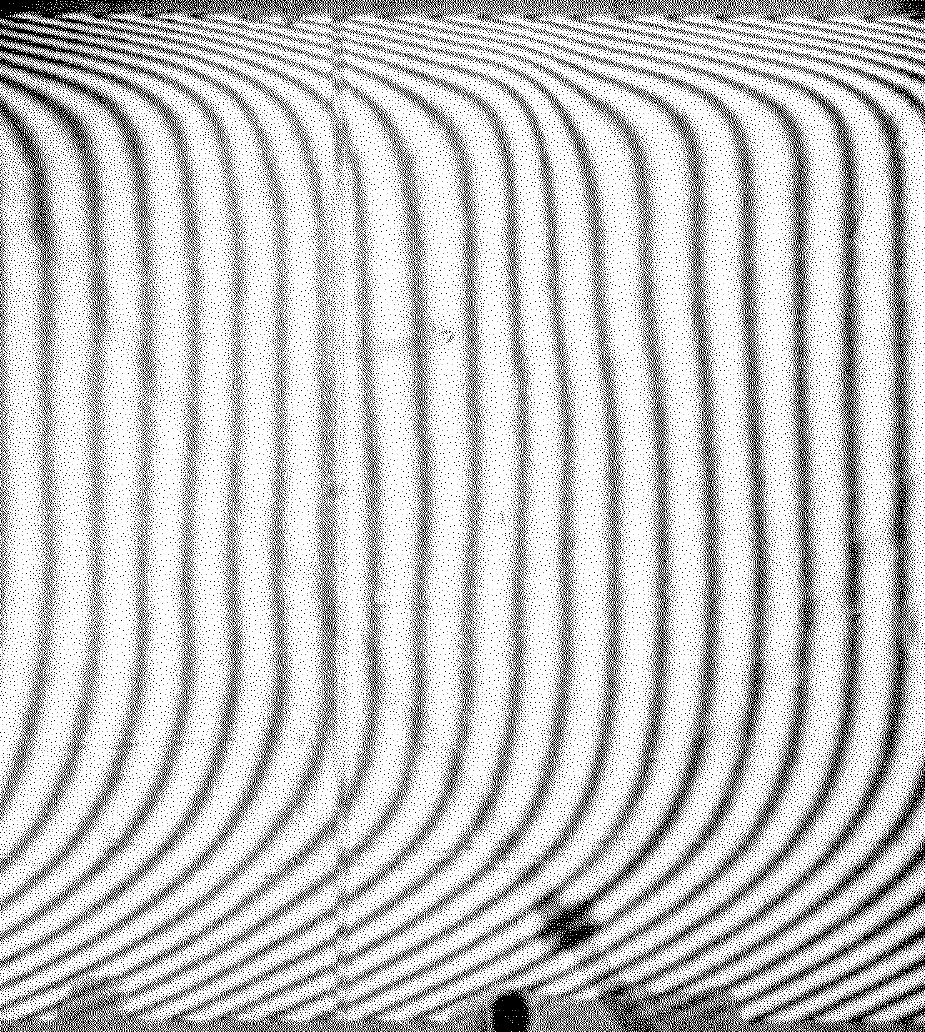}
\end{center}
\caption{(a) Mean temperature profiles, $\oT(z)$, from the 3D DNS of Goluskin \& van der Poel \cite{Goluskin2015b} for a fluid with $Pr=1$ between no-slip boundaries. The rightmost profile is that of the static state. The others, from right to left, are for $R=10^6$, $10^7$, $10^8$, $10^9$, and $10^{10}$. (b) An interferogram from the experiments of Kulacki and Goldstein \cite{Kulacki1972} with $R=1.5\e5$ and $Pr=5.8$. Any curve of constant color acts approximately as a graph of instantaneous temperature averaged over one horizontal direction.}
\label{fig: IH1 profiles}
\end{figure}
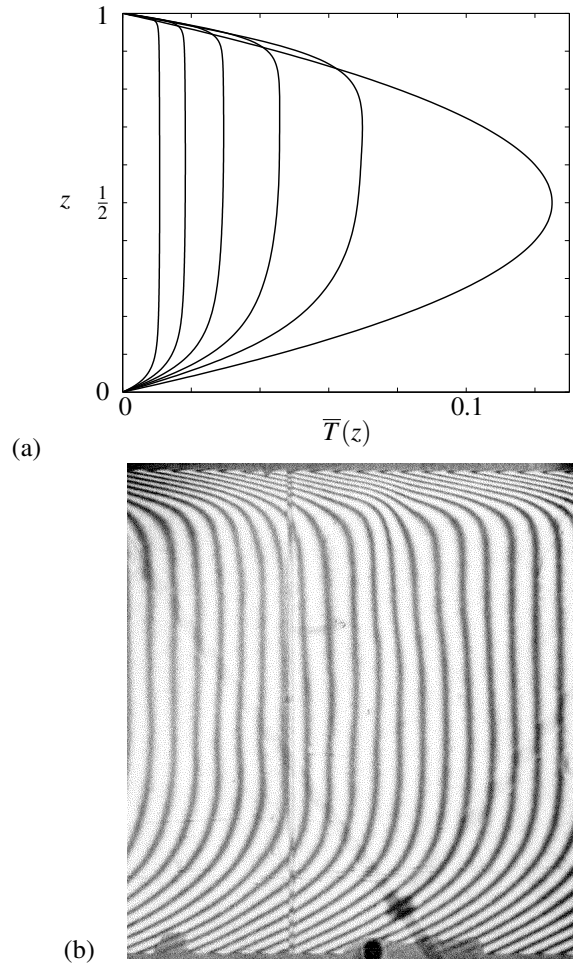

Figure \ref{fig: IH1 profiles}a shows $\oT(z)$ profiles from the 3D DNS data of Goluskin \& van der Poel \cite{Goluskin2015b}. As in the IH3 configuration, raising $R$ strengthens convection, which decreases the dimensionless temperature and brings the interior closer to isothermality. When $R$ is large enough for thermal boundary layers to be discernible, the top boundary layer is visibly thinner than the bottom one. This reflects the up-down asymmetry of heat fluxes; more of the produced heat flows outward across the top boundary than across the bottom one, as quantified in the next subsection. The same basic features are evident in Figure \ref{fig: IH1 profiles}b, which shows an interferogram from the experiments of Kulacki \& Goldstein \cite{Kulacki1972}. The interferogram measures horizontally averaged optical properties of the fluid that vary with its temperature, and a line of constant color can be interpreted as a temperature profile.

The IH1 configuration stands out from the other RB and IH models we have discussed in that there is a stably stratified thermal boundary layer. The configuration thus provides a simple instance of \emph{penetrative convection}, wherein buoyancy forces in an unstably stratified region drive motions that penetrate into a stably stratified region. The temperature field of figure \ref{fig: T examples}b reflects the dissimilarity between the unstably stratified upper boundary layer and the stably stratified lower one. Mixing of the cold upper layer with the warmer interior is accomplished by buoyantly driven cold plumes. At large $R$, the cold \emph{lower} layer also can mix with the warmer interior. This mixing is driven by shear forces, rather than by buoyancy, and it occurs when the interior turbulence pulls cold eddies off the bottom boundary layer.

\subsection{Maximum temperatures, mean temperatures,\\and asymmetry}
\label{sec: IH1 dT}

The mean fluid temperature, $\dT$, behaves in IH1 convection much like it does in IH3 convection, assuming its maximum value in the static state and falling as $R$ is raised. On the other hand, the mean temperature change between the boundaries, $\DT$, differs completely between the two configurations. Whereas in IH3 $\DT$ behaves rather like $\dT$, in IH1 it is identically zero. The role that $\DT$ plays in IH3 is instead approximated in IH1 by $\oT_{max}$, the maximum value that $\oT(z)$ assumes over the layer. Whereas $\DT$ equals the mean upward conduction across the entire layer, $\oT_{max}$ captures the mean \emph{outward} conduction, as described in \S\ref{sec: N def}. The quantity $\oT_{max}$ has been reported in many studies of IH1 since it is easier to estimate in the laboratory than $\dT$. However, $\oT_{max}$ does not arise as easily as $\dT$ in analytical expressions.

In addition to $\dT$ and $\oT_{max}$, IH1 convection is naturally characterized by the extent to which the flow creates asymmetry between upward and downward heat fluxes. This asymmetry can be simply conveyed by the mean fractions of produced heat leaving across the top or bottom boundaries---$\fT$ or $\fB$. As described in \S\ref{sec: additional constraints}, these fractions are related to the dimensionless convective flux, $\wT$, by
\begin{align}
\fT&=\tfrac{1}{2}+\wT & \fB&=\tfrac{1}{2}-\wT.
\end{align}
One can speak equivalently  in terms of $\wT$, $\fT$, or $\fB$. Here we focus on $\fB$ because it comes the closest to having a power-law dependence on $R$ as in the regimes studied.

Despite their simplicity and analytical attractiveness, neither $\dT$ nor $\fB$ has received much attention, although both quantities have been mentioned. Most authors have instead spoken in terms of top and bottom Nusselt numbers, here called $N_T$ and $N_B$. The most common definitions of these numbers are
\begin{align}
N_T &:= \frac{\fT}{\oT_{max}} & N_B &:= \frac{\fB}{\oT_{max}}.
\label{eq: NT NB}
\end{align}
The above expressions are not normalized to be unity in the static state; instead, both are equal to 4. Data on $N_T$ and $N_B$ are typically fit to powers of $R$.

To keep measures of asymmetry and temperature as separate as possible, we prefer not to examine $N_T$ and $N_B$. Instead, we use $\fB$ as a measure of asymmetry and use $\oT_{max}$ and $\dT$---or their inverses, $N$ and $\wt N$---as measures of temperature. One undesirable feature of $N_B$ is that it can initially drop below its static value as $R$ is raised since $\fB$ can initially fall faster than $\oT_{max}$. For instance, this occurs in the data of Kulacki \& Goldstein \cite{Kulacki1972}. Such behavior prevents $N_B$ from being well fit by a power of $R$ near onset, and it is unlike the behavior of the RB Nusselt number, which cannot be smaller than its static value. Another disadvantage of using $N_T$ and $N_B$ is that their $R$-dependence differs only in regimes where $\fB$ is changing significantly. If the decay of $\fB$ stops, as in the 2D simulations of Goluskin \& Spiegel \cite{Goluskin2012}, then $N_T$ and $N_B$ will both be dominated by the scaling of $1/\oT_{max}$, and slight changes in the asymmetry will not be captured well.

We would like to summarize past data on $R$-dependence with fits of the form
\begin{align}
\oT_{max}&\sim a\,R^{-\alpha} & \dT&\sim b\,R^{-\beta} & \fB&\sim e\,R^{-\epsilon}.
\label{eq: IH1 fits}
\end{align}
Fits of the above form have been reported for all three quantities in \cite{Goluskin2015b} and for $\dT$ in \cite{Goluskin2012}. In two other studies where the original data are available to us \cite{Kulacki1972, Worner1997}, we have calculated fits to $\oT_{max}$ and $\fB$. For the remaining studies, only fits to $N_T$ and $N_B$ are available. In these cases, we use the relations
\begin{align}
\oT_{max} &= \dfrac{1}{N_T+N_B} & \fB &= \dfrac{N_B}{N_T+N_B}.
\label{eq: conversions}
\end{align}
The reported power-law fits to $N_T$ and $N_B$ define curves for $\oT_{max}$ and $\fB$ that are not pure powers of $R$, so we have re-fit pure power laws to the latter curves.

\def\arraystretch{1.3}
\begin{sidewaystable}
\begin{center}
\vspace{20pt}
\begin{tabular}{rC{45pt}C{80pt}R{50pt}R{55pt}R{55pt}R{55pt}R{55pt}}
& $Pr$ & $R$ & $\oT_{max}$ fit & $N$ fit &
	$\dT$ fit & $\wt N$ fit & $\fB$ fit \\
\hline
\multicolumn{1}{l}{Laboratory experiments}\\
\hline
Kulacki \& Goldstein \cite{Kulacki1972} &
	$5.7-6.3$ & $\hspace{18.5pt}R_L-2.4\e7$ & $1.71\,R^{-0.180}$ & $0.0958\,Ra^{0.219}$ &
	& & $1.21\,R^{-0.0848}$ \\
Jahn \& Reineke \cite{Jahn1974, Mayinger1975} &
	$\approx 7$ & $10^5-10^9$ & $1.96\,R^{-0.194}$ & $0.0778\,Ra^{0.240}$ &
	& & $1.36\,R^{ -0.0988}$ \\
Ralph \emph{et al.}\ \cite{Ralph1977} &
	$6-7$ & $\hspace{3pt}3.7\e8-1.1\e{12}$ & $5.39\,R^{-0.224}$ & $0.0191\,Ra^{0.289}$ &
	& & $0.692\,R^{-0.0494}$ \\
Lee \emph{et al.}\ \cite{Lee2007} &
	$0.71-0.74$ & $1.1\e{10}-3.7\e{11}$ & $3.86\,R^{-0.209}$ & $0.0315\,Ra^{0.264}$ &
	& & $2.48\,R^{-0.0947}$ \\
\hline
Simulations (3D DNS)\\
\hline
W\"orner \emph{et al.} \cite{Worner1997} &
	$7$ & $10^5-10^8$ & $1.86\,R^{-0.186}$ & $0.0847\,Ra^{0.229}$ & 
	& & $1.16\,R^{-0.0845}$ \\
Goluskin \& van der Poel \cite{Goluskin2015b} &
	$1$ & $\hspace{3pt}5\e7-2\e{10}$ & $1.62\,R^{-0.217}$ & $0.0379\,Ra^{0.277}$ &
	$1.11\,R^{-0.204}$ & $0.0386\,\wt{Ra}^{0.256}$ & $0.803\,R^{-0.0554}$ \\
\hline
Simulations (2D DNS)\\
\hline
Jahn \& Reineke \cite{Jahn1974, Mayinger1975} &
	$7$ & $1\e5-1\e9$ & $2.20\,R^{-0.192}$ & $0.0678\,Ra^{0.238}$ &
	& & $1.19\,R^{-0.0854}$ \\
\rowcellR{Peckover \& Hutchinson \cite{Peckover1974}\\(free-slip, steady)} &
	$8$ & $5.1\e4-1.4\e6$ & $ 0.575\,R^{-0.104}$ & $0.182\,Ra^{0.116}$ & 
	& & $0.953\,R^{-0.0752}$ \\
\rowcellR{Straus \cite{Straus1976}\\(free-slip, steady)} &
	$\infty$ & $\hspace{9pt}10^5-3\e5$ & $1.82\,R^{-0.217}$ & $0.0795\,Ra^{0.277}$ &
	& & $1.23\,R^{-0.100}\hspace{3pt}$ \\
Emara \& Kulacki \cite{Emara1980} &
	$6.5$ & $5\e4-5\e7$ & $1.96\,R^{-0.186}$ & $0.0795\,Ra^{0.229}$ &
	& & $1.02\,R^{-0.0672}$ \\
Goluskin \& Spiegel \cite{Goluskin2012} &
	$1$ & $\hspace{12pt}10^8-2\e{10}$ & & & 
	$1.13\,R^{-0.200}$ & $0.0384\,\wt{Ra}^{0.250}$
\end{tabular}
\end{center}
\caption{Summary of IH1 experiments and simulations reporting approximate power-law dependence of $\oT_{max}$, $\DT$, or $\fB$ on $R$. Internal heating was achieved by heating elements in one laboratory experiment \cite{Lee2007} and by electric current in the others. Fits to $\oT_{max}$ and $\fB$ are computed directly from data for a few studies \cite{Kulacki1972, Worner1997, Goluskin2012, Goluskin2015b}, while for other studies we have computed them from reported fits to $N_T$ and $N_B$ (see text). Simulations employ no-slip boundary conditions, except when specified otherwise.}
\label{tab: IH1 fits}
\end{sidewaystable}
\def\arraystretch{1}

Table \ref{tab: IH1 fits} summarizes power-law fits to the $R$-dependence of $\oT_{max}$, $\dT$, and $\fB$. The fits to $\oT_{max}$ are also stated in terms of $N$ and $Ra$, and the fits to $\dT$ are also stated in terms of $\wt N$ and $\wt{Ra}$. For the IH1 configuration, we have defined these diagnostic quantities in \S\S\ref{sec: Nusselt numbers}-\ref{sec: diagnostic Rayleigh numbers} as
\begin{align}
N &= 1/8\,\oT_{max} & Ra &= R/N \\
\wt N &= 1/12\,\dT & \wt{Ra} &= R/\wt N.
\label{eq: IH1 N and Ra}
\end{align}
The fits (\ref{eq: IH1 fits}) to $\oT_{max}$ and $\dT$ imply fits to and $N$ and $\wt N$ of the form (\ref{eq: N fits}), where $\gamma=\alpha/(1-\alpha)$, $\delta=\beta/(1-\beta)$, $c=(8a)^{-1/(1-\alpha)}$, and $d=(12b)^{-1/(1-\beta)}$.

\subsubsection{Maximum temperatures}

Fits of the form $\oT_{max}\sim aR^{-\alpha}$ are shown in Table \ref{tab: IH1 fits}. In all laboratory experiments and all simulations with no-slip boundaries, the exponent $\alpha$ lies between 0.180 and 0.224. This means that the dimensional maximum temperature, $\oT_{max}\hspace{1pt}\Delta$, grows with the volumetric heating at rates between $H^{0.776}$ and $H^{0.820}$. In the sole study for which both $\oT_{max}$ and $\dT$ are reported \cite{Goluskin2015b}, the decay of $\oT_{max}$ is slightly faster than the decay of $\dT$, the fit exponents being $\alpha=0.217$ and $\beta=0.204$, respectively. This makes sense since $\oT_{max}$ initially must `catch up' to $\dT$ as the temperature profile flattens. When the $\oT_{max}$ fits are restated in the form $N\sim c\,Ra^\gamma$, the exponents range from $\gamma=0.220$ to $\gamma=0.289$. The bottom end of this range is smaller than any exponents found for the ordinary RB1 Rayleigh number, except at very small $Pr$ \cite{Grossmann2000}.

\subsubsection{Mean temperatures}

The quantity $\dT$ has been reported only in two numerical studies, and each gives a fit of the form $\dT\sim b\,R^{-\beta}$ for $Pr=1$. Despite one study being 3D and the other 2D, the growth rates of $\dT$ with $R$ are very similar, having exponents of $\beta=0.204$ in 3D \cite{Goluskin2015b} and $\beta=0.200$ in 2D \cite{Goluskin2012}. This is reminiscent of the Nusselt number in RB convection, which is not much affected by dimensionality unless $Pr$ is small \cite{Schmalzl2004, VanderPoel2013}.

The dimensional mean temperature, $\dT\hspace{-1pt}\Delta$, grows with the volumetric heating proportionally to $H^{0.796}$ in 3D and to $H^{0.800}$ in 2D. When the $\dT$ fits are restated in the form $\wt N\sim d\wt{Ra}^\delta$, the exponents are $\delta=0.256$ in 3D and $\delta=0.250$ in 2D. These $\delta$ values are within the range of Nusselt number growth rates seen in RB1 convection, though they are at the lower end of that range (cf.\ \S\ref{sec: IH3 DT}). We cannot yet draw comparison with IH3 convection, for which no data on $\dT$ have been reported.

\subsubsection{Asymmetry}
\label{sec: asymmetry}

The asymmetry between upward and downward heat fluxes in IH1 convection, as quantified by the fraction of heat that flows downward, $\fB$, seems to have no analogues in our other five IH or RB configurations. First, this fraction changes with $R$ much more slowly than any other integral quantity we have discussed. Second, the $R$-dependence of $\fB$ can differ greatly between 2D and 3D, even when $Pr$ is not small. This is because shear, rather than buoyancy, is the mechanism responsible for mixing the cooler lower boundary layer with the warmer interior. The asymmetry is generally greater in 3D than in 2D because the shear-driven mixing, which helps heat escape across the bottom boundary, is less effective in 3D \cite{Goluskin2015b}.

A particularly simply question without an obvious answer is: as $R\to\infty$, what is the limit of $\fB$? The extreme possibilities of either 0 or 1/2 seem most likely, although intermediate values are also plausible. In the highest-$R$ simulation data available in 3D, $\fB$ falls monotonically as R is raised \cite{Goluskin2015b}. The highest-$R$ data available in 2D are quite different, except perhaps at large $Pr$ \cite{Goluskin2012}. For instance, in the 2D simulations of Goluskin \& Spiegel \cite{Goluskin2012} with $Pr=1$, the fraction $\fB$ reaches a minimum of 0.33 near $R=10^9$ and then increases as $R$ is raised further. This non-monotonic $R$-dependence in 2D is yet another way that $\fB$ stands apart from other quantities we have considered.

When $\fB$ decreases monotonically as $R$ is raised, as in all past 3D studies and some 2D ones, we can seek fits of the form $\fB\sim eR^{-\epsilon}$. Table \ref{tab: IH1 fits} summarizes these fits, all of whose decay rates are quite small. The decay exponents range from $\epsilon=0.0494$ to $\epsilon=0.0988$. It remains a mystery whether such decay will continue or reverse at larger~$R$

The dependence of $\fB$ on $Pr$ has been examined in two studies \cite{Goluskin2012, Goluskin2015b}. The value of $\fB$ seems to fall monotonically as $Pr$ is raised, meaning that the asymmetry increases, until saturating at large $Pr$. The effect of $Pr$ on the asymmetry is fairly strong---stronger than its effect on $\dT$ or $\oT_{max}$.

\subsection{Scaling arguments}
\label{sec: scaling arguments}

Several scaling arguments have been put forth to explain the parameter-dependence of mean temperatures in IH convection \cite{Cheung1977, Cheung1980, Cheung1987, Goluskin2012}. In RB convection, the Nusselt number displays a wide diversity of scaling behavior in different regions of parameter space \cite{Stevens2013}. It is likely that the same is true of mean temperatures in IH convection since (inverses of) these temperatures have many parallels to the RB Nusselt number. This remains to be confirmed by a wider exploration of parameter space. If such a diversity of scaling can indeed be found in IH convection, then any broadly applicable scaling arguments must reflect this. For the standard RB1 configuration, the only arguments that attempt to capture the full range of scaling behavior are those put forth by Grossman \& Lohse in \cite{Grossmann2000} and subsequent papers (see \cite{Stevens2013}). The arguments of \cite{Grossmann2000} carry through analogously for IH convection \cite{Goluskin2012}. When the predicted scalings are phrased in terms of $N$ and $Ra$, or $\wt N$ and $\wt{Ra}$, they are the same as the scalings predicted for the Nusselt number in the RB1 case. However, further work on scaling arguments is perhaps premature until data are available across a wider swath of parameter space.

\section{Future directions}
\label{sec: future}

In the future study of IH convection, the main task accessible to mathematical analysis is proving parameter-dependent bounds on key integral quantities. The only results of this kind are the $R$-dependent lower bounds on volume-averaged temperatures described in \S\ref{sec: bounds}. We have conjectured in \S\ref{sec: wT and T} that there should also exist $R$-dependent upper bounds on the mean convective flux, $\wT$. These would amount to lower bounds on the fraction of heat flowing downward in IH1 and on the mean temperature difference between the boundaries in IH2 and IH3. Bounds are lacking also for the maximum horizontally averaged temperature, $\oT_{max}$, that has often been measured in IH1 experiments. Bounds depending analytically on the Prandtl number are highly desirable as well.

There is much fertile ground for physical and computational experiments on IH convection. This is especially true for computation since most prior results are several decades old, so modern computers would be able to probe unexplored parameter regimes with relative ease. Neither the IH2 nor IH3 configuration has been simulated in 3D, and the DNS carried out in 2D has not approached the large $R$ that are now computationally accessible. The IH1 configuration has been the subject of two DNS studies in 3D \cite{Worner1997, Goluskin2015b}, but a much wider exploration of parameter space is called for. The asymmetry between upward and downward heat fluxes in IH1 is particularly hard to predict; even its value as $R$ approaches infinity is not certain. In each of the three IH configurations, simulating a wide range of $R$ and $Pr$ would produce a more global picture of how key integral quantities depend on the control parameters. The complicated parameter-dependence of Nusselt numbers in RB convection \cite{Ahlers2009, Stevens2013} suggests that fitting integral quantities to pure powers of $R$ will not suffice.

A combination of mathematical analysis, simulation, and physical experimenta- tion will lead to a better understanding of the three internally heated configurations we have studied in this SpringerBrief. We hope that this, in turn, will lead to a better understanding of more complicated occurrences of IH convection. The many past studies of RB convection should prove useful in guiding future studies of IH convection, and to this end we have described a number of analogies between the two classes of flows. Still, the analogies are not perfect, and some consequences of internal heating cannot be foreseen. Judging by the complexity of RB convection, we expect that these novel aspects of IH convection will remain rich areas of inquiry for many more years.

\backmatter

\addcontentsline{toc}{chapter}{References}
\renewcommand\bibname{References}
\bibliographystyle{style}
\bibliography{book.bbl}

\end{document}